\documentclass[
superscriptaddress,
twocolumn,
preprintnumbers,
nofootinbib,
 amsmath,amssymb,
 aps,prd,
floatfix,
]{revtex4-2}

\usepackage{booktabs}
\usepackage{xspace}

\usepackage{aas_macros}
\usepackage{xcolor}
\usepackage{graphicx}
\usepackage{dcolumn}
\usepackage{bm}
\usepackage[colorlinks,linkcolor=blue,citecolor=blue,urlcolor=blue ]{hyperref}
\usepackage[normalem]{ulem}

\usepackage{newtxtext,newtxmath}

\usepackage{lineno}

\newcommand{\zph}{z_{\rm ph}\xspace}
\newcommand{\zeff}{z_{\rm eff}\xspace}
\newcommand{\lcdm}{$\Lambda$CDM\xspace}
\newcommand{\wcdm}{$w$CDM\xspace}
\newcommand{\nucdm}{$\nu \Lambda$CDM\xspace}
\newcommand{\kcdm}{$k\Lambda$CDM\xspace}
\newcommand{\ok}{\ensuremath{\Omega_k}\xspace}
\newcommand{\om}{\ensuremath{\Omega_{\rm m}}\xspace}

\newcommand{\olambda}{\ensuremath{\Omega_\Lambda}\xspace}
\newcommand{\ob}{\ensuremath{\Omega_{\rm b}}\xspace}

\newcommand{\tu}{\ensuremath{t_{\rm U}}\xspace}
\newcommand{\hubble}{\ensuremath{H_0}\xspace}
\newcommand{\kmsMpc}{km s$^{-1}$Mpc$^{-1}$}
\newcommand{\hunit}{\xspace km\,s$^{-1}$Mpc$^{-1}$\xspace}
\newcommand{\appendixcite}[1]{\hyperref[#1]{Appendix \ref*{#1}}}

\newcommand{\thetastar}{\ensuremath{\theta_ \star}\xspace}
\newcommand{\wacdm}{$w_0$$w_a$CDM\xspace}

\newcommand{\shoes}{{SH0ES}\xspace}
\newcommand{\planck}{\textit{Planck}\xspace}

\newcommand{\neutrinomass}{\ensuremath{\sum m_\nu}\xspace}
\newcommand{\ev}{${\rm eV}$\xspace}

\usepackage{array}
\newcolumntype{Z}{>{\setbox0=\hbox\bgroup}c<{\egroup}@{}}

\newcommand{\leftparbox}[2]{\parbox{#1}{\begin{flushleft} #2 \end{flushleft}}}
\newcommand{\twoonesig}[4][\pbwidth]{
\begin{equation}
\left.
 \begin{aligned}
#2 \\ #3
 \end{aligned}
\ \right\} \ \ \mbox{\text{\leftparbox{#1}{#4}}}
\end{equation}
}
\newcommand{\twoonesignonum}[4][\pbwidth]{
\begin{equation*}
\left.
 \begin{aligned}
#2 \\ #3
 \end{aligned}
\ \right\} \ \ \mbox{\text{\leftparbox{#1}{#4}}}
\end{equation*}
}

\begin{document}

\preprint{DES-2024-0849}
\preprint{FERMILAB-PUB-25-0127-PPD}

\title{Dark Energy Survey: implications for cosmological expansion models from the final DES Baryon Acoustic Oscillation and Supernova data}



\author{T.~M.~C.~Abbott}
\affiliation{Cerro Tololo Inter-American Observatory, NSF's National Optical-Infrared Astronomy Research Laboratory, Casilla 603, La Serena, Chile}

\author{M.~Acevedo}
\affiliation{Department of Physics, Duke University Durham, NC 27708, USA}

\author{M.~Adamow}
\affiliation{Center for Astrophysical Surveys, National Center for Supercomputing Applications, 1205 West Clark St., Urbana, IL 61801, USA}

\author{M.~Aguena}
\affiliation{Laborat\'orio Interinstitucional de e-Astronomia - LIneA, Av. Pastor Martin Luther King Jr, 126 Del Castilho, Nova Am\'erica Offices, Torre 3000/sala 817 CEP: 20765-000, Brazil}

\author{A.~Alarcon}
\affiliation{Institute of Space Sciences (ICE, CSIC),  Campus UAB, Carrer de Can Magrans, s/n,  08193 Barcelona, Spain}

\author{S.~Allam}
\affiliation{Fermi National Accelerator Laboratory, P. O. Box 500, Batavia, IL 60510, USA}

\author{O.~Alves}
\affiliation{Department of Physics, University of Michigan, Ann Arbor, MI 48109, USA}

\author{F.~Andrade-Oliveira}
\affiliation{Physik-Institut, University of Zürich, Winterthurerstrasse 190, CH-8057 Zürich, Switzerland}

\author{J.~Annis}
\affiliation{Fermi National Accelerator Laboratory, P. O. Box 500, Batavia, IL 60510, USA}

\author{P.~Armstrong}
\affiliation{The Research School of Astronomy and Astrophysics, Australian National University, ACT 2601, Australia}

% ...

\collaboration{DES Collaboration}\email{des-publication-queries@fnal.gov}


\date{\today}

\begin{abstract}
The Dark Energy Survey (DES) recently released the final results of its two principal probes of the expansion history: Type Ia Supernovae (SNe) and Baryonic Acoustic Oscillations (BAO). In this paper, we explore the cosmological implications of these data in combination with external Cosmic Microwave Background (CMB), Big Bang Nucleosynthesis (BBN), and age-of-the-Universe information. The BAO measurement, which is $\sim2\sigma$ away from \planck's \lcdm predictions, pushes for low values of \om compared to \planck, in contrast to SN which prefers a higher value than \planck. We identify several tensions among datasets in the \lcdm model that cannot be resolved by including either curvature (\kcdm) or a constant dark energy equation of state (\wcdm). By combining BAO+SN+CMB despite these mild tensions, we obtain \ok=$-5.5^{+4.6}_{-4.2}\times10^{-3}$ in \kcdm, and $w=-0.948^{+0.028}_{-0.027}$ in \wcdm. In \wcdm, BAO and SN push again in different directions of parameter space, favoring, respectively $w<-1$ and $w>-1$.
If we open the parameter space to \wacdm (where the equation of state of dark energy varies as $w(a)=w_0+(1-a)w_a$), all the datasets are mutually more compatible, and we find concordance in the $[w_0>-1,w_a<0]$ quadrant, with BAO pushing for $w_a<0$ and SN for $[w_0>-1,w_a<0]$. 
For DES BAO and SN in combination with \planck-CMB, we find a $3.2\sigma$ deviation from \lcdm, with $w_0=-0.673^{+0.098}_{-0.097}$, $w_a = -1.37^{+0.51}_{-0.50}$, a Hubble constant of $\hubble=67.81^{+0.96}_{-0.86}$\hunit, and an abundance of matter of \om$=0.3109^{+0.0086}_{-0.0099}$. For the combination of all the background cosmological probes considered (including CMB's angular acoustic scale \thetastar), we still find a deviation of $2.8\sigma$ from \lcdm in the $w_0-w_a$ plane. 
Assuming a minimal neutrino mass, this work provides tentative evidence for non-\lcdm physics, which is consistent with recent claims in support of evolving dark energy, or a source of unknown systematics.
\end{abstract}

\maketitle




\section{Introduction}
\label{sec:introduction}

The Dark Energy Survey (DES) was designed as a multi-probe experiment to constrain properties of the dark energy and other cosmological parameters~\cite{DES05,DESoverview,Flaugher,DES_all_probes_Y1}. 
For that, it observed the DES wide field, nearly 5,000 deg$^2$ of the southern sky, over six years (2013-2019) with 5 filters ($grizY$) to a depth of $i=23.8$~\cite{DESDR2}. In parallel, the DES supernova program (DES-SN) was repeatedly observed with a 5/6 night cadence of ten $\sim$3 deg$^2$ fields for five $\sim$6 month seasons. 

Two analyses of the main cosmology probes from the final DES dataset were recently published: the Baryonic Acoustic Oscillations (BAO) and type Ia Supernova (SNe~Ia) analyses. DES measured the angular BAO feature from the clustering of $\sim$16 million galaxies to determine the ratio of the angular distance to the sound horizon with a precision of $2.1\%$ at $\zeff = 0.85$
~\cite{des-y6-bao,des-y6-bao-sample}. Additionally, DES measured the luminosity distance to redshift relation from $1,635$ high-redshift photometrically classified SNe~Ia, hereafter referred to as DES-SN5YR \cite{des-y5-sn}. That paper along with~\cite{Camilleri24b} explored in detail the implications of DES SNe for the cosmological model. 
In the near future, DES will release results from other principal cosmological probes using the final dataset, such as the combination of galaxy clustering and weak lensing, galaxy cluster number counts, and cross-correlation with external datasets. For many of those probes, DES work based on previous data releases are considered the state-of-the-art \cite{y3-cosmicshear1, y3-cosmicshear2, y3-galaxyclustering, y1-clusters-kp, y3-2x2ptaltlensresults,y3-2x2ptredmagic,y3-3x2ptkp,y3-3x2ptkp_ext}.
Both SN and BAO represent measurements of the homogeneous properties of the Universe and are only sensitive to the expansion history (as opposed to properties that probe the distribution and evolution of inhomogeneities). We refer to these probes as measurements of the cosmological background.
On the other hand, the remaining DES probes will be sensitive to spatial perturbations, and thus to the history of growth of structure in the Universe. 

In the last year (2024), a lot of attention has been paid to tensions of current data with the standard cosmological model, (flat) \lcdm,  at the background level, in particular when allowing the equation of state of dark energy ($w$) to evolve. Typically, this is reported via the CPL parametrization (named for the authors of~~\cite{CPL1,CPL2}), where the equation of state is assumed to evolve linearly with the scale factor $a$: 
$w(a)=w_0 + w_a (1-a)$.
First, DES-SN5YR reported a $\sim$2$\sigma$  deviation in the $w_0$-$w_a$ parameter plane~\cite{des-y5-sn}. Furthermore, the DES-Y6-BAO measurement of  $D_M/r_d$ at $\zeff=0.85$ showed a 2.1$\sigma$ deviation from the value predicted by \planck in the (flat) \lcdm model, though that paper did not evaluate the impact on cosmological parameters. 
Finally, the DESI 2024 BAO results showed a tension with \lcdm \cite{desi2024vi}: when DESI BAO~\cite{desi2024iii,desi2024iv} is combined with DES SN and Planck CMB, the reported tension in the $w_0-w_a$ plane increases to $3.9\sigma$ with \lcdm.
The significance of the deviation from \lcdm remains at a similar level when considering DESI's full-shape analysis \cite{desi2024vii}, which includes information from the growth of structure.   

In this context, this paper has several goals. First, we aim to understand how the $2.1\sigma$ deviation of the DES-Y6-BAO translates to cosmological parameters in \lcdm and simple extensions (\kcdm, \wcdm, \nucdm). Second, we demonstrate the power of combining the two DES background probes, BAO and SN, as well as what insight they together can contribute to the current investigations of evolving dark energy.\footnote{A recent paper, \cite{Notari24}, also explored this combination. The two main differences with that analysis are that we do not include CMB-lensing (see \autoref{sec:CMB})
and we consider a baseline metric for deviations from \lcdm that is sensible for non-Gaussian likelihoods, as opposed to that based on $\Delta \chi^2$, as discussed in \autoref{sec:deviations}.
In addition, we explore many probe combinations (beyond the baseline BAO+SN+CMB) and several expansion history models.} For this, we combine and compare our data with other background probes such as Big Bang Nucleosynthesis (BBN), direct $\hubble$ measurements from \shoes, the age of the Universe, and the angular scale of the sound horizon, \thetastar, as seen by \planck. We will also combine our data with primary CMB probes (temperature and polarization: TT+TE+EE).
The CMB is the only probe considered here that contains information beyond the background expansion level. 
External probes that are more sensitive to nonlinear structure growth, such as CMB-lensing and redshift space distortions, are not considered in this work and will be compared in the future to other DES growth of structure probes (cosmic shear, galaxy clustering, cluster counts, etc).

The paper is organized as follows. We describe the different probes considered in \autoref{sec:data}. The methodology for cosmology inferences in described in \autoref{sec:methods}. We then present the results in \autoref{sec:results}. Discussion and conclusion are presented in \autoref{sec:discussion} and \autoref{sec:conclusions}, respectively.

\section{Data}
\label{sec:data}

\begin{figure}[!h]
    \centering
    \includegraphics[width=\linewidth]{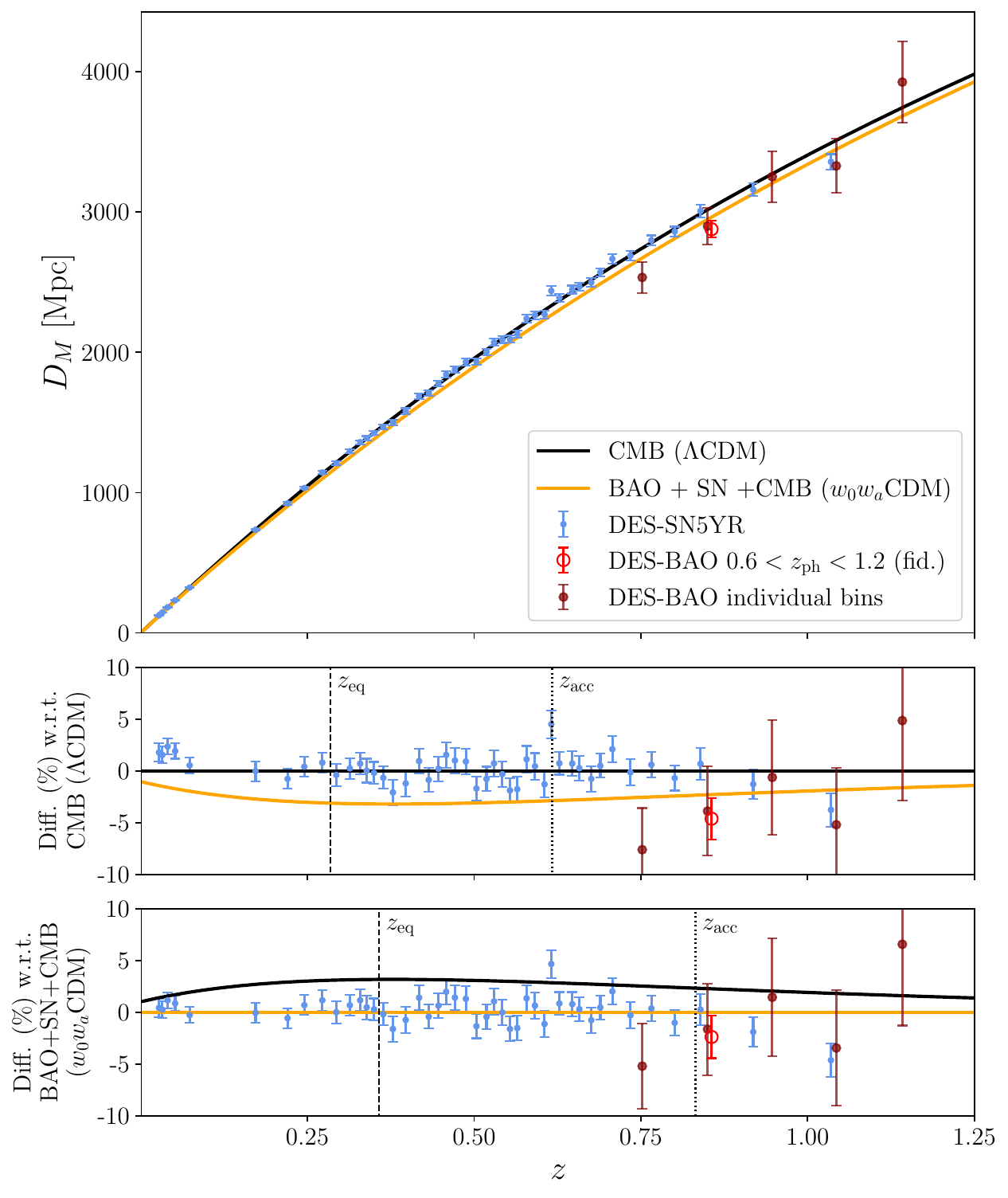}
    \caption{ 
    Illustration of the distance-redshift relation from DES compared to the best-fit CMB-\lcdm and BAO+SN+CMB-\wacdm predictions.
    We show the (comoving) angular distance $D_M(z)$ from DES BAO results both from our fiducial single-bin measurement (evaluated at an effective redshift of $\zeff=0.85$ by fitting all the data in $0.6<\zph<1.2$), but also the alternative 5-bin split measurements (with individual bins of $\Delta \zph=0.1$). In this figure, for BAO, we assume a value of  $r_d = 147.46$ Mpc (see \autoref{sec:rd}). We also show the SN binned results, for which we plot the luminosity distance transformed to angular distance using \autoref{eq:distance_equivalence} for the 1829 SNe in DES-SN5YR (1635 DES SNe + 194 low-$z$ SNe from external samples). To obtain the SN distances we calibrate the SN absolute magnitude $\mathcal{M}$ such that the residuals with respect to the given cosmology average to zero. This calibration is different by $\delta \mathcal{M} \sim 0.06$ for the two cosmologies and hence we show in the lower panels two residual plots, one with SNe calibrated to CMB-only \lcdm (this calibration is also used in the upper panel) and the other calibrated to the best-fit \wacdm model for the BAO+SN+CMB data combination. The residual plot shows the percentage difference in $D_M(z)$ compared to the best fit. The 1829 SNe are binned with equal numbers in each bin (with the $D_M$ and $z$ shown being the average weighted by the inverse variance of $D_M$). 
    The \wacdm model fits better the $z<0.1$ SNe and the $z\gtrsim0.75$ BAO and SN data. 
    We also include two vertical lines to indicate the redshift of matter-dark energy equality ($z_{\rm eq}$, dashed) and the redshift when acceleration starts ($z_{\rm acc}$, dotted) for each the two models.
    This figure illustrates in a simplified way how BAO and SN together constrain the expansion history models, however, in our analyses both $\mathcal{M}$ and $r_d$ are varied.
    }
    \label{fig:distance-redshift}
\end{figure}

\subsection{DES Y6 BAO}

The Dark Energy Survey released the Baryonic Acoustic Oscillations analysis from the final dataset (Y6, which spans six years of data) in~\cite{des-y6-bao}, which builds upon the Y1~\cite{des-y1-bao} and Y3 analyses~\cite{y3-baokp}. For that work, we built a BAO-optimized sample in the redshift range $0.6\lesssim z \lesssim1.2$, where DES could obtain the most competitive constraints~\cite{des-y6-bao-sample}. The BAO sample was defined using a red and bright selection and is made up of $\sim$16 million galaxies from a $\sim$4,300 deg$^2$ area. This sample was split into 6 tomographic bins based on photometric redshift $\zph$\footnote{$\zph$ is the main redshift estimate of the DNF photometric redshift algorithm.} with width $\Delta \zph=0.1$. The redshift distribution of each bin is estimated using a combination of Directional Neighboring Fitting (DNF~\cite{DNF}), clustering redshifts~\cite{cawthon22}, and direct calibration with spectra from VIPERS~\cite{vipers}. Spurious density correlations with foregrounds and observing conditions are corrected with linear weights via the Iterative Systematic Decontamination (ISD) method~\cite{y1_lss_sys_kp,y3-galaxyclustering}. More details about the BAO sample and its calibration can be found in~\cite{des-y6-bao-sample}, which uses a  methodology based on that from Y1~\cite{DESY1baosample} and Y3~\cite{y3-baosample}.

\subsubsection{Fiducial BAO: angular BAO in $0.6<\zph<1.2$ (BAO)}

The DES Y6 BAO analysis measured the angular BAO feature from three different estimators of galaxy clustering: the angular correlation function (ACF or $w(\theta)$, \cite{des-y1-acf}), the angular power spectrum (APS or $C_\ell$, \cite{Camacho-des-y1-aps}) and the projected correlation function (PCF or $\xi_p(s_\perp)$, \cite{Ross-des-y1-pcf,Chan21-pcf-method,PCF_Y3_BAO}). In all three cases, we analyzed the 6 tomographic bins simultaneously by fitting a single angular BAO shift parameter
\begin{equation}
    \alpha=(D_M/r_d)/(D_M/r_d)_{\rm fid} \;\;.
\end{equation}
The methods were validated against $\sim$2,000 ICE-COLA mock catalogs (built using the methodology from~\cite{ice-cola,cola,DESY1baomocks,y3-baomocks}) and checked for robustness against variations in assumptions about the galaxy redshift distributions. From this validation, we obtained a small contribution of systematic error that was added in quadrature to the final statistical error bars. 

The angular BAO shift parameter $\alpha$ was measured in a blind analysis: its inferred value was concealed and only revealed after performing a series of tests. These tests included showing that the $\alpha$ inference was robust to partial data removal (e.g., redshift splits), method variations (e.g., scale choices), different data calibration choices (e.g., redshifts, systematic weights) and between the three summary statistics (ACF, APS, PCF and their combination).

Finally, the likelihood of the angular BAO shift from the three estimators (ACF, APS and PCF) is combined into a single likelihood for $\alpha$, which is close to a Gaussian approximation yielding
\begin{equation}
    D_M(\zeff=0.85)/r_d = 19.51\pm0.41 \, .
    \label{eq:BAO}
\end{equation}
For the purposes of this paper, we employ the full likelihood reported in~\cite{des-y6-bao}, in the form of an interpolated table of $\chi^2$ versus $\alpha$. All results using this likelihood are labeled as ``BAO''. All the data products associated to this paper are publicly available.\footnote{\url{https://des.ncsa.illinois.edu/releases/y6a2/Y6bao}}

This is our fiducial BAO constraint and is represented in red in \autoref{fig:distance-redshift} at the effective redshift of our sample, $\zeff=0.85$. This value is found to be $2.1\sigma$ and $4.3\%$ below the prediction of $ D_M(\zeff=0.85)/r_d =20.39$ given by the \planck \lcdm best fit, and represented by the black line in \autoref{fig:distance-redshift}.

\subsubsection{Alternative BAO: individual $\Delta \zph=0.1$ tomographic bins (BAO-5)}


As a supplement to our main BAO analyses, we also consider an alternative BAO likelihood with shift parameters $\alpha$ estimated separately for each of the individual $\Delta \zph=0.1$ tomographic bins. This builds on~\cite{des-y6-bao}, in which these binned $\alpha$ values were determined.  
Although each of the individual bins will have a lower signal-to-noise ratio than their combination, having several measurements of the distance-redshift relation could potentially resolve features in the expansion history not revealed by a single point. This motivates us to investigate the possibility in this paper, though these are not considered fiducial results because they were not subject to the same level of validation as the single-bin BAO studies.

More specifically, the alternative version of the BAO likelihood is based on the ACF in the thin $\Delta \zph =0.1$ bins. The methodology to obtain this likelihood is explained in detail in \appendixcite{app:5bins-method}. There, we compute the systematic errors (from modeling and redshifts) associated with each redshift bin and the covariance of their $\alpha$ measurements, since they overlap slightly in redshift. We do not consider the APS and PCF methods in this case, since they are found to be less robust for individual redshift bins. The resulting likelihood consists of five angular BAO points represented in orange in \autoref{fig:distance-redshift} with a correlation matrix shown in \autoref{fig:cov_plot}. There are five rather than six measurements because the lowest redshift bin ($0.6<\zph<0.7$) does not show a detection of BAO. As studied in~\cite{des-y6-bao}, given the lower SNR of the individual bins, finding one non-detection is not unexpected. From our $\sim$2000 simulations, we found that $>25\%$ of them show at least one bin with a non-detection.  

We find that the redshift-binned $\alpha$ measurements fluctuate with redshift, with some apparent trend with redshift. Although these were found to be compatible with statistical fluctuations, in order to test whether they could be hinting at a feature in the expansion history of the Universe, we ran a few analyses substituting our fiducial BAO likelihood ($0.6<\zph<1.2$) with the alternative individual-bin BAO, labeled as BAO-5. As we found a negligible impact on the cosmological parameters, we will not discuss these individual-bin BAO measurements in the main text, but we do report the results in \appendixcite{app:5bins-cosmo}.

\subsection{DES-SN5YR (SN)}
\label{sec:SN}

The Dark Energy Survey released the Hubble diagram from the sample of SNe Ia discovered and measured during the full five years of the DES-SN program in~\cite{des-y5-sn}.
The DES-SN survey 
detected over 30,000 SN candidates over 5 years of observations and made a classification based only on photometry. From these, 1,635 
were deemed high-quality SNe Ia-like, had spectroscopically measured redshifts from their host galaxies, and are included in the Hubble diagram with a weight inversely proportional to their probability of being a Type Ia SN, as given by the machine-learning classifier SuperNNova~\citep{SuperNNova,moller22,vincenzi24}. 
The Hubble diagram also includes 194 spectroscopically confirmed low-$z$ ($z < 0.1$) SNe Ia from external surveys. For the cosmological analyses in this paper, these two subsamples are used together, simply labeled as SN, and are shown in redshift bins as blue points in \autoref{fig:distance-redshift}. 

A series of papers describe the details of the DES-SN5YR analysis, itself built upon the interim DES-SN3YR analysis of spectroscopically confirmed SNe \cite{DES-SN3YR}. The processing and calibration of DES SN light curves are described by \cite{brout22,Sanchez24}. SN light-curve fitting and standardization as well as the estimation of the final SN distance moduli are presented by \cite{taylor23,vincenzi24}.
Host galaxy properties are presented in \cite{Kelsey23} based on deep coadd imaging of the SN fields \cite{wiseman20}. Finally, \cite{vincenzi24} 
provides a detailed description of all the sources of systematics included in the final DES-SN5YR uncertainty covariance matrix. 

 In the DES-SN5YR analysis, SN standardized distances
$\mu_\mathrm{obs}$ are estimated as 

\begin{equation}
     \mu_\mathrm{obs} = m_x + \alpha\ x_1 - \beta\ c + \gamma\ G_{\rm host} (M_\star) - M - \Delta\mu_\mathrm{bias},
     \label{eq:SN}
 \end{equation} 
where $m_x$,\ $x_1$\ and $c$ are the fitted amplitude, stretch and color of each SN. $\alpha$, $\beta$ and $\gamma$ are global nuisance parameters. $\Delta\mu_{\rm bias}$ are distance corrections applied to take into account the effects of sample selection, calibrated as a function of the properties of the SN and host galaxy. $G_{\rm host}$ is an environmental adjustment taking the form of a step function depending on the stellar mass of the SN host, $M_\star$. These parameters are calibrated and validated in the characterization of the SN sample in the studies cited above.
Finally, the SN fiducial absolute magnitude $M$ is degenerate with the Hubble constant $\hubble$ and is combined into a single parameter $\mathcal{M} = M + 5\log_{10}{(c/\hubble)}$, which is (analytically) marginalized over in the SN likelihood \cite{des-y5-sn}.

The SN distances $\mu_\mathrm{obs}$ and associated statistical and systematic covariance matrices are publicly available.\footnote{\url{https://github.com/des-science/DES-SN5YR}}

\subsection{\planck CMB}
\label{sec:CMB}

\subsubsection{Temperature and polarization anisotropy (CMB)}

We incorporate measurements of the CMB temperature and polarization anisotropies using the \planck 2018 likelihood~\cite{plancklike}, which we will subsequently refer to using the label ``CMB''. Specifically, for temperature and polarization spectra for $\ell\geq 30$, we employ the {\tt Plik-lite} likelihood, which incorporates the effects of marginalizing over \planck foreground and nuisance parameters and includes measurements of spectra up to $\ell_{\rm max}=2508$ for TT, and $\ell_{\rm max}=1996$ for TE and EE. Following the standard \planck analysis, at low multipoles ($2\leq\ell<30$ ) we use the {\tt Commander} likelihood for the TT spectrum and the {\tt SimAll} likelihood for the EE polarization spectrum. We do not include CMB lensing constraints. 

When including this CMB likelihood, we fit several additional parameters compared to our background-only studies. These include $A_\mathrm{s}$ and $n_\mathrm{s}$, the amplitude and slope of the primordial power curvature spectrum, as well as the optical depth $\tau$. We additionally marginalize over the total \planck calibration $a_{\rm \planck}$ as a nuisance parameter.  The priors used for these parameters can be found in Table~\ref{tab:params}.  
\footnote{Note that this implementation of the CMB likelihood is slightly different than what was used in the DES-SN5YR analysis of~\cite{des-y5-sn}, which uses a python implementation of the \planck likelihood described in~\cite{planck-lite-python}. At $\ell\geq 30$ that implementation is identical to what is used in this paper, but it differs at low-$\ell$. The DES-SN5YR analysis uses a Gaussian approximation for low-$\ell$ CMB temperature and includes a Gaussian prior on $\tau$ instead of low-$\ell$ polarization data. While we expect these choices to have a negligible impact on our conclusions, they may induce slight differences between SN+CMB constraints reported in this paper and in~\cite{des-y5-sn}.}

\subsubsection{Angular scale of the acoustic peak (\thetastar)}
\label{sec:thetastar}
In order to isolate geometric/background information from the CMB, in some cases we instead consider a constraint on 
\begin{equation}
    \thetastar = r_s(z_\star)/D_M(z_\star) \;\;,
\end{equation}
the ratio between the baryon-photon sound horizon and the comoving distance at the redshift of recombination, 
$z_\star$. We incorporate this via a Gaussian likelihood taken from the same \planck 2018 temperature and polarization data described above, \cite{Planck}, having
\begin{equation}
100\,\thetastar=1.04109\pm 0.00030\, .
\label{eq:thetastar}
\end{equation}
For ease of comparison, we note that the $\thetastar $ likelihood used in DESI analyses~\cite{desi2024vi} has a nearly identical mean based on \planck 2018 constraints which include lensing, but DESI additionally increased this width by 75\% to account for possible modeling uncertainties.

\subsubsection{Comoving scale of the acoustic peak ($r_d$)}
\label{sec:rd}

Along with constraints on the parameters of specific cosmological models, we will consider a cosmographic expansion to measure $\hubble$ (\autoref{sec:cosmographic}). For this study, we incorporate a measurement of the comoving scale of the acoustic peak, $r_d$. This quantity is the maximum distance that sound waves could travel in the early Universe before photons and baryons decoupled from each other. It depends on the baryon density and total matter density in the early Universe. 

When this constraint is applied, we use a Gaussian prior on the sound horizon given by 

\begin{equation}
    r_d = 147.46 \pm 0.28\ {\rm Mpc}\ .
    \label{eq:rd_value}
\end{equation}

This value was determined using chains from~\cite{Lemos23}, which are based on \planck PR4 data and incorporate information from the CMB in a way that removes late-time cosmology dependence associated with the late-Integrated Sachs-Wolfe effect, the optical depth to reionization, CMB lensing and foregrounds. 

\subsection{Age of the Universe (\boldmath \tu)}
\label{sec:age}
To further inform our \wacdm analysis, particularly for understanding how its additional degrees of freedom affect the expansion history, we additionally consider a prior on the age of the Universe, $t_{\rm U}$, inferred from observations of globular clusters. 
The age of the Universe, $t_{\rm U}$, has been historically used to provide a prior on cosmological parameters, primarily through its inverse proportionality to $\hubble$, although there is dependence on other cosmological parameters (see \autoref{eq:age}).
It is interesting to revisit this in light of new data and extended cosmological models. We use a result derived from the color-magnitude diagrams of globular clusters in the Milky Way. 

Globular clusters are metal-poor stellar systems that formed early in the Universe. Historically, their ages (by definition younger than the age of the Universe) were determined by examining the main sequence turnoff and comparing it to results from stellar modeling codes \citep{Haselgrove1956}. Systematics arise from uncertainty in stellar modeling codes (especially in relation to convection), extinction, metallicity, blending effects, uncertain distances to the clusters, nuclear reaction rates, and whether they arise from single or multiple populations. 

Modern methodologies attempt to control for these systematics by fitting the entire color-magnitude diagram in a Bayesian framework, with marginalization over parameters describing some of the systematics \citep{Valcin2020}. In this way, a simultaneous determination is made of age, metallicity, distance and absorption to each cluster, which may be checked for consistency with other data (for example, parallax-derived distances to the cluster). Whether the cluster contains a single or multiple populations (which in principle could have different ages), may be checked by examining the posterior distribution of the metallicity of the cluster as populations originating at different times may have different metallicities. 

We use the results of \cite{Valcin2020}, who combined the posteriors of the ages of the 38 most metal-poor clusters (presumed to be the oldest). This was convolved with a reasonable prior probability density of its formation time, derived from assuming the clusters formed at redshift $z>11$ \cite{Jimenez2019}. Although there is some cosmological dependence implicit in the conversion from formation redshift to formation time, it is small compared to other sources of error, and therefore it may be used to constrain non-standard cosmologies in addition to \lcdm. We therefore use the Gaussian prior given by
\begin{equation}
t_{\rm U} = 13.5 \pm 0.52\ {\rm Gyr} \, ,
\label{eq:tu_value}
\end{equation}
where the error combines statistical and systematic sources, with the highest error contribution arising from uncertain nuclear reaction rates. This is compatible with the \planck estimate of $t_{\rm U} = 13.797 \pm 0.023$ Gyr within \lcdm (Table 2 of \cite{Planck}).  
When implementing it, we impose this as a Gaussian prior as a post-processing of the equivalent chains without this prior. We describe the methodology further in \autoref{sec:inference}.

\subsection{Big Bang Nucleosynthesis (BBN)}
\label{sec:BBN}

Big Bang Nucleosynthesis (BBN) theory predicts the abundance of light elements in the early Universe, such as deuterium and helium, as well as their relation to the baryon-to-photon ratio. Therefore, the observational determination of the primordial deuterium abundance and the helium fraction can be used to compute the physical baryon density parameter at present, $\ob h^2$, where $h\equiv \hubble / (100\ {\rm km}\ {\rm s}^{-1}{\rm Mpc}^{-1})$ (see~\cite{cooke2018one,aver2022comprehensive} for further details).

The resulting constraints on $\ob h^2$ depend on the modeling of underlying nuclear interactions. We employ a BBN constraint from a recent analysis~\cite{schoneberg20242024} that recalculates the predictions while marginalizing over uncertainties in reaction rates. It reports a conservative constraint of
\begin{equation}
    \ob h^2=0.02218 \pm 0.00055 \, .
\end{equation}
This is the same  BBN Gaussian prior on $\ob h^2$ used for the cosmological inference from the BAO measurement in the DESI 2024 analysis~\cite{desi2024vi}.

\subsection{\boldmath{\bf Direct  measurement of the Hubble constant (direct \hubble)}}
\label{sec:sH0es}

The \shoes collaboration has recently used observations from the Hubble Space Telescope (HST) of Cepheid variable stars in the host
galaxies of 42 type Ia SNe to calibrate the Hubble constant. This analysis is described in~\cite{riess2022comprehensive}, where they find 
\begin{equation}
    \hubble = 73.04 \pm 1.04\ {\rm km\, s}^{-1}\, {\rm Mpc}^{-1} \, 
\end{equation}
as their baseline result. 
Here, we do not include this value in any of our analyses, but we rather consider its comparison to our derived $\hubble$ values on several occasions throughout the paper.

There is an ongoing debate about the value of \hubble. In particular, direct calibration methods such as the one quoted above tend to prefer a higher value than those using CMB or other calibrations from the early Universe (). The measurement shown above is typically regarded as the consensus direct-\hubble measurement. A recent review on this topic can be found in \cite{Verde24}.

\section{Methods}   
\label{sec:methods}

\subsection{Background evolution}
\label{sec:background}

The main focus of this paper is to characterize the constraints of DES probes of the expansion history of the Universe: BAO and SN.  We will often refer to these as {\it background} probes because they measure the background expansion upon which all other cosmological processes rest. 
Here we briefly review the basic concepts around these observables and the expansion history of the Universe. 

We consider four different models for the late-time expansion of the Universe. The first is the standard model, \lcdm, which assumes flatness and a constant dark energy density given by $\olambda = (1-\om)$. Second, we consider the \kcdm model, with free curvature given by $\ok$. Third, the \wcdm model includes a constant dark energy equation of state $w$ as a free parameter, assuming flat geometry.  
Finally, we study the 
\wacdm model, where the dark energy equation of state evolves linearly with the scale factor, 
%
\begin{equation}
    w(a)=w_0 + w_a (1-a)\, .
\end{equation}
Each of these models has a different parametrization of the late-time expansion history of the Universe as given by the Friedmann equation: \footnote{We are neglecting in this equation the radiation term $\Omega_r a^{-4}$, which is only relevant for the early Universe. However, we do include radiation in calculations for the CMB constraints.}

\begin{equation}
\frac{H(a)^2}{\hubble^{\ 2}}= 
\left \{  \begin{array}{ll}
\om a^{-3} + (1-\om) & \hspace{-2.3cm}\mbox{for \lcdm}\\[0.2cm]
\om a^{-3} + (1-\om)a^{-3(1+w)} & \hspace{-2.3cm}\mbox{for \wcdm}\\[0.2cm]
\om a^{-3} + (1-\om)a^{-3(1+w_0+w_a)}e^{-3w_a(1-a)} & \\[0.1cm]
\hspace{4.5cm}\mbox{for \wacdm} & \\[0.2cm]
\om a^{-3} + (1-\om -\ok)+\ok a^{-2} & \\[0.1cm]
\hspace{4.5cm}
\mbox{for \kcdm}\\[0.2cm]
\end{array} \right .
\label{eq:Hz}
\end{equation}
with $\hubble$ the Hubble constant. The parenthesis in the second term represents the density parameter of dark energy (\olambda in \lcdm and \kcdm), but this is not a free parameter, since we always need to ensure $\sum \Omega_i =1$.

We additionally consider constraints on a flat \lcdm cosmology with a non-minimal sum of neutrino masses, \nucdm. As was recently highlighted in~\cite{loverde_weiner_neutrinos}, while cosmological neutrinos are more commonly discussed in terms of how they impact structure formation, background probes like those we consider here are also sensitive to their mass. This is because the mass of neutrinos determines when they become non-relativistic, and thus when they contribute to \autoref{eq:Hz} as matter as opposed to radiation. Relatedly, the fact that neutrinos are relativistic at the time of recombination means that CMB constraints on matter density are primarily sensitive to CDM and baryons, while BAO and SN probe the total matter density --- including neutrinos. Thus, together the CMB and probes of late-time expansion provide complementary information about neutrino mass. For the other cosmological models described above, we model a single massive neutrino species with its mass set to 0.06 eV, the minimum allowed by neutrino oscillation experiments. For \nucdm,  we model neutrinos with three degenerate mass species, varying the sum of their masses \neutrinomass as a free parameter.

The two key probes considered here rely on the distance-redshift relation, which depends on an integral over \autoref{eq:Hz}.
The BAO feature serves as a standard ruler that constrains the comoving angular distance, which is given by:
\begin{equation}
    D_M = \frac{c}{\hubble \sqrt{\lvert \ok\lvert }} S_k\left[ \int \sqrt{\lvert \ok \lvert } \frac{dz}{H(z)/\hubble} \right],
    \label{eq:Dm}
\end{equation}
with $S_{k=0}[x]\equiv x$, $S_{k<0}[x] \equiv {\rm sin}[x]$, and $S_{k>0}[x]\equiv {\rm sinh}[x]$ where $k$ is the sign of \ok.

BAO observations constrain the ratio between $D_M$ and the sound horizon $r_d$ as follows.
Baryonic Acoustic Oscillations 
are generated by sound waves in the early Universe that propagate in the photo-baryon plasma until they decouple at the drag epoch, $z_d$. This leaves a preferred scale in the distribution of matter in the Universe, given by the sound horizon at that epoch: 
\begin{equation}
    r_d =  \int_{z_d}^{\infty} \frac{c_s(z;\ob h^2)}{H(z)}dz 
    \label{eq:rs}, 
\end{equation}
where $c_s$ is the sound speed. 
The fact that the sound speed depends on the physical density of baryons,  $\ob h^2$ means it will be interesting to combine BAO information with BBN constraints (\autoref{sec:BBN}).
Complementary information can also be provided by the acoustic peak as detected in the CMB, which is sensitive to the sound horizon at a slightly different epoch (see \autoref{sec:thetastar}). 

SNe Ia are standardizable candles, which constrain the luminosity distance, related to  angular distance by: 
\begin{equation}
    D_L(z) = (1+z)D_M(z).    \label{eq:distance_equivalence}
\end{equation}
This is typically transformed to the distance modulus $\mu(z) = 5 {\rm log}_{10}\big(D_L(z)/10{\rm pc}\big)$, which appears in \autoref{eq:SN}. One thing to bear in mind is that we do not know {\it a priori} the actual absolute magnitude of SNe Ia, hence this quantity ($\mathcal{M}$ in \autoref{eq:SN}) is fully degenerate with $\hubble$ in \autoref{eq:Hz}.

When using the priors on the age of the Universe (\autoref{sec:age}), from the definition of $H(t) = {\dot a}/a$, we compute 
\begin{equation} 
t_{\rm U}= \int_0^1 \frac{da}{a H(a) }\;.
\label{eq:age}
\end{equation}
Considering this expression, we note that the $w_0w_a$CDM model is a parameterization that allows one to test whether time-varying dark energy is a better fit than a cosmological constant but is unlikely to be a true model down to $a\to0$. 
However, since the impact on the age is largest at later times when the matter density is low, this calculation remains a useful framework in which to constrain the evolution of the expansion history $H(t)$ that would be too extreme in terms of observed stellar ages.

\subsection{Parameter inference}
\label{sec:inference}

\begin{table}
        \renewcommand{\arraystretch}{1.2} 
        \setlength{\tabcolsep}{15pt} 
	\centering	
	\caption{	\label{tab:params}
	Sampled parameters and priors used in the  \lcdm, \kcdm, \wcdm, \wacdm, and \nucdm analyses. When including  CMB data, we additionally vary the parameters listed in the bottom section. Square brackets denote a flat prior, while parentheses denote a Gaussian prior of the form $\mathcal{N}(\mu,\sigma)$, with $\mu$ and $\sigma$ being the mean and standard deviation, respectively. The parameter \neutrinomass is fixed to 0.06 \ev for all models other than \nucdm. 
    } 
	\vspace{0.2cm}

    \begin{tabular}{cZc}
        \toprule\toprule
        Parameter & Fiducial & Prior \\\hline
    
        \multicolumn{3}{c}{\bf \boldmath\lcdm} \\
        $\hubble$ [\kmsMpc] & 69 & [55, 91] \\
        $\om$ & 0.3 & [0.1, 0.9] \\ 
        $\ob$ & 0.048 & [0.03, 0.07] \\
        \hline
        
        \multicolumn{3}{c}{\bf \boldmath 
 \kcdm} \\
        $\ok$ & 0 & [-0.25, 0.25] \\
        \hline
    
        \multicolumn{3}{c}{\bf \boldmath\wcdm} \\
        $w$ & -1.0 & [-3, -0.33] \\
        \hline
    
        \multicolumn{3}{c}{\bf \boldmath\wacdm} \\
        $w_0$ & -1.0 & [-3, -0.33] \\
        $w_a$ & 0.0 & [-3, 3] \\
        \hline
    
        \multicolumn{3}{c}{$\boldsymbol{\nu \Lambda}$\textbf{CDM}} \\
        $\neutrinomass [eV]$ & 0.06 & [0, 1] \\
        \hline\hline
        \multicolumn{3}{c}{\textbf{Chains that include CMB}} \\
        $\tau$ & 0.067 & [0.04, 0.15]\\
        $A_\mathrm{s} \times 10^{9}$ & 2.19 & [0.5, 5.0] \\ 
        $n_{\rm s}$ & 0.97 & [0.87, 1.07] \\
        $a_{\rm Planck}$ & 1.0 & (1.0, 0.0025) \\
        \bottomrule\bottomrule
    \end{tabular}

\end{table}

To infer constraints on the parameters $\mathbf{p}$ given the data $\mathbf{D}$, we construct a posterior probability distribution following the Bayes' Theorem:
\begin{equation}
    P(\mathbf{p}\, |\, \mathbf{D}, M) \propto \mathcal{L}(\mathbf{D}\,|\, \mathbf{p}, M )P(\mathbf{p}\,|\,M),
    \label{eq: posterior}
\end{equation}
where $M$ is the assumed theoretical model, $P(\mathbf{p}\,|\,M)$ is a prior probability distribution on the parameters, and $\mathcal{L}$ is the likelihood function of the parameters given the data. The proportionality constant in \autoref{eq: posterior} is given by the inverse of the Bayesian Evidence
\begin{equation}
    P(\mathbf{D}\,|\,M) = \int \mathrm{d}\mathbf{p} \mathcal{L}(\mathbf{D}\,|\, \mathbf{p}, M )P(\mathbf{p}\,|\,M).
    \label{eq:evidence}
\end{equation}
Under the Gaussian likelihood approximation, we can define $\mathcal{L}$ as
\begin{equation}
    \mathcal{L}(\mathbf{D}\,|\, \mathbf{p}, M ) \propto  e^{-\frac{1}{2}\chi^2},
\end{equation}
where $\chi^2$ is the goodness of fit of the model $M$ to the data $\mathbf{D}$, given the data covariance. Throughout this work, we use the $\chi^2$ or likelihood functions publicly released from each data set described in \autoref{sec:data}. We do not assume any correlation between likelihood functions from different data sets. 

The list of parameters sampled and the priors assumed for them is included in \autoref{tab:params}. In addition to the priors listed, we impose the condition $w_0+w_a<0$ when considering the $w_0w_a$CDM model. This prior ensures $w(a)<0$ at all redshifts, avoiding the parameter space for which there is no radiation domination era.

To implement a prior for the age of the Universe $t_{\rm U}$  (see Section \ref{sec:age}), we exploit the fact that the prior $P$ and the likelihood appear in combination in Equations \ref{eq: posterior} and \ref{eq:evidence}. We may therefore implement the age prior by either multiplying $P$ or the likelihood  $\mathcal{L}$ by $P(t_{\rm U}) \sim \mathcal{N} (13.5, 0.52)$ Gyr (it is also required to re-normalize $P$ such that $\int \mathrm{d} \mathbf{p}\,P(\mathbf{p}\,|\,M) = 1$). It is computationally convenient to adopt the former for posterior estimation (by re-weighting chains) and the latter for evidence calculation (by adjusting the likelihood).

We sample the posterior distributions of the parameters using two different Monte Carlo nested samplers, \textsc{Polychord} \cite{polychord2} and \textsc{Nautilus} \cite{nautilus}, finding equivalent constraints with the two methods and using the former by default. 
We use the implementation of these two samplers available in the \textsc{CosmoSIS}\footnote{\url{https://cosmosis.readthedocs.io/}} framework \cite{cosmosis}, which we use as our main inference pipeline throughout this work.\footnote{With the exception of \autoref{sec:cosmographic}, which uses the methodology described in \cite{Camilleri24}.} We use the CAMB Boltzmann solver \cite{camb1,camb2} to compute the underlying background quantities and its \textsc{Halofit} Takahashi implementation for the nonlinear matter power spectrum \cite{halofit-takahashi,halofit-bird} when including the CMB likelihood. For \kcdm chains including the CMB likelihood, since CAMB calculations can become significantly slower for $\ok\neq 0 $ we initially run the \textsc{Nautilus} sampler using lower-resolution CAMB settings, then importance sample the result with our fiducial pipeline to obtain a final posterior estimate. 

When providing constraints, we report the mean in each parameter and use the \textsc{GetDist}\footnote{\url{https://github.com/cmbant/getdist}} package \cite{getdist} to obtain equal-posterior credible regions (c.~r.) and to plot the posterior distributions. The procedure for reporting credible regions is the following. 

We examine the 68\% credible regions and use their distance from the mean to determine (potentially asymmetric) $1\sigma$ errors. We then determine whether those bounds are close to the prior boundary, with closeness defined by assessing whether the distance between the boundary and the 68\% credible region is smaller than that $1\sigma$ error bar. There are three different scenarios:
\begin{enumerate}
    \item \textbf{both bounds are far from the prior boundaries:} 
    If the $2\sigma$ region does not overlap with the prior boundary, we report two-sided errors.
    \item \textbf{one bound is close to a prior boundary:} 
    If one of the $2\sigma$ bounds is close to the prior boundary, we report a one-sided 95\% bound.
    \item \textbf{both bounds are close to the prior boundaries:} 
    If both $2\sigma$ bounds are close to the prior boundaries, we report no constraint.
\end{enumerate}

\begin{table*}[]
    \setlength{\tabcolsep}{8pt} 
    \centering
    \begin{tabular}{l c c c}
        \midrule
          & \multicolumn{3}{c}{Deviations from {\bf \boldmath\lcdm} ($\sigma$)} \\ 
        \midrule
        Dataset & {\bf \boldmath\kcdm} & {\bf \boldmath\wcdm}  & {\bf \boldmath\wacdm}  \\
        \midrule
        
        BAO + SN + BBN  & 1.4 &  1.4 & 1.8 \\
        BAO + SN + BBN + \tu & -- & -- & 2.0  (2.7)  \\
        BAO + SN + $\theta_\star$  & 2.5 & 2.7  & 2.3 \\
        BAO + SN + $\theta_\star$ + BBN  & 2.8 & {3.1}   & 2.8 \\
        BAO + SN + $\theta_\star$ + BBN  + \tu & -- & --  & 2.9 (2.8) \\
        SN & 1.3 & 1.6 & 2.0 \\
        CMB & { 3.0} & 1.7  & 2.5 \\
        SN + CMB & 2.9 & 2.0  & 2.2 \\
        BAO + CMB & 0.6 & 2.8  & { 3.4} \\
        {\bf BAO + SN + CMB} & 1.2 & 1.8  & {\bf 3.2} \\
        \bottomrule
    \end{tabular}
    \caption{
    Statistical significance, in $\sigma$s, of deviations from \lcdm based on shifts in the additional parameter(s) in the extended model: \ok in \kcdm, $w$ in \wcdm and $\{w_0,w_a\}$ in \wacdm.  
    See the methodology described in \autoref{sec:deviations}. For the case of \tu priors, we consider the fiducial Gaussian prior case and, in parenthesis, the case where only a lower bound on \tu is set. We highlight in bold the case for BAO+SN+CMB, our most constraining combination, which in the \wacdm model shows a $3.2\sigma$ deviation from \lcdm.  
    }
    \label{tab:deviations}
\end{table*}

\begin{table*}[]
    \setlength{\tabcolsep}{12pt} 
    \centering
    \begin{tabular}{l c c c}
        \midrule
          & \multicolumn{3}{c}{$\Delta\chi^2$ improvement compared to {\bf \boldmath\lcdm} } \\ 
        \midrule
        Dataset & {\bf \boldmath\kcdm} & {\bf \boldmath\wcdm}  & {\bf \boldmath\wacdm}  \\
        \midrule
        BAO + SN + BBN  & 1.0 \:(1.0) &  1.6 \:(1.3)  & 5.8 \:(1.9)\\
        BAO + SN + $\theta_\star$  & 2.9 \:(1.7) &  3.8 \:(2.0) & 5.1 \:(1.8)\\
        BAO + SN + $\theta_\star$ + BBN \quad & 9.3 \:(3.1) & 10.4 \:(3.2) & 10.9 \:(2.9)\\
        SN & 1.0 \:(1.0) & 1.6 \:(1.3) &  5.9 \:(1.9) \\
        CMB & 8.9 \:(3.0) & 3.3 \:(1.8) & 4.2 \:(1.5) \\
        SN + CMB & 9.0 \:(3.0) &  3.7 \:(1.9) &  7.8 \:(2.3) \\
        BAO + CMB & 0.9 \:(0.9) & 8.4 \:(2.9) & 8.7 \:(2.5) \\
        {\bf BAO + SN + CMB} & 1.6 \:(1.3) &  3.5 \:(1.9) &  {\bf 12.1 \:(3.0)} \\
        \bottomrule
    \end{tabular}
    \caption{ 
    Improvement in goodness-of-fit from freeing additional model parameters computed via the difference between the minimum $\chi^2$ estimated for \lcdm and that for each extended model. Positive values indicate an improved fit in the extended model. Numbers in parentheses indicate the statistical significance in $\sigma$s assuming a Gaussian approximation for the posterior, which may not be accurate for less constraining data combinations.
    }
    \label{tab:chi2}

\end{table*}

\begin{table*}[]
    \setlength{\tabcolsep}{8pt} 
    \centering
    \begin{tabular}{l c c c c c}
        \midrule
          & \multicolumn{5}{c}{Tension ($\sigma$)} \\ 
        \midrule
        Datasets & {\bf \boldmath\lcdm} & {\bf \boldmath\kcdm} & {\bf \boldmath\wcdm} & {\bf \boldmath\wacdm} & {\bf \boldmath\nucdm} \\
        \midrule
        BAO vs SN & 0.5 & 0.0 & 0.0 & 0.3 & 0.2\\
        CMB vs SN & 1.7 & 1.5 & 1.3 & 1.1 & 1.2\\
        CMB vs BAO & 2.0 & 3.2 & 0.6 & 0.1 & 2.0\\
        \midrule
        SN vs BAO + \thetastar & 2.4 & - & - & - & - \\
        CMB vs BAO + SN + BBN & 2.2 &  3.3 & 2.2 & 1.2 & -  \\
        SN vs BAO + BBN  & 0.4 & - & - & - & - \\
        SN vs BAO + BBN + \thetastar & 2.9 & 0.5 & 0.0 & 0.9 & 2.6 \\
        BAO + CMB vs SN & 2.1 & 1.5 & 2.5 & 1.6 & 2.1 \\
        CMB vs BAO + SN + BBN + \tu & 1.5 (0.8)  & - & - &  0.9 (0.9)  & -\\
        \bottomrule
    \end{tabular}
    \caption{Tensions, in $\sigma$s, among independent (combinations of) probes for a given model. See the methodology described in \autoref{sec:tension_probes}. For the case of \tu priors, we consider the fiducial Gaussian prior case and, in parenthesis, the case where only a lower bound on \tu is set. We note that these tensions are reported in the whole parameter space unlike deviations in \autoref{tab:deviations}, which refer to the parameter additional to \lcdm. } 
    \label{tab:tensions}
\end{table*}

\subsection{Tension metrics}
\label{sec:tensions}

Given the recent debate about possible evidence favoring dynamical dark energy over a cosmological constant, in this work we will be especially interested in computing  (1) deviations from the reference model, \lcdm, and (2) tension between datasets. The methods to quantify these are laid out below. 

\subsubsection{Quantifying deviations from \lcdm}
\label{sec:deviations}

To quantify preferences for an extended model relative to \lcdm we compare constraints on cosmological parameters.
To do so, we compute the probability of a shift in the alternative model's added cosmological parameters relative to their corresponding \lcdm values. 
This probability is defined as:
\begin{equation}
    \Delta(D, M) \equiv \int_{P(\mathbf{p}\, |\, D, M)>P(\mathbf{p}^*\, |\, D, M)} P(\mathbf{p}\, |\, D, M)  \text{d} \mathbf{p}\,,
    \label{eq:DeviationsFromLCDM}
\end{equation}
where $\mathbf{p}$ represents the additional parameters of the model $M$ with respect to \lcdm (e.g., $w_0$ and $w_a$ in \wacdm), and $\mathbf{p}^*$ denotes the \lcdm values of these parameters (e.g. $w_0=-1$, $w_a=0$). 
This integral quantifies the posterior mass exceeding the iso-density contour defined by the \lcdm posterior value, $P(\mathbf{p}^*\, |\, D, M)$. 
Note that if the extra parameters have flat priors, as it is in the cases considered here,
this result is parameter invariant.

To compute the integral in \autoref{eq:DeviationsFromLCDM}, we use the Kernel Density Estimate (KDE) method described in~\cite{tensions}. In the remaining part of this section, we use the shorthand notation $P(\mathbf{p}) \equiv P(\mathbf{p}\, |\, D, M)$. Since we have posterior samples from $P$, the integral in \autoref{eq:DeviationsFromLCDM} can be estimated as a Monte Carlo (MC) volume integral:
\begin{equation} 
\hat{\Delta} = \frac{1}{\sum_{i=1}^{n} w_i} \sum_{i=1}^{n} w_i S( \hat{P}(\mathbf{p}_i)-\hat{P}(\mathbf{p}^*_i) ),
\label{Eq:KDEMCMCParamShiftProbability}
\end{equation}
where $S(x)$ is the Heaviside step function, equal to unity for $x > 0$ and zero otherwise, $n$ is the number of weighted samples $\mathbf{p}_i$ from $P$, and $\hat{P}$ represents their KDE estimates. Given that we are analyzing one- or two-dimensional posterior distributions, KDE evaluations can be efficiently performed using Fast Fourier Transforms (FFT) on a discrete grid, making the computation effectively instantaneous.

We always report results as the effective number of standard deviations.
Given an event of probability $\Delta$, it is given by ~\cite{PhysRevD.99.043506}:
\begin{equation} 
n_\sigma \equiv \sqrt{2}\,{\rm Erf}^{-1}(\Delta)\,.
\label{Eq:EffectiveSigmas}
\end{equation}
This corresponds to the number of standard deviations that an event with the same probability would have had if it had been drawn from a Gaussian distribution.

In ~\autoref{tab:deviations} we present statistical deviations from \lcdm for different datasets and some combination of them at each extension model considered in the analysis. 

In interpreting these results, it is worth considering how this parameter shift metric differs from the primary model comparison statistics used in DES-SN5YR \cite{des-y5-sn}, which employs Bayesian evidence ratios, and in the DESI Y1 BAO analysis~\cite{desi2024vi}, which reports frequentist $\Delta\chi^2$ goodness-of-fit improvements. Compared to evidence ratios, our parameter shift metric is less directly sensitive to the choice of parameters' prior ranges; however, for weakly constrained posteriors, its reported significance may be impacted by how prior bounds (including in other parameter directions) shape the marginalized posteriors of the beyond-\lcdm parameters. In the limit of Gaussian posteriors, the significance of deviations reported based on parameter shifts and $\Delta\chi^2$ is equivalent, as explored e.g. in~\cite{lemosraveri_tensions}. In contrast to parameter shifts, the value of $\Delta\chi^2$ is less sensitive to the prior and related projection effects, though it is subject to uncertainty due to noise in the $\chi^2$-minimization procedure. Also, translating $\Delta\chi^2$ estimates to model comparison significances relies on assumptions of Gaussianity --- both of the likelihood in data space and of the posterior in parameter space --- which may not hold for all data combinations we consider.  

Since these metrics provide complementary information,  in \autoref{tab:chi2} we additionally report $\Delta\chi^2$ values for most of the same model and data combinations shown in \autoref{tab:deviations}. (Data combinations with $t_U$ are excluded because the method used to include that prior complicates the process of estimating $\chi^2$.) 
For each combination, we estimate the minimum $\chi^2$ by performing an ensemble of optimization searches launched from the 50 highest posterior samples from the associated chain. In parentheses, we also report the significance of these changes in goodness-of-fit using likelihood ratio tests and Wilks' theorem~\cite{Wilks1938}, 
\footnote{We compute the probability to exceed our $\Delta \chi^2$ for a $\chi^2$ PDF with the d.o.f. given by the number of beyond-\lcdm parameters, and transform this via \autoref{Eq:EffectiveSigmas} to deviations in terms $\sigma$ for a Gaussian PDF considering both tails.} though we caution that for less constraining data combinations (less Gaussian posteriors) the focus should be put on the actual improvement of fit ($\Delta \chi^2$).

A common pattern in \autoref{tab:chi2} is that for the most constraining combinations, there is a reasonable agreement on the significance of deviation to that reported in \autoref{tab:deviations} (up to $\sim 0.5\sigma$, well within the expected difference between methods; see \cite{lemosraveri_tensions}). As discussed above, for data combinations that are not particularly constraining, these methods are expected to differ, and both should be interpreted with caution. For most of the discussion in this paper, when quoting the significance of a deviation from \lcdm, we will refer to the parameter shift metric of \autoref{tab:deviations}, which is better suited for capturing non-Gaussian features of the posteriors.

\subsubsection{Tensions among probes in a given model}
\label{sec:tension_probes}

To make sure that a deviation from the \lcdm model is robust, we want to quantify the agreement of different probes within a given model before combining such datasets. To assess the consistency of parameter determinations from two posterior distributions, we calculate the probability of observing a parameter difference using the method described in ~\cite{tensions, PhysRevD.99.043506}.
We start by building the posterior distribution of parameter differences.
We consider each dataset, and in particular the two data sets denoted $1$ and $2$, to be independent. 
Under the assumption that the two-parameter sets that describe the datasets, $\mathbf{p}_1,\mathbf{p}_2$, differ, the joint distribution of their parameter determinations is given by the product of their posteriors:
\begin{equation}  
P(\mathbf{p}_1,\mathbf{p}_2|d_1, d_2) = P_1(\mathbf{p}_1|d_1)P_2(\mathbf{p}_2|d_2). 
\label{Eq:join.parameter.posterior}
\end{equation}
To compute the distribution of parameter differences, we change variables by defining $\Delta\mathbf{p} \equiv \mathbf{p}_1 - \mathbf{p}_2$, including all parameters shared by the two datasets. The distribution of $\Delta \mathbf{p}$ is obtained by marginalizing over one of the parameters:

\begin{equation}   P(\Delta \mathbf{p}) = \int P_{1}(\mathbf{p}) P_{2}(\mathbf{p}-\Delta \mathbf{p}) \text{d}\mathbf{p}. 
\label{Eq:param.diff.pdf} 
\end{equation}
The distribution of parameter differences, $P(\Delta \mathbf{p})$, provides insight into whether the parameter determinations from two datasets are consistent. 
Intuitively, if $P(\Delta \mathbf{p})$ has most of its support when $\Delta \mathbf{p}$ has large deviations from zero, the two parameter sets are incompatible, indicating a tension between the datasets.
To quantify the probability of a parameter shift, we calculate the same integral as in \autoref{eq:DeviationsFromLCDM} but for the parameter difference distribution defined in \autoref{Eq:param.diff.pdf}:
\begin{equation}  \Delta \equiv \int_{P(\Delta \mathbf{p}) > P(0)} P(\Delta \mathbf{p}) \, \text{d}\Delta \mathbf{p}, 
\label{Eq:ParamShiftProbabilityDifference} 
\end{equation}
which measures the posterior probability above the iso-density contour corresponding to no parameter shift ($\Delta \mathbf{p} = 0$).
Since the distribution of parameter differences, $P(\Delta \mathbf{p})$, is an $n$-dimensional distribution, with $n$ corresponding to the total number of parameters describing the assumed theoretical model, the integral in \autoref{Eq:ParamShiftProbabilityDifference} is computationally more expensive to estimate than \autoref{eq:DeviationsFromLCDM}. For this reason we use the machine-learning based method described in ~\cite{tensions}. The first step is to train a normalizing flow on samples from $P(\Delta \mathbf{p})$, then we evaluate the tension integral as a Monte Carlo integral analogously to \autoref{Eq:KDEMCMCParamShiftProbability}. We convert the probability of a shift into a number of standard deviations as in \autoref{Eq:EffectiveSigmas}.

In ~\autoref{tab:tensions} we show tension results for pairs of independent datasets in each cosmological model considered in this work.

\section{Results}

\label{sec:results}

\begin{figure*}
    \centering
    \includegraphics[width=\linewidth]
    {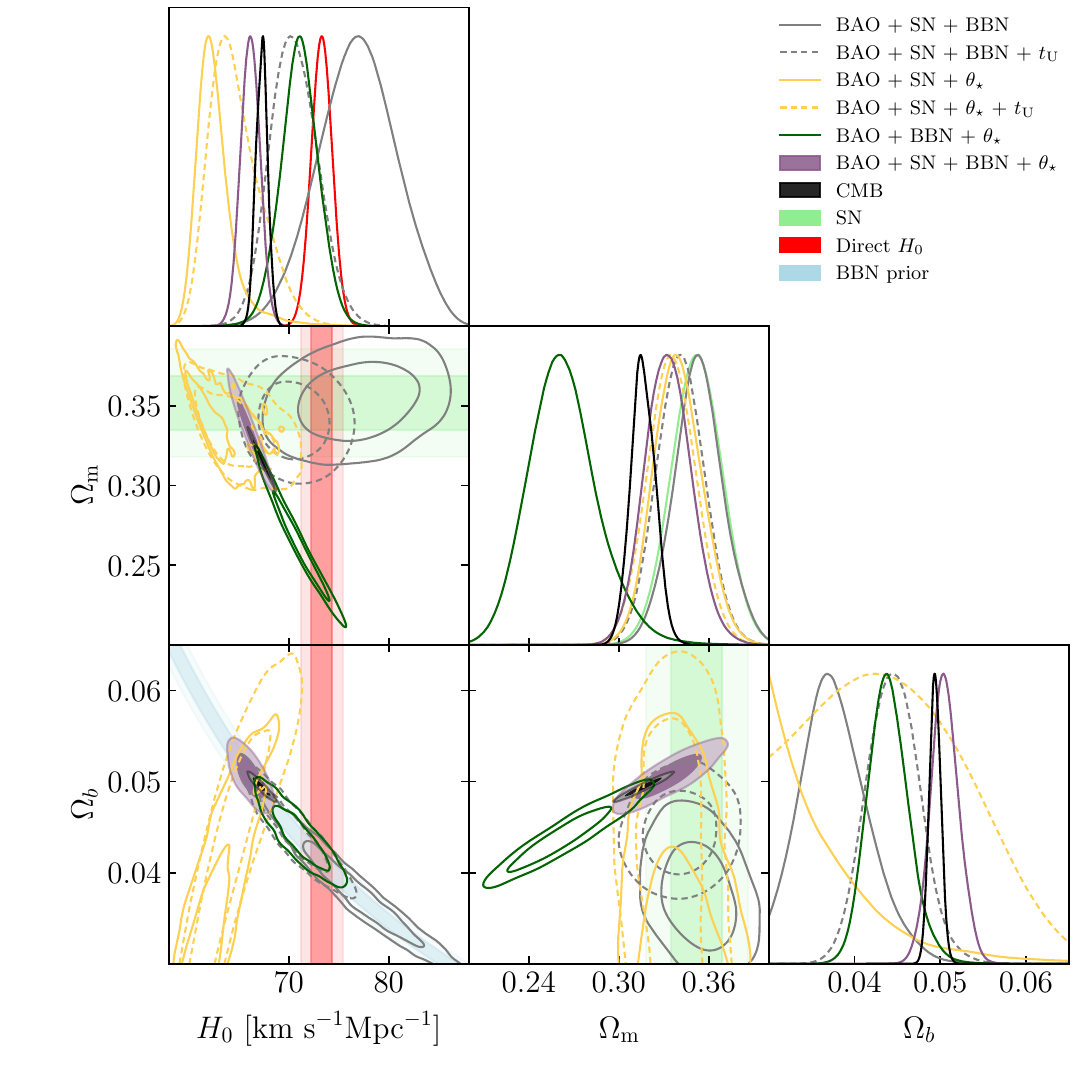}
    \caption{{\bf \boldmath\lcdm}. 68\% (darker) and 95\% (lighter) credible regions of the posteriors of different probe combinations within \lcdm. Tensions between constraints are apparent. They are further discussed in the text and quantified in \autoref{tab:tensions}. We zoom in to the more constraining combinations (including CMB and BAO+SN+BBN+\thetastar) in \autoref{fig:lcdm_h_Om}. 
    }
    \label{fig:lcdm}
\end{figure*}

\begin{figure}
    \centering
    \includegraphics[width=\linewidth]{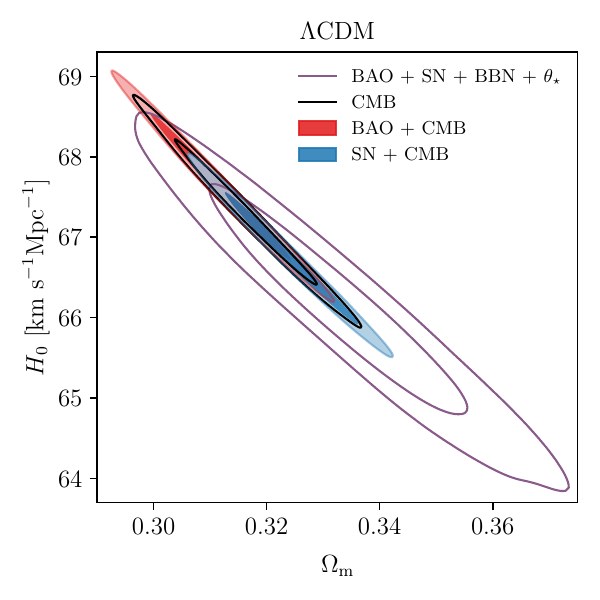}
    \caption{{\bf \boldmath\lcdm} zoom-in of the $\hubble$-\om plane to show the most constraining data combinations. We show the 68\%  and 95\% credible regions of the posteriors. As we describe in \autoref{sec:lcdm}, BAO (red) and SN (blue) tend to push in different directions of the parameter space when combined with CMB. The background probe combination (purple) is in agreement with the CMB constraints  besides coming from the combination of datasets in tension with CMB and/or among them, as discussed in the text. 
    }
    \label{fig:lcdm_h_Om}
\end{figure}

\subsection{{\bf \boldmath \lcdm}}
\label{sec:lcdm}

We start by exploring constraints on the standard model, \lcdm, in \autoref{fig:lcdm} and \autoref{fig:lcdm_h_Om}, where the latter shows a zoomed-in view of the region preferred by the CMB. For \lcdm and all other models considered, we report the marginalized constraints on parameters in \autoref{tab:parameters}, and additionally show $\hubble$ and \om constraints in \autoref{fig:H0_comparison} and \autoref{fig:Om_comparison}, respectively. 

One measurement of angular BAO on its own does not strongly constrain any individual parameter of the \lcdm space, but it excludes parts of the parameter space (by combining Equations \ref{eq:BAO}, \ref{eq:Hz}, \ref{eq:Dm}, \ref{eq:rs}). 
Even if we add a BBN prior, the contours (not shown) do not close: we have two data points and three free parameters: \om, \ob and $\hubble$. Similarly, combining BAO+\thetastar is not sufficient to close the contours (not shown) in our
prior volume, but offers us a measurement of \om$=0.255^{+0.021}_{-0.035}$. This bound is lower than that of CMB (\om$=0.3049^{+0.0082}_{-0.0090}$) and SN (\om$=0.353\pm0.017$), although we caution that it will be sensitive to the choice of prior on \ob and $h$ since constraints on both of those parameters are degenerate with \om and prior bounded.  
 
Once we combine BAO with both BBN and \thetastar, the contours close (dark green in \autoref{fig:lcdm}) and we obtain competitive constraints, with $\hubble=71.1\pm1.9$ \hunit and \om$=0.263^{+0.020}_{-0.025}$. This is between the $\hubble$ values from \shoes and from \planck, and remains compatible with either measurement. 

SN on its own tightly constrains matter abundance with a preference for high values, \om$=0.353\pm0.017$. However, SN alone does not constrain $\hubble$ unless their absolute magnitude is calibrated. One interesting way to perform that calibration is by combining BAO with information from the early Universe about $r_d$, e.g., from BBN, \thetastar, or a CMB-based inference of $r_d$ itself as we use in \autoref{sec:cosmographic}. This allows us to use BAO to infer the distance to a given redshift (in our case, $D_M(z=0.85)$), which calibrates the distance to supernovae. 
Fitting BAO+SN+\thetastar together results in a very low value of $\hubble$, with $62.7^{+1.0}_{-2.1}$\hunit. On the other hand, BAO+SN+BBN gives a high value of $\hubble=76.9\pm3.9$\hunit.
If we combine all of these probes (BAO+SN+BBN+\thetastar) we get $\hubble=66.15\pm0.96$\hunit, which is closer to \planck's value (\autoref{tab:parameters}). However, we discuss later in this section that some of these datasets are in some level of tension and, hence, all these measurements should be taken with caution.

It is interesting to consider how these constraints change when we include the prior on the age of the Universe from globular clusters measurements in~\cite{Valcin2020}  (see \autoref{sec:age} and \autoref{sec:inference}). We find that the \tu prior does not have any effect on posteriors when added to any inference including the CMB power spectra, since the CMB strongly constrains the age of the Universe. Similarly, it also had little impact when added to BAO+SN+BBN+\thetastar, demonstrating that this data combination also bounds \tu and that those constraints agree with the added prior. However, \tu does add significant information when removing either BBN or \thetastar. For BAO+SN+\thetastar, adding \tu moves the contour away from the \ob prior edge by increasing its $\hubble$ value. The contours are larger when including the \tu prior, but this is due to the BAO+SN+\thetastar case being artificially truncated by the lower bound of the \ob prior. In the case of BAO+SN+BBN, the \tu prior causes the contour to move to lower $\hubble$ values and shrink. We note that the discrepancy between the $\hubble$ values previously found between BAO+SN+BBN and BAO+BBN+\thetastar is reduced, but not completely ameliorated, when \tu is added to both data combinations. 
 
The CMB measurements from \planck are very constraining in \lcdm. Hence, even though BAO's $D_M/r_d$ is $\sim 2\sigma$ below \planck's prediction, when we combine BAO+CMB (red in \autoref{fig:lcdm_h_Om}) we only find a small shift in the cosmological parameters relative to \planck alone. Similarly, if we add SN to CMB, the shift is small. However, it is noticeable in \autoref{fig:lcdm_h_Om} that the different \om values preferred by BAO and SN push the combined constraints in opposite directions.

In \autoref{tab:tensions}, we check the consistency of different probes with the methodology from~\cite{tensions}, summarized in \autoref{sec:tensions}. We find a mild-low level of tension between CMB and either SN ($1.7\sigma$) or BAO ($2.0\sigma$). Whereas BAO and SN seem very compatible, we find that the combination of BAO+\thetastar is in tension ($2.4\sigma$) with SN. This was already hinted by \autoref{fig:lcdm} and has implications for the interpretation of differences between $\hubble$ constraints from BAO+SN+BBN versus BAO+SN+\thetastar: the latter comes from a combination of datasets in tension, and so is a less trustworthy inference. The fact that BAO+SN+\thetastar hits the bounds of the (wide) \ob prior range can also be seen as an indication that this combination did not work well in \lcdm. We also find the BAO+SN+BBN combination to be in tension with CMB ($2.2\sigma$), however, this tension is alleviated when the \tu prior is added to the former ($1.5\sigma$, see \autoref{tab:tensions}).

In conclusion, we find a number of tensions between datasets in \lcdm. From $D_M/r_d$ comparisons in our previous DES BAO analysis~\cite{des-y6-bao}, CMB and BAO are known to have some level of discrepancy in \lcdm. If we isolate just the geometric information from the CMB measurement of \thetastar and combine it with BAO, then this dataset becomes discrepant with SN. 
On the other hand, the BAO+SN+BBN combination is also in tension with CMB, but pulling in the opposite direction of $\hubble$ to BAO+SN+\thetastar. When adding \tu priors, both BAO+SN+\thetastar and BAO+SN+BBN come closer both to each other and to the CMB inferences, reducing some of the tensions. Besides those relative tensions, we note that the combination BAO+SN+BBN+\thetastar coming from combinations of datasets in tension) is compatible with the CMB constraints, and in agreement with \tu priors. 
Finally, BAO tends to favor lower values of \om, but SN prefers higher values of \om. 
All of these discrepancies in \lcdm provide an interesting context for our explorations of extended cosmological models.

\subsection{{\bf \boldmath\kcdm}}
\label{sec:klcdm}

\begin{figure}
    \centering\includegraphics[width=\linewidth]{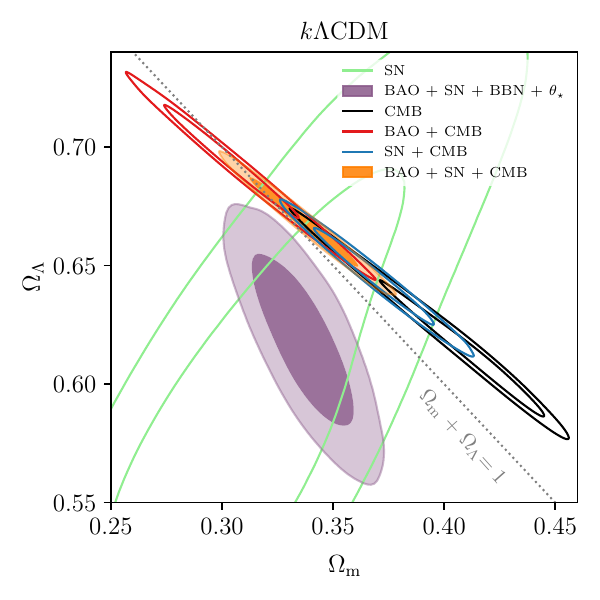}
    \caption{{\bf \boldmath\kcdm}. 1 and 2 $\sigma$ contours of the 2D posterior of \om and \olambda $\equiv (1-$\ok$-$\om$)$ in \kcdm. The tension among probes in this model is manifest. BAO+CMB and SN+CMB differ by $\sim3\sigma$ in \ok and the background probe combination (purple) is in tension with the CMB constraints. See \autoref{sec:klcdm} for discussion.
    }
    \label{fig:kLCDM}
\end{figure}

A natural extension of the \lcdm model is \kcdm, in which we allow curvature to vary. Results for our \kcdm analysis are shown in \autoref{fig:kLCDM} and the second block of \autoref{tab:parameters}. It is well known that constraints from the CMB alone exhibit a strong geometric degeneracy between \ok and \om~\cite{Planck}, which translates to a degeneracy between \om and \olambda=$1-$\om$-$\ok in \autoref{fig:kLCDM}. From the CMB alone, we find $\ok=-23.6^{+4.2}_{-7.9}\times10^{-3}$, in $\sim3\sigma$ tension with flatness according to the model-comparison metric described in \autoref{sec:deviations} and reported in \autoref{tab:deviations}. This is consistent with previous findings that have been extensively discussed in the literature (e.g., ~\cite{Planck,2020Efstathiou_ok,2021Handley_ok,2021DiValentino_ok,2022Glanville_ok,y3-3x2ptkp_ext}).

If we add BAO to CMB, we recover $\ok=1.4^{+5.8}_{-4.0}\times10^{-3}$, compatible with flat-\lcdm.
On the other hand, if we add SN to CMB, the constraints are in tension with flatness at a $2.9\sigma$ significance, on the negative side of  $\ok=-14.2^{+5.3}_{-4.9}\times10^{-3}$.\footnote{Note, as discussed in \autoref{sec:CMB}, that the CMB implementation used here is different to that in the DES SN paper \cite{des-y5-sn}, where we found \ok$=(-10\pm5)\times10^{-3}$ for CMB+SN. We also note that in earlier versions of \cite{des-y5-sn} there was a typo in the sign of this constraint.}
When BAO+SN+CMB are all combined, we find \ok$=-5.5^{+4.6}_{-4.2}\times10^{-3}$, within $\sim1\sigma$ of flat-\lcdm, though the fact that BAO+CMB and SN+CMB prefer different values of \ok by $\sim3\sigma$ means this combined result should be interpreted with caution. 

If we consider a purely background data combination of CMB's \thetastar with BAO and SN, we obtain \ok$=45^{+18}_{-14}\times10^{-3}$, again, mildly away from flatness ($\sim2.5\sigma$, see \autoref{tab:deviations}), but on the $\ok>0$ side. Similar and tighter results are recovered when we add BBN, with BAO+SN+BBN+\thetastar shown in purple in \autoref{fig:kLCDM}, giving \ok=$45^{+15}_{-14}\times10^{-3}$, $2.8\sigma$ from flatness.

Looking individually at each dataset, SN constrains one direction in the \om-\olambda plane relatively well. In our setup, the posterior on \ok hits the upper side of our prior (\ok=0.25), but if that prior was wider, as it is in \cite{des-y5-sn}, the contours would eventually close. BAO, although not shown here, disfavors the upper-right part of \autoref{fig:kLCDM}. These two probes intersect the CMB at different points along its $\Omega_{\Lambda}$-\om degeneracy direction. 
%
\autoref{tab:tensions} shows that the tensions between CMB and either BAO and SN do not decrease by varying curvature.
Interestingly, while in \kcdm, we find that the tension between SN and BAO+BBN+\thetastar goes away (\autoref{tab:tensions}), in \autoref{fig:kLCDM} we find that the background-only constraints (BAO+SN+BBN+\thetastar) are now discrepant with constraints based on CMB power spectra.

In summary, in \kcdm we do not find a general alleviation of the tensions among the probes, and some tensions actually increase.  
Additionally, BAO+SN+BBN+\thetastar and BAO+SN+CMB give very different posteriors on \ok. We, therefore, conclude that adding curvature to our model does not relieve the difficulty in reconciling constraints from the different observables we consider. 

\begin{table*}[]
    \centering
    \begin{tabular}{l |c|cZ|c|c|c|c}
        & $\hubble$  &  \om  &  $\ob$ & $10^3\ok$ & $w_0$ & $w_a$ & $\neutrinomass$\\
        \midrule
        {\bf \boldmath\lcdm} \\
        \midrule
        SN & -- & $0.353\pm 0.017$ & -- & -- & -- & -- & -- \\
        BAO + SN + BBN & $76.9\pm 3.9$ & $0.353\pm 0.016$ & $< 0.045$ & -- & -- & -- & -- \\
        BAO + SN + BBN + $t_\mathrm{U}$ & $70.5^{+2.2}_{-2.5}$ & $0.341^{+0.015}_{-0.016}$ & $0.0445\pm 0.0030$ & -- & -- & -- & -- \\
        BAO + $\theta_\star$ & $> 64.4$ & $0.255^{+0.021}_{-0.035}$ & $0.050^{+0.014}_{-0.008}$ & -- & -- & -- & -- \\
        BAO + SN + $\theta_\star$ & $62.7^{+1.0}_{-2.1}$ & $0.340^{+0.015}_{-0.017}$ & $< 0.052$ & -- & -- & -- & -- \\
        BAO + SN + $\theta_\star$ + $t_\mathrm{U}$ & $64.7^{+1.8}_{-3.1}$ & $0.335\pm0.016$ & $< 0.059$ & -- & -- & -- & -- \\
        BAO + BBN + $\theta_\star$ & $71.1\pm 1.9$ & $0.263^{+0.020}_{-0.025}$ & $0.0440^{+0.0021}_{-0.0025}$ & -- & -- & -- & -- \\
        BAO + SN + BBN + $\theta_\star$ & $66.15\pm 0.96$ & $0.333^{+0.015}_{-0.016}$ & $0.0506^{+0.0015}_{-0.0017}$ & -- & -- & -- & -- \\
        \shoes  & $73.04\pm 1.04$ & -- & -- & -- & -- & -- & -- \\
        CMB & $67.30^{+0.57}_{-0.61}$ & $0.3163^{+0.0084}_{-0.0080}$ & $0.0494^{+0.0007}_{-0.0007}$ & -- & -- & -- & -- \\
        BAO + CMB & $67.62^{+0.59}_{-0.59}$ & $0.3118^{+0.0080}_{-0.0082}$ & $0.0490^{+0.0007}_{-0.0007}$ & -- & -- & -- & -- \\
        SN + CMB & $66.74^{+0.54}_{-0.55}$ & $0.3242^{+0.0079}_{-0.0076}$ & $0.0500^{+0.0007}_{-0.0006}$ & -- & -- & -- & -- \\
        BAO + SN + CMB & $67.03^{+0.53}_{-0.55}$ & $0.3200^{+0.0074}_{-0.0079}$ & $0.0497^{+0.0006}_{-0.0006}$ & -- & -- & -- & -- \\
        \midrule
        {\bf \boldmath\kcdm} \\
        \midrule
        SN & -- & $0.317^{+0.032}_{-0.052}$ & -- & $> -100$ & -- & -- & -- \\
        BAO + SN + $\theta_\star$ & $> 65.9$ & $0.336^{+0.014}_{-0.015}$ & $0.043^{+0.007}_{-0.008}$ & $45^{+18}_{-14}$ & -- & -- & -- \\
        BAO + SN + BBN + $\theta_\star$ & $74.75^{+3.1}_{-3.0}$ & $0.337^{+0.014}_{-0.015}$ & $0.0400^{+0.0030}_{-0.0036}$ & $45^{+15}_{-14}$ & -- & -- & -- \\
        CMB & $< 63.5$ & $0.408^{+0.031}_{-0.017}$ & $> 0.0563$ & $-23.6^{+4.2}_{-7.9}$  & -- & -- & -- \\
        BAO + CMB & $68.3^{+2.5}_{-2.1}$ & $0.307^{+0.017}_{-0.024}$ & $0.0482^{+0.0027}_{-0.0038}$ & $1.4^{+5.8}_{-4.0}$  & -- & -- & -- \\
        SN + CMB & $62.1\pm 1.6$ & $0.369^{+0.018}_{-0.018}$ & $0.0585^{+0.0030}_{-0.0032}$ & $-14.2^{+5.3}_{-4.9}$  & -- & -- & -- \\
        BAO + SN + CMB & $65.1\pm 1.6$ & $0.338^{+0.015}_{-0.017}$ & $0.0530^{+0.0026}_{-0.0029}$ & $-5.5^{+4.6}_{-4.2}$  & -- & -- & -- \\      
        \midrule
        {\bf \boldmath\wcdm} \\
        \midrule
        SN & -- & $0.264^{+0.081}_{-0.065}$ & -- & -- & $-0.82^{+0.15}_{-0.11}$ & -- & -- \\
        BAO + SN + BBN & $<81.5$ & $0.283^{+0.067}_{-0.059}$ & $0.049^{+0.0087}_{-0.013}$ & -- & $-0.85^{+0.15}_{-0.10}$ & -- & -- \\
        BAO + SN + $\theta_\star$ & $68.9^{+4.6}_{-6.6}$ & $0.278^{+0.023}_{-0.029}$ & $> 0.030$ & -- & $-0.826^{+0.062}_{-0.047}$ & -- & -- \\
        BAO + SN + BBN + $\theta_\star$ & $67.0\pm 1.0$ & $0.281^{+0.018}_{-0.020}$ & $0.0495\pm 0.0016$ & -- & $-0.828^{+0.049}_{-0.043}$ & -- & -- \\
        CMB & $>65.1$ & $0.244^{+0.016}_{-0.052}$ & $< 0.0520$ & -- & $-1.32^{+0.12}_{-0.25}$ & -- & -- \\
        BAO + CMB & $>72.0$ & $0.223^{+0.011}_{-0.031}$ & $< 0.0427$ & -- & $-1.41^{+0.08}_{-0.17}$ & -- & -- \\
        SN + CMB & $65.66^{+0.76}_{-0.75}$ & $0.3326^{+0.0086}_{-0.0088}$ & $0.0519^{+0.0012}_{-0.0012}$ & -- & $-0.946\pm 0.028$ & -- & -- \\
        BAO + SN + CMB & $65.97^{+0.79}_{-0.77}$ & $0.3282^{+0.0090}_{-0.0092}$ & $0.0515\pm 0.0012$ & -- & $-0.948^{+0.028}_{-0.027}$ &  -- & -- \\
        \midrule
        {\bf \boldmath\wacdm} \\
        \midrule
        SN & -- & $0.377^{+0.066}_{-0.022}$ & -- & -- & $-0.82^{+0.13}_{-0.11}$ & $< 0.18$ & -- \\
        BAO + SN + BBN & $>62.8$ & $0.362^{+0.062}_{-0.025}$ & $< 0.055$ & -- & $-0.79^{+0.12}_{-0.10}$ & $< 0.11$ & -- \\
        BAO + SN + BBN + \tu & $69.6^{+2.4}_{-2.5}$ & $0.308^{+0.029}_{-0.036}$ & $0.0460\pm 0.0034$ & -- & $-0.76\pm 0.11$ & $-0.79^{+0.87}_{-0.67}$ & -- \\
        BAO + SN + $\theta_\star$ & $<79.6$ & $0.298^{+0.029}_{-0.037}$ & -- & -- & $-0.74\pm 0.10$ & $-0.76^{+0.86}_{-0.61}$ & -- \\
        BAO + SN + $\theta_\star$ + \tu & $68.4^{+3.1}_{-3.3}$ & $0.293^{+0.024}_{-0.031}$ & $0.0502^{+0.0091}_{-0.0076}$ & -- & $-0.74^{+0.09}_{-0.10}$ & $-0.72^{+0.82}_{-0.58}$ & -- \\
        BAO + SN + BBN + $\theta_\star$ & $67.5\pm 1.2$ & $0.295^{+0.020}_{-0.025}$ & $0.0487\pm 0.0018$ & -- & $-0.74^{+0.09}_{-0.10}$ & $-0.72^{+0.77}_{-0.55}$ & -- \\
        BAO + SN + BBN + $\theta_\star$ + \tu & $67.8^{+1.1}_{-1.2}$ & $0.296^{+0.020}_{-0.025}$ & $0.0486^{+0.0017}_{-0.0019}$ & -- & $-0.74^{+0.09}_{-0.10}$ & $-0.78^{+0.75}_{-0.54}$ & -- \\
        CMB & $>65.4$ & $0.247^{+0.040}_{-0.056}$ & $< 0.0518$ & -- & $>-1.6$ & $< 0.52$ & -- \\
        BAO + CMB & $>67.1$ & $0.242^{+0.019}_{-0.050}$ & $< 0.0488$ & -- & $>-1.5$ & $< 0.089$ & -- \\
        SN + CMB & $67.3\pm 1.0$ & $0.317^{+0.010}_{-0.011}$ & $0.0495^{+0.0013}_{-0.0015}$ & -- & $-0.73\pm 0.11$ & $-1.09^{+0.57}_{-0.51}$ & -- \\
        {\bf BAO + SN + CMB} & \boldmath $67.81^{+0.96}_{-0.86}$ & \boldmath $0.3109^{+0.0086}_{-0.0099}$ & \boldmath $0.0488^{+0.0012}_{-0.0014}$ & -- & \boldmath $-0.673^{+0.098}_{-0.097}$ & \boldmath $-1.37^{+0.51}_{-0.50}$ & -- \\
        \midrule
        {\bf \boldmath\nucdm} \\
        \midrule
        CMB & $66.9^{+1.3}_{-0.7}$ & $0.321^{+0.009}_{-0.017}$ & $0.0499^{+0.0008}_{-0.0017}$ & -- & -- & -- & $< 0.28$ \\
        BAO + CMB & $67.70^{+0.80}_{-0.64}$ & $0.311^{+0.008}_{-0.011}$ & $0.0489^{+0.0007}_{-0.0010}$ & -- & -- & -- & $< 0.15$ \\
        SN + CMB & $65.8^{+1.1}_{-0.9}$ & $0.336^{+0.012}_{-0.016}$ & $0.0514^{+0.0012}_{-0.0018}$ & -- & -- & -- & $< 0.37$ \\
        BAO + SN + CMB & $66.66^{+0.96}_{-0.72}$ & $0.325^{+0.009}_{-0.013}$ & $0.0503^{+0.0009}_{-0.0014}$ & -- & -- & -- & $< 0.27$ \\
        \midrule
        \bottomrule
    \end{tabular}
    \caption{We report the 68\% credible region (1$\sigma$) or 95\% of the upper/lower limit of cosmological parameters (in columns) under different cosmological models (in 5 tiers) and different data combinations. See the methodology in \autoref{sec:inference}. $\hubble$ is given in units of \kmsMpc\ and neutrino mass in \ev. We highlight in bold what we consider our main constraints.}
    \label{tab:parameters}
\end{table*}

\subsection{{\bf \boldmath \wcdm}}
\label{sec:wcdm}

\begin{figure*}
    \centering
    \includegraphics[width=0.5\linewidth]{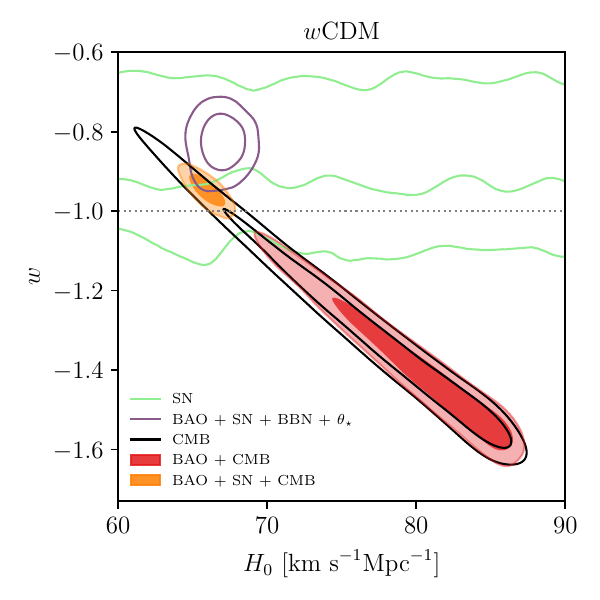}\includegraphics[width=0.5\linewidth]{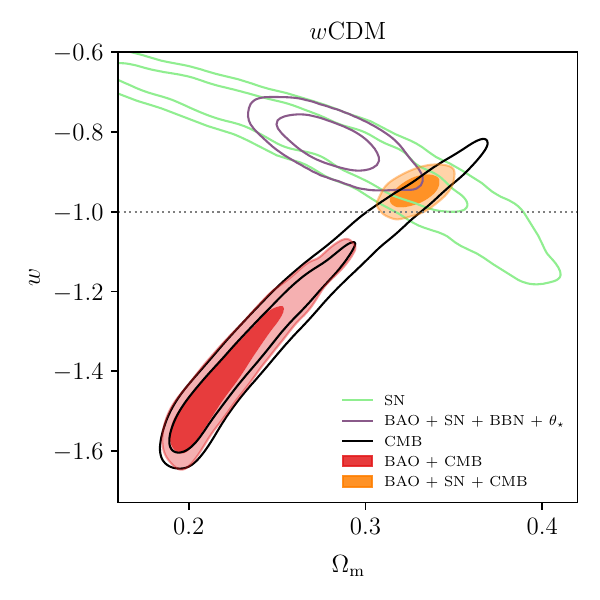}
    \caption{{\bf \boldmath\wcdm}. 1 and 2 $\sigma$ contours of the 2D posterior of $w$-$\hubble$ (left) and $w$-\om (right). BAO and SN still push for different regions of parameter space, $w<-1$ and $w>-1$, respectively. Nevertheless, SN dominates the constraints on $w$. Some tensions among probes are still apparent, as discussed in \autoref{sec:wcdm}. 
    }
    \label{fig:wCDM}
\end{figure*}

Next, we consider a one-parameter extension of \lcdm in which we constrain a constant equation of state of dark energy $w$. We present these \wcdm results in \autoref{fig:wCDM}. 

As in  \kcdm, the CMB has difficulties constraining this additional parameter on its own. Nevertheless, it gives a (wide) bound of  $w=-1.32^{+0.12}_{-0.25}$, at $1.7\sigma$ from \lcdm  ($w=-1$), again with that deviation's significance evaluated using the method from \autoref{sec:deviations} and reported in \autoref{tab:deviations}. If we add BAO to CMB, the contours tighten to $w=-1.41^{+0.08}_{-0.17}$, resulting in a $2.8\sigma$ deviation from \lcdm. Such a deviation from \lcdm is not unexpected, as we recall that the BAO $D_M/r_d$ constraint is $\sim2\sigma$ away from the CMB-\lcdm prediction. Compared to \lcdm, the tension between the CMB and BAO measurements reduces in significance from $2.0\sigma$ to $0.6\sigma$ (see \autoref{tab:tensions}), i.e., effectively removing the tension. 

SN on its own prefers a higher-than-standard value of the equation of state $w=-0.82^{+0.15}_{-0.11}$, which is $\sim1.6\sigma$ away from \lcdm. 
If we combine SN+CMB results we obtain
$w=-0.946\pm0.028$, obtaining a $\sim$3\% precision in $w$ and a $2.0\sigma$ deviation from \lcdm. Adding BAO to that barely changes the results, giving $w=-0.948^{+0.028}_{-0.027}$ and slightly decreasing the deviation \lcdm to $1.8\sigma$, though we caution that the combination of BAO+CMB is in 2.5$\sigma$ tension with SN. 
An interesting feature is that whereas SN and CMB have a large degeneracy in the  $\{w, \om\}$ plane (right panel of \autoref{fig:wCDM}), they overlap relatively close to the \lcdm value ($w=-1$)

Background-only constraints on $w$ are largely driven by the SN measurements. Like SN-alone, the background combination of BAO+SN+\thetastar prefers $w>-1$ at a significance of $2.7\sigma$ ($w=-0.826^{+0.062}_{-0.047}$), increasing to $3.1 \sigma$ when adding BBN ($w=-0.828^{+0.049}_{-0.043}$). 
The non-CMB combination of BAO+SN+BBN also prefers this region ($w=-0.85^{+0.15}_{-0.10}$), but with small significance ($1.4 \sigma$). All of these combinations give relatively similar $w$ constraints to SN alone, but the added observables additionally allow us to constrain the Hubble constant. The combination of all the background constraints (BAO+SN+BBN+\thetastar)  results in $\hubble=67.0\pm1.0$\hunit and  \om=$0.337^{+0.014}_{-0.015}$, much tighter constraints than the SN-only value of \om$=0.264^{+0.081}_{-0.065}$. It is also worth noting that extending to \wcdm fully alleviates the tension seen in \lcdm between SN and BAO+BBN+\thetastar.  

To sum up, as we have seen for other models, BAO and SN tend to pull constraints in different directions of the \wcdm parameter space. While extending to \wcdm reduces some tensions --- between CMB and BAO, as well as between SN and BAO+BBN+\thetastar --- it does not reconcile all of our observables. Notably, when combined with the CMB, non-negligible tensions remain between BAO and SN inferences. Since BAO and SN probe different redshifts, this motivates extending the model further with a time-dependent equation of state for dark energy. 


\subsection{ {\bf \boldmath \wacdm}}
\label{sec:w0wa}

\begin{figure*}
    \centering
    \includegraphics[width=0.5\textwidth]{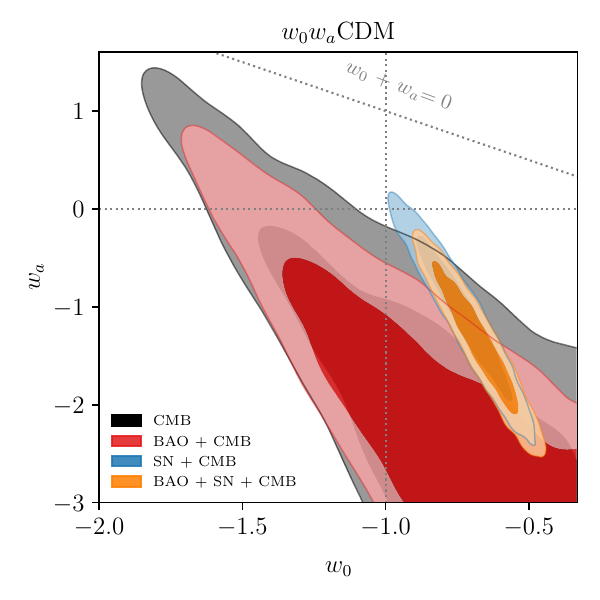}\includegraphics[width=0.5\textwidth]{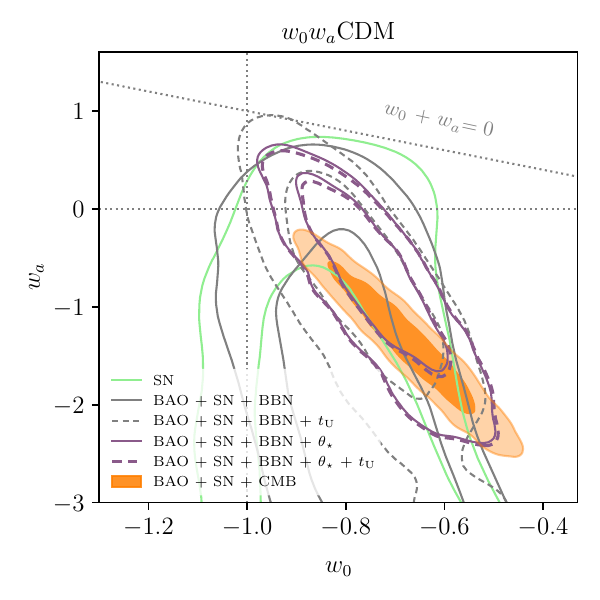}
    \caption{
    Constraints on the {\bf \boldmath \wacdm } model. {\bf Left:}  Constraints from the CMB and its combinations with SN and BAO, with our main constraint, BAO+SN+CMB, in orange. {\bf Right:} We {zoom-in and} compare the main constraint from BAO+SN+CMB (orange) to other inferences that do not rely on CMB power spectra. All probes tend to prefer the lower-right (high $w_0$, low $w_a$) quadrant and tensions between different observables are much relaxed relative to \lcdm (see \autoref{tab:tensions}). Our main constraint, BAO+SN+CMB (orange), disfavors \lcdm ({$w_0=-1$, $w_a=0$}) at a 3.2$\sigma$ significance (\autoref{tab:deviations}). The background-only (dashed-purple) and non-CMB (dashed-gray) cases also show $2.8\sigma$ and $2.0\sigma$ deviations from \lcdm, respectively. See \autoref{sec:w0wa} for more details and discussion. The gray-dotted line indicate the \lcdm case ($w_0=-1$, $w_a=0$), and the limit of the $w_0+w_a<0$ prior.
    }
    \label{fig:w0wa_zoom}
\end{figure*}

Constraints on \wacdm are shown in \autoref{fig:w0wa_zoom}.
On the left, we show results from the CMB alone, which is not strongly constraining this parameter space and finds a $2.5\sigma$ deviation from the \lcdm case (as always, quantified using the method described in \autoref{sec:tensions} and reported in \autoref{tab:deviations}). Adding BAO pushes the contours to lower values of $w_a$, resulting in a $3.4 \sigma$ tension with \lcdm, but do not dramatically change the constraints. 
In the right panel of \autoref{fig:w0wa_zoom}, we see that SN on their own 
can give 2-sided constraints on $w_0$ well within our priors. Whereas SN can also give 2-sided constraints on $w_a$ if one chooses broad priors (with the 1-$\sigma$ bound reaching $w_a\sim-15$, see \cite{des-y5-sn}), the constraints hit our lower limit described in \autoref{tab:params}.
When combining SN with CMB, the contours close well within our priors in the $w_0-w_a$ plane (shown in the left panel) and leave a $2.2\sigma$ deviation from \lcdm. 

When we add BAO to SN and CMB, the contours shrink slightly and move towards lower values of $w_a$, resulting in a $3.2 \sigma$ deviation from \lcdm and 
\twoonesig[4cm]{w_0 &= -0.673^{+0.098}_{-0.097}}{w_a &= -1.37^{+0.51}_{-0.50}}{BAO + SN + CMB. \label{eq:CMB_BAO_SN_w0wa}}

On the right-hand side of \autoref{fig:w0wa_zoom}, we compare the tightest constraint (BAO+SN+CMB, in orange in both panels) to other data combinations that do not rely on the CMB power spectra.
Compared to SN alone, the CMB-independent combination of BAO+SN+BBN gives similar constraints on $w_0$-$w_a$.

To further characterize these results, we now consider adding a prior on the age of the Universe, \tu, as described in \autoref{sec:age}. Most notably, adding this prior to BAO+SN+BBN has a large impact, as seen in the comparison between solid and dashed gray lines in the right panel of \autoref{fig:w0wa_zoom}. We find 
\twoonesig[4cm]{w_0 &= -0.76\pm0.11}{w_a &= -0.79^{+0.87}_{-0.67}}{BAO + SN + BBN + \tu. \label{eq:BAO_SN_BBN_w0wa}}
In contrast to the BAO+SN+BBN constraint, BAO+SN+BBN+\tu produces closed contours in both $w_0$ and $w_a$, fully independent of the CMB (which did not provide close contours on its own). This case is $2\sigma$ away from \lcdm. 
Another interesting case is when we also add \thetastar from the CMB to obtain our tightest constraint from the background expansion probes alone: 
\twoonesig[4cm]{w_0 &= -0.74^{+0.09}_{-0.10}}{w_a &= -0.78^{+0.75}_{-0.54}}{BAO + SN + BBN + \thetastar + \tu. \label{eq:background_w0wa}}
We find that once \thetastar is added to BAO+SN (with or without BBN), the \tu prior does not add much information. Hence, we conclude that BAO+SN+\thetastar already determines the age of the Universe and that determination agrees with the \tu prior we are considering.

Whereas in \lcdm and 1-parameter extensions discussed above, we find that SN and BAO tend to push parameter constraints in different directions, this is not the case in \wacdm. This is partially due to the model's increased flexibility since BAO moves mostly $w_a$ and SN mostly constraints $w_0$, as can be seen in the left and right panels of \autoref{fig:w0wa_zoom}, respectively. Regarding the tension metrics reported in \autoref{tab:tensions}, we find that the tension between CMB and BAO seen in \lcdm is completely alleviated ($0.1\sigma$ for \wacdm versus $2.0\sigma$ for \lcdm), and other tension metrics, e.g., between SN and either the CMB or BAO+CMB, are also reduced.
These tension metrics indicate that all the data combinations we consider agree within this model. This observation is reinforced by the fact that their 1-$\sigma$ confidence regions overlap in the lower-right quadrant of the $w_a$ vs $w_0$ plane. We also remark that, even though we impose a $w_0+w_a<0$ prior (see \autoref{sec:inference}), this region of space is also naturally excluded by the data, since none of the data combinations highlighted (except, marginally, BAO + SN + BBN +\tu) hit this prior within the $2\sigma$ contours.

Our tightest constraint, BAO+SN+CMB, disfavors \lcdm at $3.2 \sigma$ significance. It is remarkable that this is at a comparable level of significance to the recent results reported by DESI \cite{desi2024vi} from their combined analysis of \planck CMB, DES SN, and DESI 2024 BAO. 
To more directly compare, in the \appendixcite{app:desi} we re-analyze the DESI BAO results using our analysis framework and priors, finding the deviation from \lcdm to be $3.6\sigma$ when combining SN+CMB+DESI2024BAO.\footnote{The deviations quoted here differ slightly from those quoted in \cite{desi2024vi} and those in \cite{Notari24}, as those paper adopt a the method based 
 on $\Delta\chi^2$, whereas we quantify deviation in terms of parameter shifts (see \autoref{sec:deviations}). 
We also use only temperature and polarization from CMB, see \autoref{sec:CMB}.} 
The combination of DESI BAO with DES BAO, DES SN and CMB could show an even higher deviation from \lcdm, as discussed in \appendixcite{app:desi}.

\subsection{{\bf \boldmath \nucdm}}
\label{sec:neutrinos}

\begin{figure}
    \centering\includegraphics[width=\linewidth]{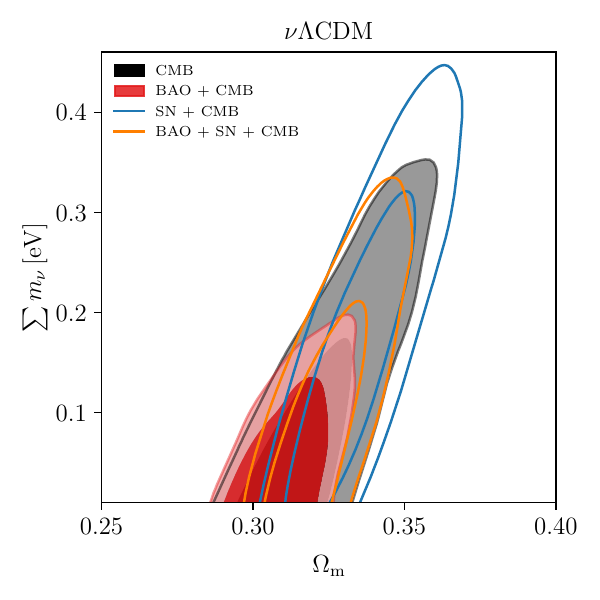}
    \caption{{\bf \boldmath \nucdm}. 1 and 2 $\sigma$ contours of the posterior of the neutrino mass vs. matter density. Given the degeneracy of these two parameters set by CMB, the preference for high-\om by SN results in a relaxation of the \neutrinomass bounds, while the low-\om preference by BAO translates to tighter constraints. }
    \label{fig:nu}
\end{figure}

Another interesting extension\footnote{One could argue that neutrinos are known to have mass and, hence, this should not be considered an extension. For example, in DES 3$\times$2pt neutrino masses are always varied in the baseline analysis \cite{y3-3x2ptkp}.} of \lcdm is the variation of neutrino mass. Whereas in all analyses described above we fixed the sum of neutrino masses to $\neutrinomass=0.06 e$V, we now let it vary as described in \autoref{sec:background}, below \autoref{eq:Hz}. We focus on constraints including CMB measurements, shown in \autoref{fig:nu}. The fact that the CMB primarily constrains CDM and baryon densities produces the degeneracy between \neutrinomass and \om seen for all contours. 

Variations between different data combinations can thus be interpreted in terms of how that degeneracy is broken via constraints on \om. The CMB on its own only places a 95\% upper limit of $\neutrinomass<0.28$\ev, with the \om information coming largely from the damping produced by lensing on the high-$\ell$ power spectra. 
As BAO prefers a lower value of \om than CMB, the BAO+CMB combination pushes that bound down to $\neutrinomass<0.15$\ev. On the other hand, because SN prefers a high value of \om, SN+CMB relaxes the limit on neutrino mass to $\neutrinomass<0.37$\ev. If we combine all of BAO+SN+CMB, we obtain $\neutrinomass<0.27$\ev. These results, with BAO tightening \neutrinomass constraints compared to the CMB alone and DES SN tending to weaken them, are in line with what has been previously seen in the DESI BAO analysis~\cite{desi2024vi} as well as~\cite{loverde_weiner_neutrinos}. We note that our BAO+CMB neutrino mass bounds are higher than those reported for DESI BAO (0.072 eV), reflecting the fact that DES BAO \om constraints are weaker and compatible with higher values than those from DESI BAO.  As discussed for DESI BAO and other BAO studies in the literature (e.g.~\cite{desi2024vi,bossdr12, negativemnu_green_meyers}), our BAO+CMB bounds are tight enough that the peak of the marginalized posterior hits the lower $\neutrinomass=0$ prior bound and would peak at $\neutrinomass<0$ if one were to fit that posterior with a Gaussian that allowed negative values. In contrast, the peak of the marginalized posteriors for SN+CMB occurs at $\neutrinomass>0$.

Finally, we remark that according to \autoref{tab:tensions}, freeing the neutrino masses does not alleviate significantly the tensions among probes.

\subsection{Cosmographic expansion}
\label{sec:cosmographic}

Next, we use a cosmographic expansion~\citep{Visser_2004, Zhang_2017}  to measure $\hubble$ using the DES BAO+SN with an external calibration of the sound horizon, $r_d$. A cosmographic expansion is a Taylor expansion of the scale factor $a$ that is agnostic about the energy contents of the Universe while maintaining the assumptions of homogeneity and isotropy. 
We follow the same definitions and methodology defined in Sec.~2 of~\cite{Camilleri24}, which assumes a spatially flat Universe. We determine our result using the $4^{\mathrm{th}}$ order expansion and a Gaussian prior on the sound horizon from~\cite{Lemos23} of $r_d\sim \mathcal{N}(147.46,\,0.28)\,$ Mpc (see \autoref{sec:rd}). We obtain
\begin{equation}
    \hubble=68.6^{+1.7}_{-1.6} \mathrm{~km} \mathrm{~s}^{-1} \mathrm{~Mpc}^{-1},
\end{equation}
which is consistent with the \planck \lcdm value along with previous inverse distance ladder measurements~\cite{Aubourg_2015, Macaulay_2019, Camilleri24}. We use the Akaike Information Criterion, ${\rm AIC}\equiv 2k-2\ln\mathcal{L}^{\max}$ \citep{1100705}, where $k$ is the number of parameters in the model, to assess whether the additional parameters used in the higher order cosmographic models are required by the data.  We find the $4^{\mathrm{th}}$ order cosmographic model used to obtain our key result quoted above,  to have a strong and moderate preference over the $2^{\mathrm{nd}}$ and $3^{\mathrm{rd}}$ order models respectively. However, we find no preference for or against the $4^{\mathrm{th}}$ and $5^{\mathrm{th}}$ order expansions and therefore focus on the model with fewer parameters.

\subsection{Comparison of \boldmath{$\hubble$} \& \boldmath{\om} across models}
\label{sec:H0}

\begin{figure*}
    \centering
    \includegraphics[width=\linewidth]{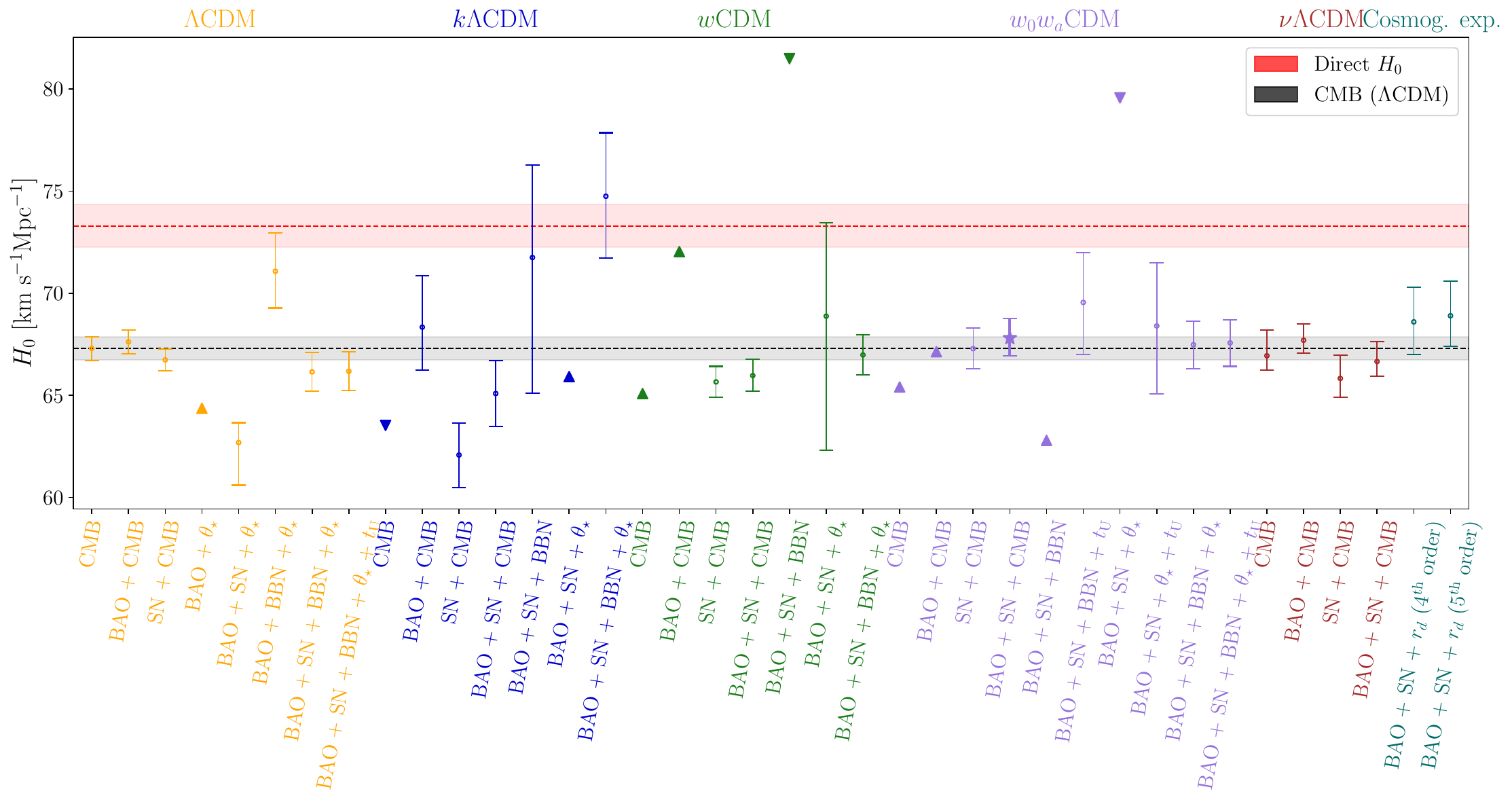}
    \caption{Hubble constant constraints for most of the combinations of models and datasets considered in this work, visualizing values reported in \autoref{tab:parameters}.  Two-sided 1-$\sigma$ (68\% c.r.) constraints are shown as points with error bars, and one-sided 95\% upper/lower limits are shown as triangular markers with the point facing down/up.  We also show the \shoes and the CMB-\lcdm constraints as shaded bands for comparison. We note that $\hubble$ anti-correlates with \om (shown in \autoref{fig:Om_comparison}) when including CMB with other probe combinations.
    Besides the variety of values, we note that the more extreme constraints are associated with tensions between datasets and that our main results (BAO+SN+CMB in \wacdm, marked with a star) agree with CMB-\lcdm (gray band). 
     See discussion in \autoref{sec:H0}.  
    }
    \label{fig:H0_comparison}
\end{figure*}

\begin{figure*}
    \centering
    \includegraphics[width=\linewidth]{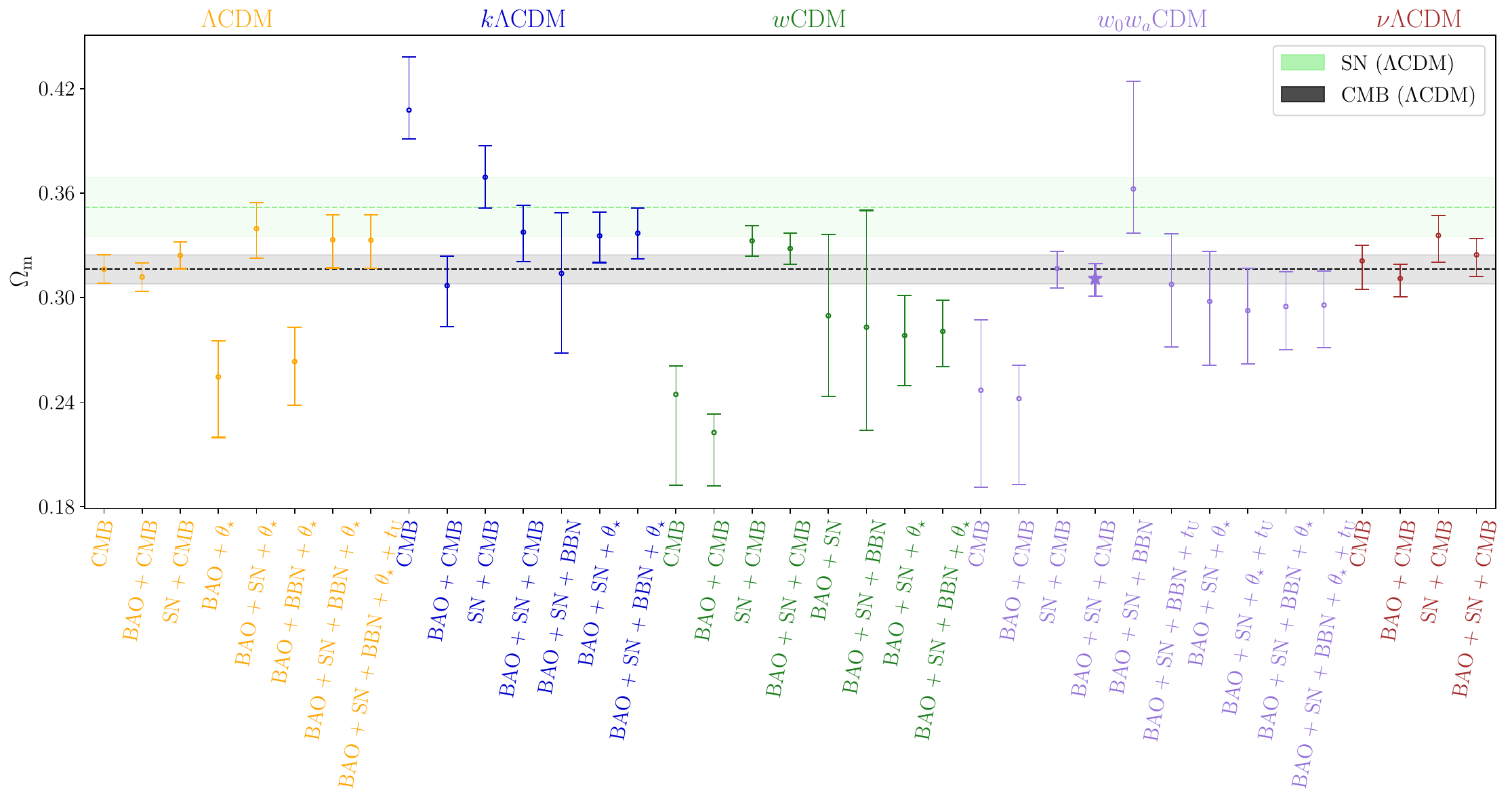}
    \caption{Density parameter of matter in the Universe, \om, with its corresponding 68\% c.r. (points with error bars) or 95\% upper/lower limit (triangles pointed down/up), given by the constraints for different combinations of models and datasets considered in this work. We also show the CMB-\lcdm and SN-\lcdm constraints, noting their full posterior are 1.7$\sigma$ apart (\autoref{tab:tensions}, \autoref{sec:tension_probes}). Note that \om anti-correlates with $\hubble$ (shown in \autoref{fig:H0_comparison}) when including CMB and other probe combinations. Some of the more extreme constraints can be related to tensions between datasets, as is discussed in \autoref{sec:H0}. Our main results, BAO+SN+CMB \wacdm, are marked with a star.
    }
    \label{fig:Om_comparison}
\end{figure*}

In this section, we compare the values of \om and $\hubble$ inferred from all relevant combinations of models and probes. We summarize these results in \autoref{fig:H0_comparison}, which shows 68\% c.r.  \hubble constraints, and  \autoref{fig:Om_comparison}, which shows 68\% c.r. \om constraints. As a reference for the \hubble constraints, \autoref{fig:H0_comparison} shows a band with the direct-\hubble (\shoes) results in red and the \planck-\lcdm results in black, highlighting the known tension between those measurements. 
For \om constraints in \autoref{fig:Om_comparison}, we show two bands corresponding to the 1-$\sigma$ intervals for CMB (black) and SN (light green) in \lcdm, whose tension is quantified as $1.7\sigma$ (\autoref{tab:tensions}).

We remind the reader that CMB tend to constrain the combination $\om \hubble^2$ very well. Hence, \om and \hubble measurements typically anti-correlate. Particularly, SN or BAO do not constrain \hubble on their own, but their constraints on \om propagate to \hubble when combining them with CMB thanks to that degeneracy. 
With this in mind, we focus our discussion below on the Hubble constant, but similar effects (in the opposite direction) can be seen in $\om$. We note the following highlights:

\begin{itemize}
    \item Within \lcdm, BAO tends to push \hubble CMB constraints to slightly higher values, whereas SN tends to push for lower values of \hubble (via the \om-$\hubble$ anti-correlation imposed by CMB). However, the constraining power of the CMB dominates when combined with SN and/or BAO. 
    
    \item In \lcdm, we can obtain \hubble constraints from only BAO+BBN+\thetastar. These bounds fall between those from \planck and \shoes, but are closer to the latter. 
    
    \item In \lcdm, very low values of \hubble are obtained with BAO+SN+\thetastar. However, these inferences are driven by tensions between SN and BAO+\thetastar, with the latter preferring extreme values for \ob (hitting our priors) and the age of the Universe. Hence, this tension relaxes when including BBN and/or \tu priors, bringing the $\hubble$ values up. 

    \item BAO pushes to higher values of $\hubble$ when added to CMB in all models considered. This is particularly significant for \wcdm, where BAO+CMB places a lower bound on \hubble that is consistent with the value from \shoes and not with \planck-alone. However, we note this combination (BAO+CMB for \wcdm) is in tension with SN at 2.5$\sigma$. The BAO+CMB combination barely constrains $\hubble$ in \wacdm. 

    \item SN pushes to lower values of $\hubble$ (via the \om-$\hubble$ degeneracy) when added to CMB in all the extensions. 
    
    \item  CMB constraints on \hubble significantly relax when extending the model beyond \lcdm. However, combining the CMB with both SN and BAO recovers tight constraints on $\hubble$. These combinations are always compatible with the \hubble value inferred for CMB-\lcdm.

    \item The cosmographic expansion (\autoref{sec:cosmographic}) BAO+SN+$r_d$ gives $\hubble=68.6^{+1.7}_{-1.6}$\hunit, compatible with CMB-\lcdm, and is robust against the order considered in the expansion. 

    \item Remarkably, in \wacdm where all the tensions among datasets disappear, we find a very tight constraint from BAO+SN+CMB, despite the flexibility of the model: $\hubble=67.81^{+0.96}_{-0.86}$\hunit.

    \item In \wacdm, when using only expansion history probes (BAO+SN+BBN+$\theta_\star$+\tu), we obtain again a well-constrained Hubble constant, $\hubble=67.8^{+1.1}_{-1.2}$\hunit. 
\end{itemize}
Hence, we conclude that the tensions seen in \lcdm, \kcdm and \wcdm among probes do not seem to hint at a resolution of the Hubble tension problem and our data combinations tend to favor \hubble values similar to \lcdm CMB constraints.

\section{Discussion}
\label{sec:discussion}
 
\subsection*{Findings and tensions in \boldmath{\lcdm} and 1-parameter extensions}


A recurring theme of this work is that in most of the cosmological models we consider, we find some level of tension between datasets, often with our two background probes, BAO and SN, pulling in different directions of parameter space. With this in mind, we summarize the main findings for \lcdm and its one-parameter extensions:

\begin{itemize}
    \item \lcdm
    \begin{itemize}
        \item When adding BAO to CMB, it pushes for lower values of \om, whereas SN pushes in the opposite direction. This is consistently found in other cosmological models. 
        \item Several tensions among data and/or inconsistencies are found. For example, BAO+BBN+\thetastar is found in 2.9$\sigma$ tension with SN.
        \item The age-of-the-Universe priors can help alleviate some tensions.
        \item The combination BAO+SN+BBN+\thetastar 
        can set competitive results to CMB, and are within $\sim1\sigma$ of it. 
    \end{itemize}
    \item \kcdm
        \begin{itemize}
        \item CMB alone is subject to significant geometric degeneracies and prefers negative \ok. When adding BAO, it becomes compatible with \ok=0, whereas SN tightens the constraints around negative values of \ok. The combination BAO+SN+CMB is within $\sim 1 \sigma$ of flatness. 
        \item BAO+SN+BBN+\thetastar, with $2.8\sigma$ evidence for positive \ok, is not compatible with CMB and its combination with SN and/or BAO. 
        \item BAO is found to be in $3.2\sigma$ tension with CMB in this model.  
        \item Considering BAO+SN+CMB, the detection of dark energy is at $\sim 40\sigma$.\footnote{Unlike the rest of the paper, here we simply divide by the $\sigma$ to estimate these tensions. \label{fn:10sigma}}
        \end{itemize}
    \item \wcdm
        \begin{itemize}
        \item CMB alone has a very large degeneracy between $w$ and \om, preferring $w<-1$ at $1.7\sigma$. When adding BAO, contours tighten,  preferring $w<-1$ at $2.8\sigma$. On the other hand, SN prefers $w>-1$ ($w=-0.82^{+0.15}_{-0.11}$), remaining nearly identical if we also add BAO.
        \item This difference in the preferred value of $w$ leads to a tension of $\sim 2.5\sigma$ between
        BAO+CMB and SN in \wcdm. This can be interpreted as mild hint for evolving dark energy, given that the effective redshifts of SN and BAO are different. 
        \item The combination of BAO+SN+CMB gives a preference for an accelerating fluid ($w<-1/3$) at a significance of $\sim20\sigma$\footref{fn:10sigma} . 
        \end{itemize}
    \item \nucdm
    \begin{itemize}
        \item In this model, tensions among probes remain at a similar level as in \lcdm (\autoref{tab:tensions}).
        \item CMB alone sets \neutrinomass $<0.28$\ev, with a positive correlation between \om and \neutrinomass. Hence, the preference for higher \om of SN results in a more relaxed neutrino constraint ($\neutrinomass<0.38$\ev) and the preference for lower \om of BAO for more stringent constraint ($\neutrinomass<0.15$\ev). BAO + SN + CMB results are very similar to CMB alone: \neutrinomass$<0.27$\ev.
    \end{itemize}
\end{itemize}

We find it difficult to reconcile all the data with these models and report our main results within \wacdm, where tensions are alleviated and different data combinations agree. We argue this choice below and present the main \wacdm conclusions in \autoref{sec:conclusions}.

\subsection*{Is \boldmath{\wacdm} preferred over \boldmath{\lcdm}?}

Our choice of \wacdm model for the main results is driven by the tensions between inferences from different datasets observed in the other models: all have at least one case of $\geq2.5\sigma$ tension (\autoref{tab:tensions}).\footnote{ Additionally, recent works like \cite{Tang24} suggest that tensions in \om in \lcdm would naturally appear if the Universe followed a \wacdm cosmology.} In addition to tensions explicitly reported in \autoref{tab:tensions} and mathematically defined in \autoref{sec:tension_probes}, there are additional significant offsets in the posteriors inferred for different data combinations shown in Figures \ref{fig:lcdm}, \ref{fig:kLCDM} and \ref{fig:wCDM}. These latter tensions cannot be quantified by the methodology in \autoref{sec:tension_probes} when part of the data is shared among the two data combinations considered.
Both types of tensions get significantly relieved in \wacdm, with a maximum tension of only $1.6\sigma$ among 7 data combinations tested. Furthermore, all data combinations tested have 1-$\sigma$ regions that overlap in the $w_0$-$w_a$ plane, and several of them report $>2\sigma$ deviations --mathematically defined in \autoref{sec:deviations}-- from \lcdm.  For our main data combination, BAO+SN+CMB, the deviation from \lcdm reaches the level of $3.2\sigma$, further supporting the choice of \wacdm(\autoref{tab:deviations}). We note that this is tentative evidence ($p$-value of 0.0014), and that $5\sigma$ ($p\sim6\times10^{-7}$) is typically required to claim a discovery. Nevertheless, when datasets show $\gtrsim 2\sigma$ tensions, the parameter estimates resulting from their combination are not very reliable. Hence, we show the main results in terms of \wacdm, where important tensions go away.

Some other works have relied on Bayesian evidence or some approximation to it to choose a preferred model. However, we argue here that these methods can strongly dilute their evidence as the priors widen, whereas the parameter difference method we employ (\autoref{sec:deviations}) is independent of that effect, provided the data are sufficiently constraining (and we consider it to be the case for \autoref{eq:CMB_BAO_SN_w0wa} and \autoref{eq:background_w0wa}). 
We note that both methods penalize adding more free parameters (as we do in \wacdm) to compensate for the additional degrees of freedom. This penalization is explicit in the Bayesian evidence ratios. For our method, it naturally appears since given a change in goodness of fit, the significance of the deviation dilutes when having more degrees of freedom. 
We also see in \autoref{tab:chi2} that \wacdm improves the goodness of fit by $\Delta\chi^2\sim12$ with respect to \lcdm for the two most constraining data combinations. Under the Gaussian approximation for the likelihood (Wilks' theorem \cite{Wilks1938}), this corresponds to a $\sim3\sigma$ preference for \wacdm, given the two additional degrees of freedom, in line with our method considering the full shape of the posterior in the $w_0-w_a$ plane ($\sim3\sigma$, \autoref{tab:deviations}).

One could potentially be concerned about projection effects when extending the parameter space, or about the effect of informative priors. This effect can be present in some instances with weak constraints such as CMB alone in \autoref{fig:w0wa_zoom}. However, probe combinations with strong constraining power such as BAO+SN+CMB or BAO+SN+BBN+\thetastar are not expected to be affected by this.
To verify the robustness of constraints and the significance of their deviation from \lcdm, we ran a series of checks on the \wacdm inferences reported in the text above for BAO+SN+CMB (\autoref{eq:CMB_BAO_SN_w0wa}), BAO+SN+BBN+\tu   (\autoref{eq:BAO_SN_BBN_w0wa}) and BAO+SN+BBN+\thetastar+\tu (\autoref{eq:background_w0wa}). These checks, which included the comparison of 2D marginalized posteriors with the position of the 10 highest-posterior samples from the chain, and with an approximate profile likelihood, support the idea that these constraints are not significantly impacted by projection effects. As validation of the reported parameter-shift deviation from \lcdm, we confirmed that the iso-density contour going through the \lcdm point in parameter space is well behaved --- i.e. that it is not dominated by noise.

Hence, we conclude that among the models tested in this paper, \wacdm is the only one capable of reconciling all the data without $\sim2.5\sigma$ tensions, and preferred over \lcdm at a significance of $\sim 3\sigma$. Whereas other models of the expansion history might offer another solution, exploring more alternative models is beyond the scope of this paper. Another possibility would be that one or several datasets have unaccounted for systematic errors. 
Both the Y6 BAO paper series \cite{des-y6-bao,des-y6-bao-sample} and the DES-SN5YR paper series \citep{vincenzi24,des-y5-sn} have already scrutinized and quantified a large ensemble of known systematic errors, which are subdominant and added in quadrature to the statistical error. Hence, all the known systematics are already accounted for, although it is always possible that a source of unknown systematics is present. Nevertheless, none of the tests point in the direction of having to exclude one particular dataset.
Furthermore, re-analyzing their systematic errors, which were thoroughly analyzed in previous studies, is not the goal of this paper.

\subsection*{The Hubble constant}

Given the choice of \wacdm results as our main focus, we can consider the implications for the Hubble tension,  referring to $\hubble$ constraints examined in \autoref{sec:H0}. We report a consensus value given by BAO+SN+CMB \wacdm of 
\begin{equation*}
    \hubble=67.81^{+0.96}_{-0.86}\, {\rm km\, s}^{-1}\, {\rm Mpc}^{-1} \, ,
\end{equation*} 
noting that this value is in good agreement with that given by CMB-\lcdm ($\hubble=67.30^{+0.57}_{-0.61}$\hunit) and BAO+SN+CMB-\lcdm ($\hubble=67.03^{+0.53}_{-0.55}$\hunit). It is also compatible with  BAO+SN+CMB constraints in all models considered, as well as with the cosmographic expansion model (BAO+SN+$r_d$ at fourth order: $\hubble=68.6^{+1.7}_{-1.6}$\hunit, see \autoref{sec:cosmographic}). This inference is also consistent with the CMB-independent data combination  BAO+SN+BBN+\tu ($\hubble=69.6^{+2.4}_{-2.5}$\hunit in \wacdm) and with  the background-probe combination BAO+SN+BBN+$\theta_\star$+\tu$ (\hubble=67.8^{+1.1}_{-1.2}$\hunit in \wacdm). Whereas we find some higher or lower values of $\hubble$ for certain other combinations of datasets and models, these values are always associated with tensions seen between probes. In summary, the inferred value of \hubble from the data we consider is fairly robust to different choices for model and data combinations, and the improved fit for \wacdm does not substantially impact considerations for the Hubble tension.

\section{Conclusions}
\label{sec:conclusions}

In this paper, we studied the cosmological parameter implications of background probes, BAO and SN, from the DES final dataset.
We did this in combination with external probes: \planck's CMB (in three different forms: temperature and polarization power spectra, angular acoustic scale, \thetastar, or comoving acoustic scale, $r_d$), BBN \cite{schoneberg20242024}, and age-of-the-Universe priors \cite{Valcin2020}, as explained in \autoref{sec:data}.   
We studied the \lcdm model, extensions on the background evolution (\kcdm, \wcdm and \wacdm, see \autoref{sec:background}), and an extension with free neutrino masses (\nucdm). 

Following from the discussion above, our main conclusion is that these datasets fail to agree in parameter space at the $\gtrsim2.5\sigma$ level (\autoref{sec:tensions})
except for the most complex model we considered here, \wacdm, where datasets are consistent (see \autoref{sec:discussion} for the discussion on findings in \lcdm, \wcdm, \kcdm and \nucdm, as well as why we choose to express our main conclusions in terms of  \wacdm). This corresponds to the model known as CPL, where the equation of state of dark energy varies linearly with the scale factor ($w(a)=w_0+(1-a)w_a$). In this parameter space, our data combination of BAO+SN+CMB prefers $\{w_0>-1,w_a<0\}$ (\autoref{eq:CMB_BAO_SN_w0wa}) over \lcdm ($\{w_0=-1,w_a=0\}$) with a significance of $\sim 3.2\sigma$ (always, as defined in \autoref{sec:deviations}):

\twoonesignonum[2cm]{w_0 &= -0.673^{+0.098}_{-0.097}}{w_a &= -1.37^{+0.51}_{-0.50}}{BAO+SN+CMB.}

Within the \wacdm model, all datasets considered are compatible with one another and the consensus lies in the $\{w_0>-1,w_a<0\}$ quadrant. This result is partly driven by SN preferring $w_0>-1$ and BAO pushing for $w_a<0$.  
We also highlight the cases where we restrict ourselves to only background probes (BAO+SN+BBN+$\theta_\star$+\tu) or non-CMB  (BAO+SN+BBN+\tu). While these constraints are looser, they are compatible with the above combination in the $\{w_0-wa\}$ plane and find a milder deviation of $2.8\sigma$ and $2.0\sigma$, respectively, from \lcdm in that plane (see \autoref{sec:deviations}).

These results could indeed be hinting at the dynamical nature of dark energy, or they could be a sign of some unknown systematic error in some (parts of the) datasets or of another type of non-\lcdm physics shaping the expansion history of the Universe. 

Certainly, this and recent works create an interesting scenario for our understanding of the cosmological model. Whereas in the recent past, the description of dark energy dynamics in \lcdm has been in reasonably good agreement with nearly all observables studied, this has shifted in the last year. First, the final DES-SN5YR results released in January 2024 \cite{des-y5-sn} reported a $\sim2\sigma$ hint of deviation from \lcdm in favor of the \wacdm model. Shortly after, DES-Y6-BAO found another $2.1\sigma$ hint of deviation between its measurement of $D_M(z=0.85)/r_d$ and that predicted by \planck-\lcdm \cite{des-y6-bao} (although we note a similar level of deviation was previously observed in DES-Y3-BAO \cite{y3-baokp}). Finally, in April of the same year, DESI reported a $3.9\sigma$ deviation from \lcdm when combining DESI 2024 BAO with \planck-CMB and DES-SN5YR SN. 
Reanalyzing that same data, we reproduce this result (\appendixcite{app:desi}), finding a 3.6$\sigma$ tension when using our model comparison methodology (\autoref{sec:tensions}) and CMB likelihood configuration (which, as we note in \autoref{sec:CMB} is slightly different than that used in the DESI paper).

This work confirms that the previously observed tension persists at a similar level ($3.2\sigma$) when considering DES Y6 BAO combined with DES-SN5YR and \planck-CMB. For comparison, we note that substituting the single DES (angular) BAO datapoint for the seven DESI BAO measurements~\cite{desi2024iv} in combination with the same SN and CMB likelihoods only slightly increases the significance of the reported deviation from \lcdm from 3.2$\sigma$ to 3.6$\sigma$. We also find that the deviation from \lcdm becomes larger when including both DESI and DES BAO, though the exact significance would depend on the unknown (but likely small) correlations between these datasets.

Overall this work adds to the growing evidence from different studies that the equation of state of dark energy could vary with time. However, before a change of paradigm of this magnitude can be established, the community should require a larger statistical significance (canonically, $5 \sigma$), and this tension should persist over several years with new datasets or new probes and careful scrutiny of data characterization and analysis methodology. Our datasets include in their uncertainties a contribution from all known systematic errors, which are in all cases smaller than the statistical one. However,  it is always possible that unknown systematic errors are present, and new validations should be pursued
\footnote{In these lines, an alternative cross-calibration between DES and external SN appeared during the review of this paper \cite{Dovekie}. The impact of the recalibration in the SN+BAO+CMB will be evaluated in \cite{Mena26}, together with the impact of the new DESI DR2 BAO data.}.
Additionally, a physical model of cosmic acceleration that is more well motivated from first principles would help establish a viable alternative to \lcdm.

This study represents the impact of DES background cosmology probes (BAO and SN) on the current cosmological paradigm from the survey's final dataset. Looking ahead, DES will soon be releasing analyses that additionally probe the growth of structure and the density perturbations in the late-time Universe. These include studies of weak gravitational lensing, galaxy clustering, cluster counts, cross-correlations among those probes and also with external datasets, such as the CMB. These upcoming results will both provide better constraints on the properties of our cosmological models and provide crucial cross-checks of whether the emerging paradigm shift is self-consistent across probes. Certainly, the legacy of DES will be a rich source of insight for state-of-the-art cosmological analyses in the coming years. 

\begin{acknowledgments}
{\it Author Contributions.}
All authors contributed to this paper and/or carried out infrastructure work that made this analysis possible. Some highlighted contributions from the authors of this paper include:
{\it Scientific management and coordination:}  S. Avila,  A. Porredon, M. Vincenzi (science working group conveners of Large-Scale Structure, Supernova).
{\it Significant contributions to project development,
including paper writing and figures:} S. Avila, T. Davis, J. Mena-Fern\'andez, J. Muir, A. Porredon, P. Shah, M. Vincenzi. 
{\it Data analysis, scientific interpretation, code development}: R. Camilleri, G. Campailla, N. Deiosso, A. Fert{\'e}, J. Mena-Fern\'andez, J. Muir, M. Raveri, P. Shah.
{\it Internal reviewing of the paper}: D. Huterer, S. Lee, P. Wiseman.
{\it Contribution to the BAO results/catalog:} S. Avila, H. Camacho, R. Cawthon, K. C. Chan, J. De Vicente, J. Elvin-Poole, I. Ferrero, G. Giannini.  J. Mena-Fern\'andez, A. Porredon, M. Rodriguez-Monroy, I. Sevilla-Noarbe,  L. Toribio
San Cipriano, E. Sanchez and N. Weaverdyck.
{\it Construction and validation of the DES-SN5YR Hubble diagram:} P. Armstrong, D. Brout, A. Carr, R. Chen, T. M. Davis, L. Galbany, S. Hinton, R. Kessler,  J. Lee, C. Lidman, A. M{\"o}ller, B. Popovic, H. Qu,  M. Sako, B. Sanchez, D. Scolnic, M. Smith, M. Sullivan, G. Taylor, M. Toy, P. Wiseman. 
\textit{Construction and validation of the DES Gold catalog}: M. Adamow, K. Bechtol, A. Carnero Rosell, H. T. Diehl, A. Drlica-Wagner, R. A. Gruendl, W. G. Hartley, A. Pieres, E. S. Rykoff, I. Sevilla-Noarbe, E. Sheldon, and B. Yanny.
The remaining authors
have made contributions to this paper that include,
but are not limited to, the construction of DECam and
other aspects of collecting the data; data processing and
calibration; developing broadly used methods, codes, and
simulations; running the pipelines and validation tests;
and promoting the science analysis.

Funding for the DES Projects has been provided by the U.S. Department of Energy, the U.S. National Science Foundation, the Ministry of Science and Education of Spain, 
the Science and Technology Facilities Council of the United Kingdom, the Higher Education Funding Council for England, the National Center for Supercomputing 
Applications at the University of Illinois at Urbana-Champaign, the Kavli Institute of Cosmological Physics at the University of Chicago, 
the Center for Cosmology and Astro-Particle Physics at the Ohio State University,
the Mitchell Institute for Fundamental Physics and Astronomy at Texas A\&M University, Financiadora de Estudos e Projetos, 
Funda{\c c}{\~a}o Carlos Chagas Filho de Amparo {\`a} Pesquisa do Estado do Rio de Janeiro, Conselho Nacional de Desenvolvimento Cient{\'i}fico e Tecnol{\'o}gico and 
the Minist{\'e}rio da Ci{\^e}ncia, Tecnologia e Inova{\c c}{\~a}o, the Deutsche Forschungsgemeinschaft and the Collaborating Institutions in the Dark Energy Survey. 

The Collaborating Institutions are Argonne National Laboratory, the University of California at Santa Cruz, the University of Cambridge, Centro de Investigaciones Energ{\'e}ticas, 
Medioambientales y Tecnol{\'o}gicas-Madrid, the University of Chicago, University College London, the DES-Brazil Consortium, the University of Edinburgh, 
the Eidgen{\"o}ssische Technische Hochschule (ETH) Z{\"u}rich, 
Fermi National Accelerator Laboratory, the University of Illinois at Urbana-Champaign, the Institut de Ci{\`e}ncies de l'Espai (IEEC/CSIC), 
the Institut de F{\'i}sica d'Altes Energies, Lawrence Berkeley National Laboratory, the Ludwig-Maximilians Universit{\"a}t M{\"u}nchen and the associated Excellence Cluster Universe, 
the University of Michigan, NSF NOIRLab, the University of Nottingham, The Ohio State University, the University of Pennsylvania, the University of Portsmouth, 
SLAC National Accelerator Laboratory, Stanford University, the University of Sussex, Texas A\&M University, and the OzDES Membership Consortium.

Based in part on observations at NSF Cerro Tololo Inter-American Observatory at NSF NOIRLab (NOIRLab Prop. ID 2012B-0001; PI: J. Frieman), which is managed by the Association of Universities for Research in Astronomy (AURA) under a cooperative agreement with the National Science Foundation.

The DES data management system is supported by the National Science Foundation under Grant Numbers AST-1138766 and AST-1536171.
The DES participants from Spanish institutions are partially supported by MICINN under grants PID2021-123012, PID2021-128989 PID2022-141079, PID2023-151307NB-I00, SEV-2016-0588, CEX2020-001058-M and CEX2020-001007-S, some of which include ERDF funds from the European Union. IFAE is partially funded by the CERCA program of the Generalitat de Catalunya.

We acknowledge support from the Brazilian Instituto Nacional de Ci\^encia
e Tecnologia (INCT) do e-Universo (CNPq grant 465376/2014-2).

This document was prepared by the DES Collaboration using the resources of the Fermi National Accelerator Laboratory (Fermilab), a U.S. Department of Energy, Office of Science, Office of High Energy Physics HEP User Facility. Fermilab is managed by Fermi Forward Discovery Group, LLC, acting under Contract No. 89243024CSC000002. 

This work used computing resources from the Duke Compute Cluster, NERSC, the Sciama cluster in Portsmouth, and Symmetry at Perimeter Institute.
The DESSN Hubble diagram and analysis pipeline was made possible by the codebases \texttt{SNANA} \cite{Kessler_2009} and \texttt{PIPPIN} \cite{pippin}.
Some of the computing for this project was performed on the Sherlock cluster. We would like to thank Stanford University and the Stanford Research Computing Center for providing computational resources and support that contributed to these research results. This research was supported in part by Perimeter Institute for Theoretical Physics. Research at Perimeter Institute is supported by the Government of Canada through the Department of Innovation, Science, and Economic Development, and by the Province of Ontario through the Ministry of Colleges and Universities.
SA acknowledges support from the RYC2022-037311-I fellowship funded by MCIN/AEI/10.13039/501100011033 (Spain) and FSE+ (Europe).
JMu thanks discussions with Z. Weiner.  
APo acknowledges support from the EU’s Horizon Europe program under the MSCA grant agreement 101068581 and the `César Nombela' Research Talent Attraction Aid from the Community of Madrid (2023-T1/TEC29011).


\end{acknowledgments}

\appendix

\section{Combination with DESI 2024 BAO}
\label{app:desi}

\begin{figure}
    \centering
    \includegraphics[width=\linewidth]{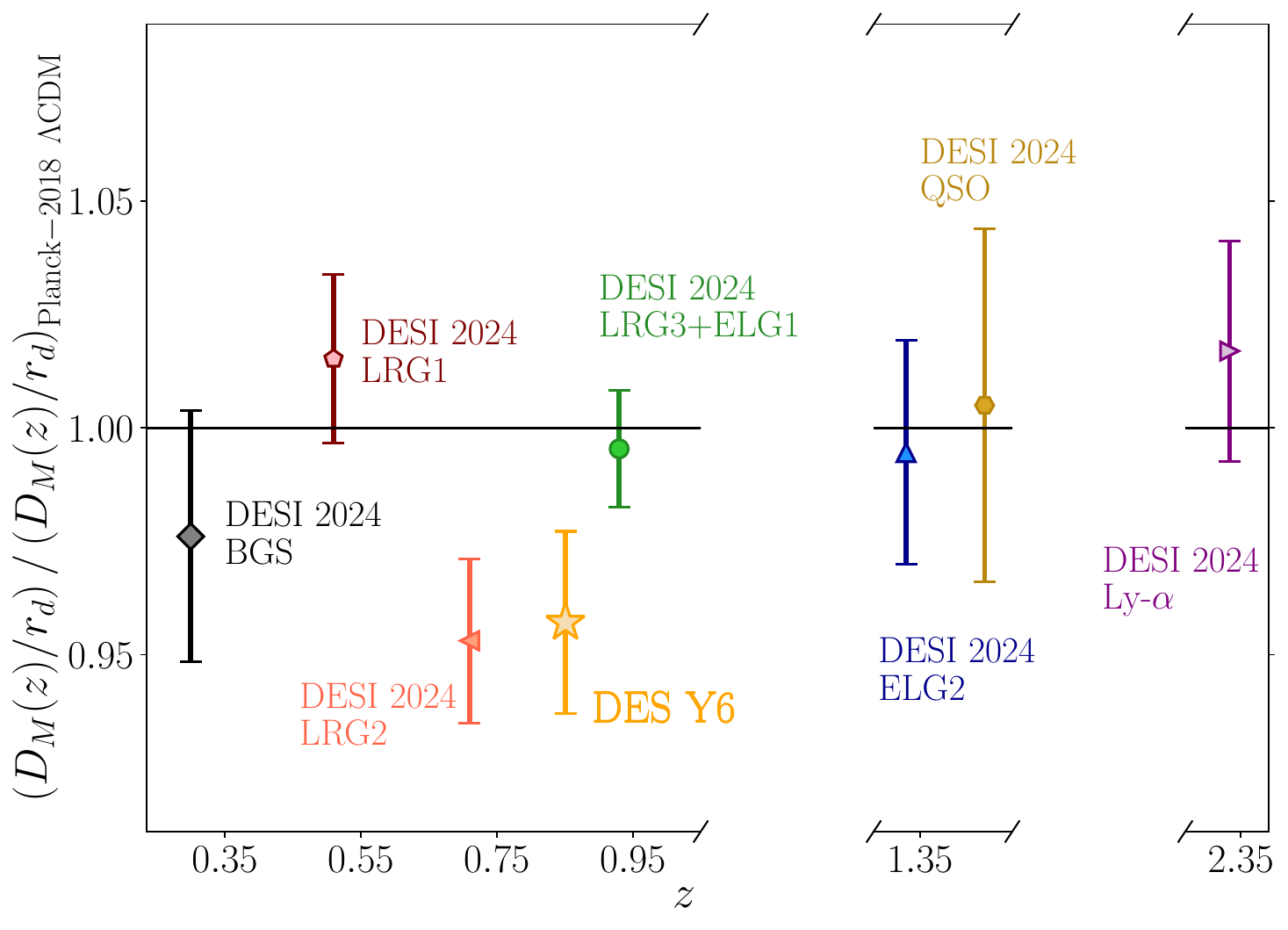}
    \caption{Ratio between the $D_M(z)/r_d$ measured using the BAO feature at different redshifts and the prediction from the cosmological parameters determined by \planck-2018, assuming $\Lambda$CDM. We include all the measurements from the DESI 2024 BAO analysis in different colors, and the DES Y6 measurement as an golden star.}
    \label{fig:bao_distance_ladder}
\end{figure}

In this appendix, we combine our BAO and SN measurements with the DESI 2024 BAO results from~\cite{desi2024iii}, and also CMB from \planck, and run chains assuming \wacdm. We include all BAO measurements from DESI, namely BGS, LRG1, LRG2, LRG3+ELG1, ELG2, QSO and LYA . In \autoref{fig:bao_distance_ladder}, we show the angular BAO distance ladder, including the results from DESI 2024 and our DES BAO measurement. For BGS and QSO, the 2D BAO fit was not carried out; therefore, there is no $D_M/r_d$ available in the table mentioned above. For these two tracers, we plot $D_V/r_d$ as the value for $D_M/r_d$ represented in the plot and $1.5\times \sigma(D_V/r_d)$ as its uncertainty.\footnote{Assuming spherical symmetry, the $D_M(z)$ constraints are 50\% less precise (see~\cite{ross2015information}) with respect to the spherically-averaged measurement.} Nevertheless, for the chains, the full likelihood of DESI 2024 BAO is used, considering $D_M/r_d$ and $D_H/r_d$ when available, and $D_V/r_d$ when not. This is labeled as DESI2024BAO. 

\begin{figure}
    \centering
    \includegraphics[width=\linewidth]{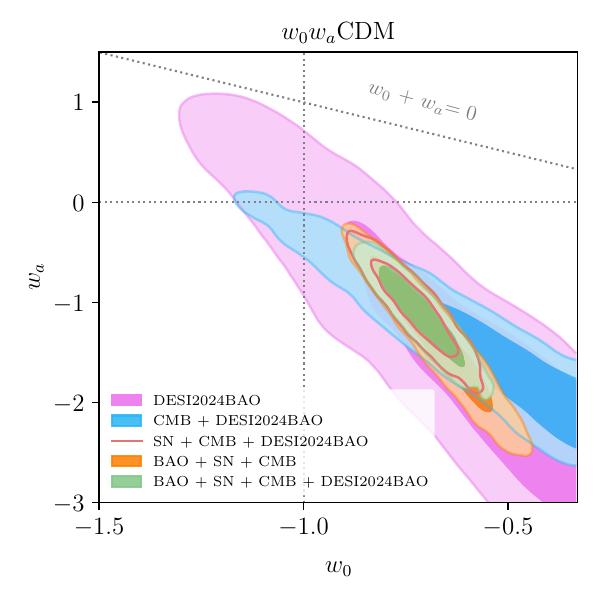}
    \caption{ \wacdm with DESI. 1 and 2 $\sigma$ contours of the 2D posterior of $w_0$-$w_a$ for several data combinations that include DESI 2024 BAO, labeled as DESI2024BAO. We also include combinations with DES BAO (simply labeled as BAO), in particular, the orange BAO+SN+CMB contour is the same one as in \autoref{fig:w0wa_zoom}. Combining CMB, SN and either BAO dataset (DES or DESI), the deviation from \lcdm is $>3\sigma$. The deviation could be larger for the green case with both BAO datasets combined, however, possible correlations among them have been neglected in this combination, see discussion in \appendixcite{app:desi}.
    }
    \label{fig:chain_w0wacdm_desi}
\end{figure}

The results of our chains are shown in \autoref{fig:chain_w0wacdm_desi}. The CMB+DESI2024BAO combination (blue) is $2.8\sigma$ away from \lcdm (see \autoref{tab:deviationsDESI}), compared to the $3.4\sigma$ deviation found with BAO+CMB (\autoref{tab:deviations}). 
When adding SN (SN+CMB+DESI2024BAO, red), the deviation rises to $3.6\sigma$ (\autoref{tab:deviationsDESI}), compared to the $3.2\sigma$ we found with DES BAO in combination with CMB and SN (\autoref{tab:deviations}). Nevertheless, we see that DESI alone does not show a strong deviation from \lcdm.

We also note that the $3.6\sigma$ deviation found here for SN+CMB+DESI2024BAO is somewhat lower than that reported by DESI ($3.9\sigma$ \cite{desi2024iv}). This is due to two main reasons. First, the CMB implementation is somewhat different (see \autoref{sec:CMB}), mainly because we do not include CMB-lensing in our likelihood, but \cite{desi2024iv} did. Second, DESI used likelihood ratios to estimate the deviation based on $\Delta \chi^2$ and Wilks' theorem, while we consider the full shape of the posterior in the $w_0$-$w_a$ parameter plane as our primary metric, see \autoref{sec:deviations}. If we use the same likelihood ratio method as DESI, we find a deviation of 4.0$\sigma$ from $\Delta \chi^2=18.9$.

Finally, we also report the combination of CMB, SN together with both BAO datasets (BAO+SN+CMB+DESI2024BAO, green in \autoref{fig:chain_w0wacdm_desi}). In this case, the parameter shift deviation from \lcdm is at the $4.2\sigma$ level while the goodness-of-fit comparison gives  
$\Delta \chi^2=18.9$ and 4.0$\sigma$. 
Nevertheless, we caution that this number relies on DES BAO and DESI BAO being fully independent. There is some small overlap in the area and redshift range of the samples that could give rise to some small correlations among these datasets. This correlation is expected to be small, but computing it is beyond the scope of this paper.

\begin{table*}[]
    \centering
    \begin{tabular}{l c}
        \midrule
         & Deviations from \lcdm ($\sigma$) \\ 
        \midrule
        Dataset & $\boldsymbol{w_0 w_a}$\textbf{CDM} \\
        \midrule
        DESI2024BAO &  1.3 \\
        CMB + DESI2024BAO &  2.8\\
        SN + CMB + DESI2024BAO & 3.6 \\
        BAO + SN + CMB + DESI2024BAO  &  4.2\\
        \bottomrule
    \end{tabular}
    \caption{Deviations from \lcdm of the constraints in the \wacdm model for data combinations that include DESI 2024 BAO results, see \appendixcite{app:desi}. This is quantified in $\sigma$s, see \autoref{Eq:EffectiveSigmas} and \autoref{sec:deviations}. We note that possible correlations between DESI BAO and DES BAO (simply labeled as BAO), are neglected and could reduce the deviation reported in the last row.}
    \label{tab:deviationsDESI}
\end{table*}

\section{Likelihood for BAO individual bins}
\label{app:5bins-method}

In \cite{des-y6-bao}, we presented measurements of the BAO shift $\alpha$ from the angular correlation function (ACF) over the entire BAO sample redshift range ($0.6<\zph<1.2$). This method was validated with 1952 mocks. Along with it, two other methods were validated: the angular power spectrum (APS) and projected correlation function. 

In this appendix, we validate the method when computed at the individual bin level, with $\Delta\zph=0.1$. In this process, we found ACF to be more robust than APS and PCF, so we will only continue with ACF.

The main validation is given by \autoref{tab:validation}, similar to Table III of \cite{des-y6-bao}. 
The method minimizes the $\chi^2$ as a function of $\alpha$, defining the best fit as the minimum $\chi^2$ and the error $\sigma_\alpha$ as the semi-width of the $\Delta \chi^2=1$ region. 
In \autoref{tab:validation}, we find the mean of the best fits, $\langle \alpha \rangle$, which is expected to be 1, since in the mocks we assume the cosmology they were generated with. We then use $\lvert\langle \alpha \rangle-1 \lvert$ as a systematic error associated to modeling in each individual bin, which we report in \autoref{tab:systematics} as $\sigma_{\rm mod, sys}$.  We also check the reasonable agreement between $\langle \sigma_\alpha \rangle$, $\sigma_{\rm std}$ and $\sigma_{68}$, which tells us about the robustness of our error bars. We refer the reader to \cite{des-y6-bao} for more details. 

In \autoref{tab:systematics}, we report our measurements of the individual bin BAO. The systematic error associated to modeling ($\sigma_{z,{\rm mod}}$) comes from \autoref{tab:validation} described in the previous paragraph. The systematic error associated with redshifts ($\sigma_{z,{\rm sys}}$) comes from Table I of \cite{des-y6-bao}. The total error is computed as the sum of the quadrature of those two with the statistical error. This is the same procedure we followed in \cite{des-y6-bao} for the combined measurement ($0.6<\zph<1.2$). Then, multiplying $\alpha$ by the fiducial $D_M/r_d$ from the cosmology assumed in the template fitted to the data, we obtain physical constraints on $D_M/r_d$.

\begin{figure}
    \centering
    \includegraphics[width=0.45\textwidth]{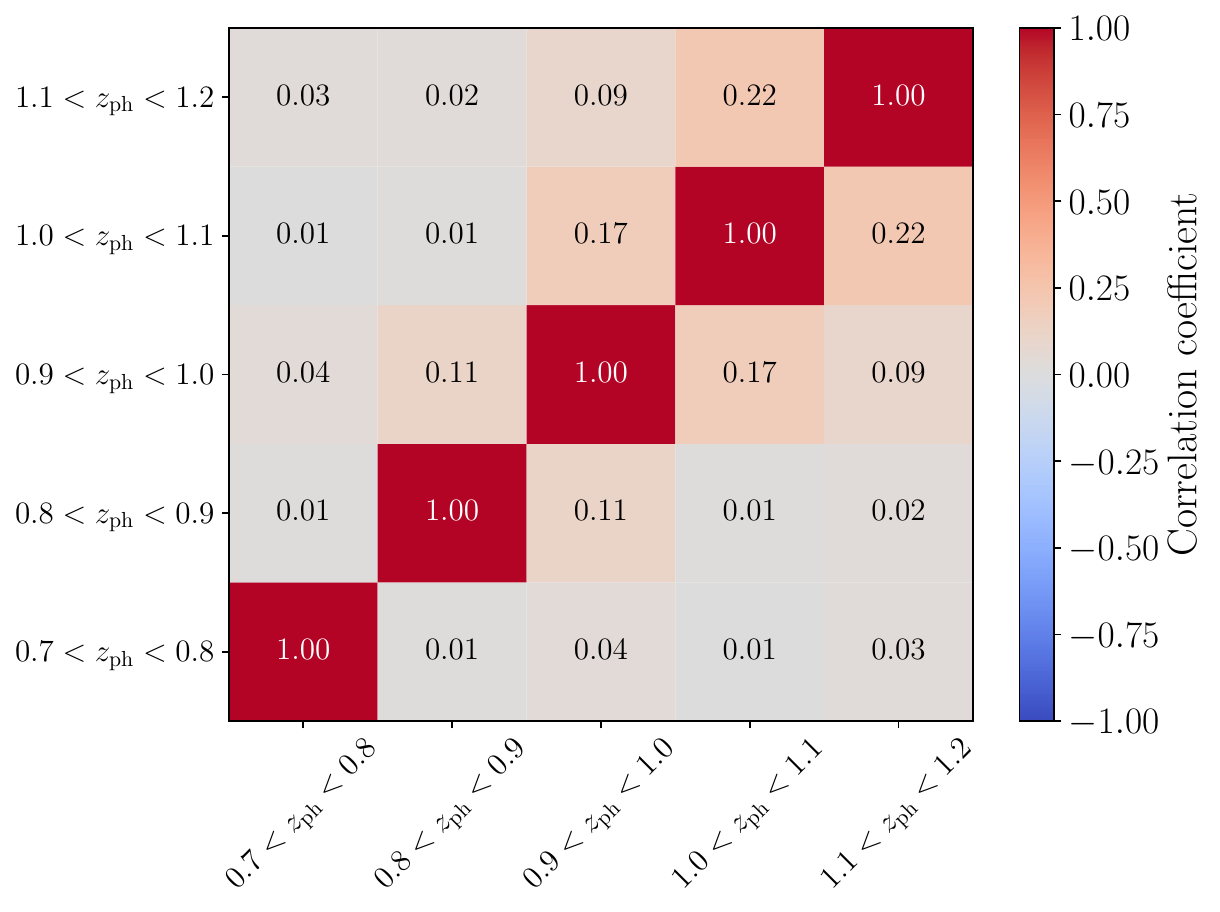}
    \caption{Correlation 
    matrix of $\alpha_i$ for the 5 redshift bins with an individual BAO detection. 
    }
    \label{fig:cov_plot}
\end{figure}

Finally, we note that these individual measurements are expected to be correlated, mostly due to their redshift overlap (Figure 2 of \cite{des-y6-bao}). Using the 1952 mock catalogues we compute the Pearson correlation coefficient ($\rho_{ij}$) among any two bins ($i$, $j$) and represent them in \autoref{fig:cov_plot}. Then, our final covariance comes from the total error ($\sigma$) reported on the last row of \autoref{tab:systematics} and the correlation from \autoref{fig:cov_plot}:

\begin{equation}
C_{i,j} = \sigma_i \sigma_j \rho_{ij} \, .
\end{equation}

With this covariance, and the mean of $\bar D\equiv D_M/r_d$ reported in \autoref{tab:systematics}, we construct a Gaussian likelihood (${\rm log} \mathcal{L}\propto (D_i({\bf p})-\bar D_i)C_{i,j}^{-1}(D_j({\bf p})-\bar D_j)$).


\begin{table*}
    \setlength{\tabcolsep}{4pt} 
    \begin{tabular}{ZcccccccZZZZZZZZ}
    \toprule
     & Bins & Redshift & $\langle\alpha\rangle$ & $\sigma_{\rm std}$ & $\sigma_{68}$ & $\langle\sigma_\alpha\rangle$ & mocks $\in \langle \alpha \rangle \pm \langle  \sigma_\alpha \rangle$ & $\langle d_{\rm norm}\rangle$ & $\sigma_{d_{\rm norm}}$ & $\langle\chi^2\rangle/$dof & mean of mocks & $\sigma_{\rm std}/\langle\alpha\rangle$ & $N\sigma$ & frac. detec. & p-value \\
    \midrule
    7 & 1  & $0.6<z_{\rm ph}<0.7$ & 1.0024 & 0.0485 & 0.0457 & 0.0454 & 67.7$\%$ & -0.0283 & 1.1162 & 16.0$/$17 & 1.0028$\pm$0.0486 & 4.84$\%$ & 0.05 & 90.32$\%$ & 0.52 \\
    8 & 2  & $0.7<z_{\rm ph}<0.8$ & 0.9999 & 0.0458 & 0.0438 & 0.0420 & 66.2$\%$ & -0.0219 & 1.1332 & 12.0$/$17 & 1.0008$\pm$0.0432 & 4.58$\%$ & 0.00 & 94.98$\%$ & 0.80 \\
    9 & 3  & $0.8<z_{\rm ph}<0.9$ & 1.0038 & 0.0407 & 0.0388 & 0.0403 & 70.0$\%$ & -0.0074 & 1.0312 & 11.3$/$17 & 1.0044$\pm$0.0396 & 4.05$\%$ & 0.09 & 97.39$\%$ & 0.84 \\
    10 & 4 & $0.9<z_{\rm ph}<1.0$ & 1.0095 & 0.0398 & 0.0370 & 0.0376 & 69.2$\%$ & -0.0189 & 1.0660 & 12.4$/$17 & 1.0100$\pm$0.0372 & 3.94$\%$ & 0.24 & 97.59$\%$ & 0.78 \\
    11 & 5 & $1.0<z_{\rm ph}<1.1$ & 1.0072 & 0.0409 & 0.0377 & 0.0416 & 72.4$\%$ & -0.0264 & 0.9839 & 10.9$/$17 & 1.0081$\pm$0.0410 & 4.06$\%$ & 0.18 & 96.67$\%$ & 0.86 \\
    12 & 6 & $1.1<z_{\rm ph}<1.2$ & 1.0067 & 0.0475 & 0.0461 & 0.0557 & 76.0$\%$ & -0.0247 & 0.8974 & 8.9$/$17 & 1.0103$\pm$0.0576 & 4.72$\%$ & 0.14 & 91.19$\%$ & 0.94 \\
    \bottomrule
    \end{tabular}
    \caption{Validation of the angular correlation function method on individual tomographic bins, when run in the 1952 COLA mocks. We show: (i) bin number, (2) redshift interval, (3) mean of the best fit BAO shift ($\alpha$) across all mocks, (4) standard deviation of best fit $\alpha$, (5) semi-width of the interval containing 68\% of the best fit $\alpha$, (6) mean of the error reported in each mock, and (7) number of mocks with a best fit within  $[\langle \alpha \rangle - \langle 
 \sigma_\alpha \rangle, \langle \alpha \rangle + \langle  \sigma_\alpha \rangle]$. This table is the equivalent of Table III of \cite{des-y6-bao} but for individual bins.}
    \label{tab:validation}
\end{table*}

\begin{table*}[]
    \centering
    \begin{tabular}{cl||r||r|r|r|r|r|r}
        \toprule
          & Bin & $0.6<\zph<1.2$ & $0.6<\zph<0.7$ & $0.7<\zph<0.8$ & $0.8<\zph<0.9$ & 
          $0.9<\zph<1.0$ & $1.0<\zph<1.1$ & $1.1<\zph<1.2$  \\
          \midrule
          & $\alpha\pm\sigma_{\rm stat}$ & 
          $0.9571\pm0.0196$
          & - & $0.9279\pm0.0404$  &  $0.9649\pm0.0434$ & $0.9974\pm0.0557$ &  $0.9511\pm0.0551$ &  $1.0515\pm0.0775$  \\
          & $\sigma_{z,{\rm sys}}$ &  - & 0.0151 & 0.0079  &  0.0082 & 0.0112 &  0.0030 &  0.0062  \\
          & $\sigma_{\rm mod, sys}$ & - & 0.0024 & 0.0001 & 0.0038  &  0.0095 & 0.0072 &  0.0067   \\
          & $\alpha\pm\sigma_{\rm tot}$ & - & - & $0.9279\pm0.0412$  &  $0.9649\pm0.0443$ & $0.9974\pm0.0576$ &  $0.9511\pm0.0556$ &  $1.0515\pm0.0780$  \\
          \midrule
           & $z_{\rm eff}$  & 0.851 & 0.652 & 0.752  &  0.850 & 0.947 &  1.043 &  1.142 \\
         & $D_M/r_d$ & $19.51\pm0.41$ & - & $17.18\pm0.75$  &  $19.66\pm0.88$ & $22.05\pm1.23$ & $22.57\pm1.31$ &  $26.62\pm1.96$  \\
         \bottomrule
    \end{tabular}
    \caption{DES BAO fits. The second column ($0.6<\zph<1.2$) point shows the main results from~\cite{des-y6-bao} using the entire BAO sample and the combination of the three clustering estimators: ACF, APS and PCF. The next 6 columns consider the individual data using only the ACF. The first bin ($0.6<\zph<0.7$) does not have a detection according to our criteria. In the first row, we show the best fit BAO shift ($\alpha$) with its associated statistical error bar, we then report two sources of systematic errors (from redshift calibration and modeling) and finally the total error bar ($\sigma_{\rm tot}$) summing in quadrature the three. In the last tier, we also report the effective redshift ($\zeff$, Equation 25 of \cite{des-y6-bao} and the final physical constraints in terms of $D_M(\zeff)/r_d$) that goes into the likelihood.
    }
    \label{tab:systematics}
\end{table*}

\section{Cosmological constraints from the BAO individual bins}
\label{app:5bins-cosmo}

\begin{figure*}[h!]
    \centering
    \begin{minipage}[b]{0.45\textwidth}
        \centering
        \includegraphics[width=\textwidth]{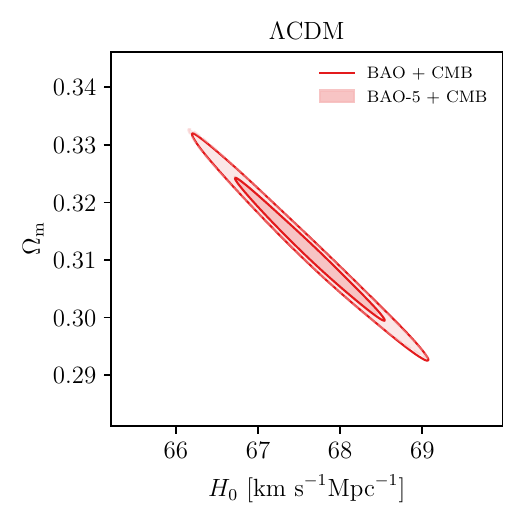}
    \end{minipage}
    \begin{minipage}[b]{0.45\textwidth}
        \centering
        \includegraphics[width=\textwidth]{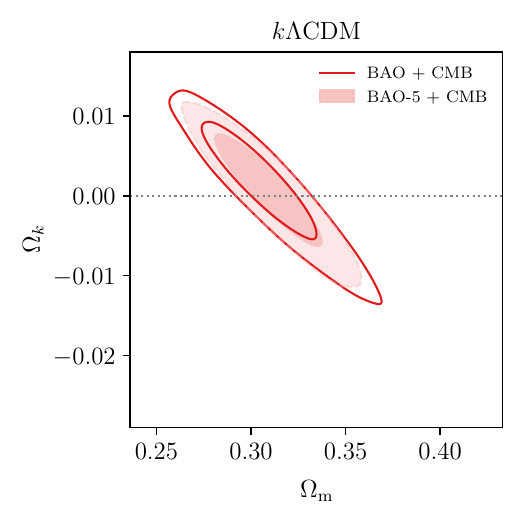}
    \end{minipage}
    \vfill
    \begin{minipage}[b]{0.45\textwidth}
        \centering
        \includegraphics[width=\textwidth]{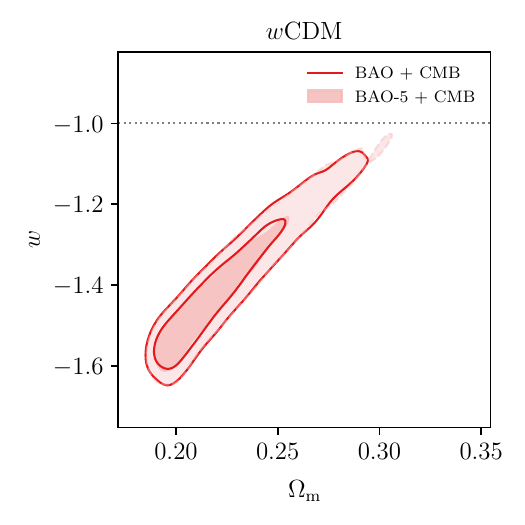}
    \end{minipage}
    \begin{minipage}[b]{0.45\textwidth}
        \centering
        \includegraphics[width=\textwidth]{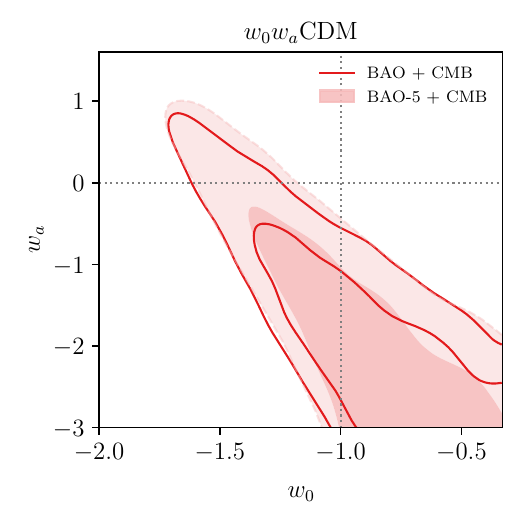}
    \end{minipage}
    \caption{Comparison of constraints from CMB in combination with the standard DES BAO constraint (BAO: 1 $D_M/r_d$ point in $0.6<\zph<1.2$) and the combination with the alternative BAO analysis based on 5 individual bins (BAO-5: $D_M/r_d$ fitted in each of the $\Delta\zph=0.1$ bins). We show constraints in \lcdm, \kcdm, \wcdm, and \wacdm, finding very small differences. }
    \label{fig:BAO5}
\end{figure*}

Once we have set up the alternative individual bin likelihood in \appendixcite{app:5bins-method}, we can re-run our chains. We will simply focus on  BAO+CMB, where the impact of BAO is expected to be more notable. Even in this case, in \autoref{fig:BAO5}, we barely see any difference with respect to the standard BAO likelihood. In other cases that were explored, but not shown, we also found a negligible effect of swapping the BAO likelihoods. 
Hence, we conclude that the different $\alpha$ values preferred by different redshift bins are likely due to statistical fluctuations (note that in \cite{des-y6-bao} we already found $\alpha_i$ redshift fluctuations consistent with that of the 1952 mocks) rather than hinting to an expansion history different to that preferred by the single BAO case. 

\bibliography{DES_cosmo_BAO}

@ARTICLE{Flaugher,
       author = {{Flaugher}, B. and {Diehl}, H.~T. and {Honscheid}, K. and others},         collaboration = {DES Collaboration},
        title = "{The Dark Energy Camera}",
      journal = {\aj},
         year = 2015,
        month = nov,
       volume = {150},
       number = {5},
          eid = {150},
        pages = {150},
          doi = {10.1088/0004-6256/150/5/150},
archivePrefix = {arXiv},
       eprint = {1504.02900},
 primaryClass = {astro-ph.IM},
       adsurl = {https://ui.adsabs.harvard.edu/abs/2015AJ....150..150F}
}

@article{pippin,
       author = {{Hinton}, Samuel and {Brout}, Dillon},
        title = "{Pippin: A pipeline for supernova cosmology}",
      journal = {The Journal of Open Source Software},
         year = 2020,
        month = mar,
       volume = {5},
       number = {47},
          eid = {2122},
        pages = {2122},
          doi = {10.21105/joss.02122},
       adsurl = {https://ui.adsabs.harvard.edu/abs/2020JOSS....5.2122H}
}

@article{Kessler_2009,
       author = {{Kessler}, Richard and {Bernstein}, Joseph P. and {Cinabro}, David and {Dilday}, Benjamin and {Frieman}, Joshua A. and {Jha}, Saurabh and {Kuhlmann}, Stephen and {Miknaitis}, Gajus and {Sako}, Masao and {Taylor}, Matt and {Vanderplas}, Jake},
        title = "{SNANA: A Public Software Package for Supernova Analysis}",
      journal = {\pasp},
         year = 2009,
        month = sep,
       volume = {121},
       number = {883},
        pages = {1028},
          doi = {10.1086/605984},
archivePrefix = {arXiv},
       eprint = {0908.4280},
 primaryClass = {astro-ph.CO},
       adsurl = {https://ui.adsabs.harvard.edu/abs/2009PASP..121.1028K}
}

@ARTICLE{DESoverview,
       author = {{DES Collaboration} },
        title = "{The Dark Energy Survey: more than dark energy -- an overview}",
      journal = {\mnras},
         year = 2016,
        month = aug,
       volume = {460},
       number = {2},
        pages = {1270-1299},
          doi = {10.1093/mnras/stw641},
archivePrefix = {arXiv},
       eprint = {1601.00329},
 primaryClass = {astro-ph.CO},
       adsurl = {https://ui.adsabs.harvard.edu/abs/2016MNRAS.460.1270D}
}

@ARTICLE{DES_all_probes_Y1,
       author = {{DES Collaboration}
                },
        title = "{Cosmological Constraints from Multiple Probes in the Dark Energy Survey}",
      journal = {\prl},
         year = 2019,
        month = may,
       volume = {122},
       number = {17},
          eid = {171301},
        pages = {171301},
          doi = {10.1103/PhysRevLett.122.171301},
archivePrefix = {arXiv},
       eprint = {1811.02375},
 primaryClass = {astro-ph.CO},
       adsurl = {https://ui.adsabs.harvard.edu/abs/2019PhRvL.122q1301A}
}

@ARTICLE{des-y6-bao,
       author = {{DES Collaboration}},
        title = "{Dark Energy Survey: A 2.1\% measurement of the angular baryonic acoustic oscillation scale at redshift $z_{\rm eff}=0.85$ from the final dataset}",
      journal = {\prd},
         year = 2024,
        month = sep,
       volume = {110},
       number = {6},
          eid = {063515},
        pages = {063515},
          doi = {10.1103/PhysRevD.110.063515},
archivePrefix = {arXiv},
       eprint = {2402.10696},
 primaryClass = {astro-ph.CO},
       adsurl = {https://ui.adsabs.harvard.edu/abs/2024PhRvD.110f3515A}
}

@ARTICLE{des-y6-bao-sample,
	    author = {{Mena-Fern\'andez}, J. and others},         collaboration = {DES Collaboration},
        title = "{Dark Energy Survey: Galaxy sample for the baryonic acoustic oscillation measurement from the final dataset}",
      journal = {\prd},
         year = 2024,
        month = sep,
       volume = {110},
       number = {6},
          eid = {063514},
        pages = {063514},
          doi = {10.1103/PhysRevD.110.063514},
archivePrefix = {arXiv},
       eprint = {2402.10697},
 primaryClass = {astro-ph.CO},
       adsurl = {https://ui.adsabs.harvard.edu/abs/2024PhRvD.110f3514M}
}

@ARTICLE{des-y5-sn,
       author = {{DES Collaboration}},
        title = "{The Dark Energy Survey: Cosmology Results with {\ensuremath{\sim}}1500 New High-redshift Type Ia Supernovae Using the Full 5 yr Data Set}",
      journal = {\apjl},
         year = 2024,
        month = sep,
       volume = {973},
       number = {1},
          eid = {L14},
        pages = {L14},
          doi = {10.3847/2041-8213/ad6f9f},
archivePrefix = {arXiv},
       eprint = {2401.02929},
       adsurl = {https://ui.adsabs.harvard.edu/abs/2024ApJ...973L..14D}
}

@article{DNF,
     author = {{De Vicente}, J. and {S{\'a}nchez}, E. and {Sevilla-Noarbe}, I.},
        title = "{DNF - Galaxy photometric redshift by Directional Neighbourhood Fitting}",
      journal = {\mnras},
         year = 2016,
        month = jul,
       volume = {459},
       number = {3},
        pages = {3078-3088},
          doi = {10.1093/mnras/stw857},
archivePrefix = {arXiv},
       eprint = {1511.07623},
 primaryClass = {astro-ph.CO},
       adsurl = {https://ui.adsabs.harvard.edu/abs/2016MNRAS.459.3078D}
}

@article{cawthon22,
	adsurl = {https://ui.adsabs.harvard.edu/abs/2022MNRAS.513.5517C},
	archiveprefix = {arXiv},
	author = {{Cawthon}, R. and others},
        collaboration = {DES Collaboration},
	doi = {10.1093/mnras/stac1160},
	eprint = {2012.12826},
	journal = {\mnras},
	month = jul,
	number = {4},
	pages = {5517-5539},
	primaryclass = {astro-ph.CO},
	title = {{Dark Energy Survey Year 3 results: calibration of lens sample redshift distributions using fclustering redshifts with BOSS/eBOSS}},
	volume = {513},
	year = 2022}

@ARTICLE{vipers,
       author = {{Guzzo}, L. and others},
        collaboration={VIPERS},
        title = "{The VIMOS Public Extragalactic Redshift Survey (VIPERS). An unprecedented view of galaxies and large-scale structure at 0.5 $< z <$ 1.2}",
      journal = {\aap},
         year = 2014,
        month = jun,
       volume = {566},
          eid = {A108},
        pages = {A108},
          doi = {10.1051/0004-6361/201321489},
archivePrefix = {arXiv},
       eprint = {1303.2623},
 primaryClass = {astro-ph.CO},
       adsurl = {https://ui.adsabs.harvard.edu/abs/2014A&A...566A.108G}
}

@article{ice-cola,
	adsurl = {https://ui.adsabs.harvard.edu/abs/2016MNRAS.459.2327I},
	archiveprefix = {arXiv},
	author = {{Izard}, Albert and {Crocce}, Martin and {Fosalba}, Pablo},
	doi = {10.1093/mnras/stw797},
	eprint = {1509.04685},
	journal = {\mnras},
	month = {Jul},
	number = {3},
	pages = {2327-2341},
	primaryclass = {astro-ph.CO},
	title = {{ICE-COLA: towards fast and accurate synthetic galaxy catalogues optimizing a quasi-N-body method}},
	volume = {459},
	year = {2016}}

@article{cola,
	adsurl = {https://ui.adsabs.harvard.edu/abs/2016MNRAS.459.2118K},
	archiveprefix = {arXiv},
	author = {{Koda}, Jun and {Blake}, Chris and {Beutler}, Florian and {Kazin}, Eyal and {Marin}, Felipe},
	doi = {10.1093/mnras/stw763},
	eprint = {1507.05329},
	journal = {\mnras},
	month = {Jun},
	number = {2},
	pages = {2118-2129},
	primaryclass = {astro-ph.CO},
	title = {{Fast and accurate mock catalogue generation for low-mass galaxies}},
	volume = {459},
	year = {2016}}

@article{y3-baomocks,
	author = {{Ferrero}, I. and others},
        collaboration = {DES Collaboration},
        title = "{Dark Energy Survey Year 3 Results: Galaxy mock catalogs for BAO analysis}",
      journal = {\aap},
         year = 2021,
        month = dec,
       volume = {656},
          eid = {A106},
        pages = {A106},
          doi = {10.1051/0004-6361/202141744},
archivePrefix = {arXiv},
       eprint = {2107.04602},
 primaryClass = {astro-ph.CO},
       adsurl = {https://ui.adsabs.harvard.edu/abs/2021A&A...656A.106F}
}

@ARTICLE{y1_lss_sys_kp,
       author = {{Elvin-Poole} and others},         collaboration = {DES Collaboration},
        title = "{Dark Energy Survey year 1 results: Galaxy clustering for combined probes}",
      journal = {\prd},
         year = 2018,
        month = aug,
       volume = {98},
       number = {4},
          eid = {042006},
        pages = {042006},
          doi = {10.1103/PhysRevD.98.042006},
archivePrefix = {arXiv},
       eprint = {1708.01536},
 primaryClass = {astro-ph.CO},
       adsurl = {https://ui.adsabs.harvard.edu/abs/2018PhRvD..98d2006E}
}

@article{y3-galaxyclustering,
	author = {{Rodr{\'\i}guez-Monroy}, M. and others},         collaboration = {DES Collaboration},
        title = "{Dark Energy Survey Year 3 results: galaxy clustering and systematics treatment for lens galaxy samples}",
      journal = {\mnras},
         year = 2022,
        month = apr,
       volume = {511},
       number = {2},
        pages = {2665-2687},
          doi = {10.1093/mnras/stac104},
archivePrefix = {arXiv},
       eprint = {2105.13540},
 primaryClass = {astro-ph.CO},
       adsurl = {https://ui.adsabs.harvard.edu/abs/2022MNRAS.511.2665R}
}

@article{Visser_2004,
    author = {{Visser}, Matt},
        title = "{Jerk, snap and the cosmological equation of state}",
      journal = {Classical and Quantum Gravity},
         year = 2004,
        month = jun,
       volume = {21},
       number = {11},
        pages = {2603-2615},
          doi = {10.1088/0264-9381/21/11/006},
archivePrefix = {arXiv},
       eprint = {gr-qc/0309109},
 primaryClass = {gr-qc},
       adsurl = {https://ui.adsabs.harvard.edu/abs/2004CQGra..21.2603V}
}

@article{Zhang_2017,
       author = {{Zhang}, Ming-Jian and {Li}, Hong and {Xia}, Jun-Qing},
        title = "{What do we know about cosmography}",
      journal = {European Physical Journal C},
         year = 2017,
        month = jul,
       volume = {77},
       number = {7},
          eid = {434},
        pages = {434},
          doi = {10.1140/epjc/s10052-017-5005-4},
       adsurl = {https://ui.adsabs.harvard.edu/abs/2017EPJC...77..434Z}
}

@article{Lemos23,
  title = {{CMB} constraints on the early Universe independent of late-time cosmology},
  author = {Lemos, Pablo and Lewis, Antony},
  journal = {\prd},
  volume = {107},
  issue = {10},
  pages = {103505},
  numpages = {9},
  year = {2023},
  month = {May},
  publisher = {American Physical Society},
  doi = {10.1103/PhysRevD.107.103505},
  url = {https://link.aps.org/doi/10.1103/PhysRevD.107.103505}
}

@ARTICLE{Camilleri24,
       author = {{Camilleri}, R. and others},         collaboration = {DES Collaboration}, 
 title = "{The Dark Energy Survey Supernova Program: an updated measurement of the Hubble constant using the inverse distance ladder}",
      journal = {\mnras},
         year = 2025,
        month = feb,
       volume = {537},
       number = {2},
        pages = {1818-1825},
          doi = {10.1093/mnras/staf122},
archivePrefix = {arXiv},
       eprint = {2406.05049},
 primaryClass = {astro-ph.CO},
       adsurl = {https://ui.adsabs.harvard.edu/abs/2025MNRAS.537.1818C}
}

@ARTICLE{Camilleri24b,
       author = {{Camilleri}, R. and others},         collaboration = {DES Collaboration},
        title = "{The dark energy survey supernova program: investigating beyond-{\ensuremath{\Lambda}}CDM}",
      journal = {\mnras},
         year = 2024,
        month = sep,
       volume = {533},
       number = {3},
        pages = {2615-2639},
          doi = {10.1093/mnras/stae1988},
archivePrefix = {arXiv},
       eprint = {2406.05048},
 primaryClass = {astro-ph.CO},
       adsurl = {https://ui.adsabs.harvard.edu/abs/2024MNRAS.533.2615C}
}

@article{Aubourg_2015,
       author = {{BOSS Collaboration}},
  title = "{Cosmological implications of baryon acoustic oscillation measurements}",
      journal = {\prd},
         year = 2015,
        month = dec,
       volume = {92},
       number = {12},
          eid = {123516},
        pages = {123516},
          doi = {10.1103/PhysRevD.92.123516},
archivePrefix = {arXiv},
       eprint = {1411.1074},
 primaryClass = {astro-ph.CO},
       adsurl = {https://ui.adsabs.harvard.edu/abs/2015PhRvD..92l3516A}
}

@ARTICLE{DES05,
       author = {{DES Collaboration}},
        title = "{The Dark Energy Survey}",
      journal = {arXiv e-prints},
         year = 2005,
        month = oct,
          doi = {10.48550/arXiv.astro-ph/0510346},
       eprint = {astro-ph/0510346},
 primaryClass = {astro-ph},
       adsurl = {https://ui.adsabs.harvard.edu/abs/2005astro.ph.10346T}
}

@article{Planck,
	adsurl = {https://ui.adsabs.harvard.edu/abs/2020A&A...641A...6P},
	archiveprefix = {arXiv},
	author = {{Planck Collaboration}},
	doi = {10.1051/0004-6361/201833910},
	eid = {A6},
	eprint = {1807.06209},
	journal = {\aap},
	month = sep,
	pages = {A6},
	primaryclass = {astro-ph.CO},
	title = {{Planck 2018 results. VI. Cosmological parameters}},
	volume = {641},
	year = 2020}

@ARTICLE{plancklike,
       author = {{Planck Collaboration}},
        title = "{Planck 2018 results. V. CMB power spectra and likelihoods}",
      journal = {\aap},
         year = 2020,
        month = sep,
       volume = {641},
          eid = {A5},
        pages = {A5},
          doi = {10.1051/0004-6361/201936386},
archivePrefix = {arXiv},
       eprint = {1907.12875},
 primaryClass = {astro-ph.CO},
       adsurl = {https://ui.adsabs.harvard.edu/abs/2020A&A...641A...5P}
}

@ARTICLE{planck-lite-python,
       author = {{Prince}, Heather and {Dunkley}, Jo},
        title = "{Data compression in cosmology: A compressed likelihood for Planck data}",
      journal = {\prd},
         year = 2019,
        month = oct,
       volume = {100},
       number = {8},
          eid = {083502},
        pages = {083502},
          doi = {10.1103/PhysRevD.100.083502},
archivePrefix = {arXiv},
       eprint = {1909.05869},
 primaryClass = {astro-ph.CO},
       adsurl = {https://ui.adsabs.harvard.edu/abs/2019PhRvD.100h3502P}
}

@ARTICLE{DESDR2,
       author = {{DES Collaboration}},
        title = "{The Dark Energy Survey Data Release 2}",
      journal = {\apjs},
         year = 2021,
        month = aug,
       volume = {255},
       number = {2},
          eid = {20},
        pages = {20},
          doi = {10.3847/1538-4365/ac00b3},
archivePrefix = {arXiv},
       eprint = {2101.05765},
 primaryClass = {astro-ph.IM},
       adsurl = {https://ui.adsabs.harvard.edu/abs/2021ApJS..255...20A}
}

@ARTICLE{1100705,
       author = {{Akaike}, H.},
        title = "{A New Look at the Statistical Model Identification}",
      journal = {IEEE Transactions on Automatic Control},
         year = 1974,
        month = jan,
       volume = {19},
        pages = {716-723},
       adsurl = {https://ui.adsabs.harvard.edu/abs/1974ITAC...19..716A}
}

@article{Macaulay_2019,
	doi = {10.1093/mnras/stz978},
	url = {https://doi.org/10.1093%2Fmnras%2Fstz978},
	year = 2019,
	month = {apr},
	publisher = {Oxford University Press ({OUP})},
	volume = {486},
	number = {2},
	pages = {2184--2196},
	author = {E Macaulay and others},         collaboration = {DES Collaboration},
	title = "{First cosmological results using Type Ia supernovae from the Dark Energy Survey: measurement of the Hubble constant}",
	journal = {\mnras}
}

@ARTICLE{riess2022comprehensive,
       author = {{Riess}, Adam G. and {Yuan}, Wenlong and {Macri}, Lucas M. and {Scolnic}, Dan and {Brout}, Dillon and {Casertano}, Stefano and {Jones}, David O. and {Murakami}, Yukei and {Anand}, Gagandeep S. and {Breuval}, Louise and {Brink}, Thomas G. and {Filippenko}, Alexei V. and {Hoffmann}, Samantha and {Jha}, Saurabh W. and {D'arcy Kenworthy}, W. and {Mackenty}, John and {Stahl}, Benjamin E. and {Zheng}, WeiKang},
        title = "{A Comprehensive Measurement of the Local Value of the Hubble Constant with 1 km s$^{-1}$ Mpc$^{-1}$ Uncertainty from the Hubble Space Telescope and the SH0ES Team}",
      journal = {\apjl},
         year = 2022,
        month = jul,
       volume = {934},
       number = {1},
          eid = {L7},
        pages = {L7},
          doi = {10.3847/2041-8213/ac5c5b},
archivePrefix = {arXiv},
       eprint = {2112.04510},
 primaryClass = {astro-ph.CO},
       adsurl = {https://ui.adsabs.harvard.edu/abs/2022ApJ...934L...7R}
}

@ARTICLE{Mena26,
       author = {{Mena-Fern{\'a}ndez}, J. and others},
        collaboration = {DES Collaboration},
        title = "{Dark Energy Survey: DESI-Independent Angular BAO Measurement}",
      journal = {arXiv e-prints},
         year = 2026,
        month = jan,
          eid = {arXiv:2601.14864},
        pages = {arXiv:2601.14864},
          doi = {10.48550/arXiv.2601.14864},
archivePrefix = {arXiv},
       eprint = {2601.14864},
 primaryClass = {astro-ph.CO},
       adsurl = {https://ui.adsabs.harvard.edu/abs/2026arXiv260114864M}
}

@article{cooke2018one,
  title={One percent determination of the primordial deuterium abundance},
  author={Cooke, Ryan J and Pettini, Max and Steidel, Charles C},
  journal={\apj},
  volume={855},
  number={2},
  pages={102},
  year={2018},
  publisher={IOP Publishing}
}

@ARTICLE{Dovekie,
        author = {{Popovic}, B. and {Shah}, P. and others},         collaboration = {DES Collaboration},
        title = "{The Dark Energy Survey Supernova Program: A Reanalysis Of Cosmology Results And Evidence For Evolving Dark Energy With An Updated Type Ia Supernova Calibration}",
      journal = {arXiv e-prints},
         year = 2025,
        month = nov,
          eid = {arXiv:2511.07517},
        pages = {arXiv:2511.07517},
          doi = {10.48550/arXiv.2511.07517},
archivePrefix = {arXiv},
       eprint = {2511.07517},
 primaryClass = {astro-ph.CO},
       adsurl = {https://ui.adsabs.harvard.edu/abs/2025arXiv251107517P}
}

@ARTICLE{aver2022comprehensive,
       author = {{Aver}, Erik and {Berg}, Danielle A. and {Hirschauer}, Alec S. and {Olive}, Keith A. and {Pogge}, Richard W. and {Rogers}, Noah S.~J. and {Salzer}, John J. and {Skillman}, Evan D.},
        title = "{A comprehensive chemical abundance analysis of the extremely metal poor Leoncino Dwarf galaxy (AGC 198691)}",
      journal = {\mnras},
         year = 2022,
        month = feb,
       volume = {510},
       number = {1},
        pages = {373-382},
          doi = {10.1093/mnras/stab3226},
archivePrefix = {arXiv},
       eprint = {2109.00178},
 primaryClass = {astro-ph.GA},
       adsurl = {https://ui.adsabs.harvard.edu/abs/2022MNRAS.510..373A}
}

@ARTICLE{schoneberg20242024,
       author = {{Sch{\"o}neberg}, Nils},
        title = "{The 2024 BBN baryon abundance update}",
      journal = {\jcap},
         year = 2024,
        month = jun,
       volume = {2024},
       number = {6},
          eid = {006},
        pages = {006},
          doi = {10.1088/1475-7516/2024/06/006},
archivePrefix = {arXiv},
       eprint = {2401.15054},
 primaryClass = {astro-ph.CO},
       adsurl = {https://ui.adsabs.harvard.edu/abs/2024JCAP...06..006S}
}

@ARTICLE{desi2024iv,
       author = {{DESI Collaboration}},
     title = "{DESI 2024 IV: Baryon Acoustic Oscillations from the Lyman alpha forest}",
      journal = {\jcap},
         year = 2025,
        month = jan,
       volume = {2025},
       number = {1},
          eid = {124},
        pages = {124},
          doi = {10.1088/1475-7516/2025/01/124},
archivePrefix = {arXiv},
       eprint = {2404.03001},
 primaryClass = {astro-ph.CO},
       adsurl = {https://ui.adsabs.harvard.edu/abs/2025JCAP...01..124A}
}

@article{desi2024vi,
  author = {{DESI Collaboration}},
       title = "{DESI 2024 VI: cosmological constraints from the measurements of baryon acoustic oscillations}",
      journal = {\jcap},
         year = 2025,
        month = feb,
       volume = {2025},
       number = {2},
          eid = {021},
        pages = {021},
          doi = {10.1088/1475-7516/2025/02/021},
archivePrefix = {arXiv},
       eprint = {2404.03002},
 primaryClass = {astro-ph.CO},
       adsurl = {https://ui.adsabs.harvard.edu/abs/2025JCAP...02..021A}
}

@article{desi2024iii,
  title="{DESI 2024 III: Baryon Acoustic Oscillations from Galaxies and Quasars}",
  author = {{DESI Collaboration}},
  journal={arXiv preprint arXiv:2404.03000},
  year={2024}
}

@ARTICLE{bossdr12,
       author = {{Alam}, Shadab and {BOSS Collaboration} },
        title = "{The clustering of galaxies in the completed SDSS-III Baryon Oscillation Spectroscopic Survey: cosmological analysis of the DR12 galaxy sample}",
      journal = {\mnras},
         year = 2017,
        month = sep,
       volume = {470},
       number = {3},
        pages = {2617-2652},
          doi = {10.1093/mnras/stx721},
archivePrefix = {arXiv},
       eprint = {1607.03155},
 primaryClass = {astro-ph.CO},
       adsurl = {https://ui.adsabs.harvard.edu/abs/2017MNRAS.470.2617A}
}

@ARTICLE{negativemnu_green_meyers,
       author = {{Green}, Daniel and {Meyers}, Joel},
        title = "{The Cosmological Preference for Negative Neutrino Mass}",
      journal = {arXiv e-prints},
         year = 2024,
        month = jul,
          eid = {arXiv:2407.07878},
        pages = {arXiv:2407.07878},
          doi = {10.48550/arXiv.2407.07878},
archivePrefix = {arXiv},
       eprint = {2407.07878},
 primaryClass = {astro-ph.CO},
       adsurl = {https://ui.adsabs.harvard.edu/abs/2024arXiv240707878G}
}

@ARTICLE{ross2015information,
       author = {{Ross}, Ashley J. and {Percival}, Will J. and {Manera}, Marc},
        title = "{The information content of anisotropic Baryon Acoustic Oscillation scale measurements}",
      journal = {\mnras},
         year = 2015,
        month = aug,
       volume = {451},
       number = {2},
        pages = {1331-1340},
          doi = {10.1093/mnras/stv966},
archivePrefix = {arXiv},
       eprint = {1501.05571},
 primaryClass = {astro-ph.CO},
       adsurl = {https://ui.adsabs.harvard.edu/abs/2015MNRAS.451.1331R}
}

@ARTICLE{Sanchez24,
       author = {{S{\'a}nchez}, B.~O. and others},         collaboration = {DES Collaboration},
        title = "{The Dark Energy Survey Supernova Program: Light Curves and 5 Yr Data Release}",
      journal = {\apj},
         year = 2024,
        month = nov,
       volume = {975},
       number = {1},
          eid = {5},
        pages = {5},
          doi = {10.3847/1538-4357/ad739a},
archivePrefix = {arXiv},
       eprint = {2406.05046},
 primaryClass = {astro-ph.CO},
       adsurl = {https://ui.adsabs.harvard.edu/abs/2024ApJ...975....5S}
}

@ARTICLE{Verde24,
       author = {{Verde}, Licia and {Sch{\"o}neberg}, Nils and {Gil-Mar{\'\i}n}, H{\'e}ctor},
        title = "{A Tale of Many H $_{0}$}",
      journal = {\araa},
         year = 2024,
        month = sep,
       volume = {62},
       number = {1},
        pages = {287-331},
          doi = {10.1146/annurev-astro-052622-033813},
archivePrefix = {arXiv},
       eprint = {2311.13305},
 primaryClass = {astro-ph.CO},
       adsurl = {https://ui.adsabs.harvard.edu/abs/2024ARA&A..62..287V}
}

@ARTICLE{Tang24,
       author = {{Tang}, Xianzhe TZ and {Brout}, Dillon and {Karwal}, Tanvi and {Chang}, Chihway and {Miranda}, Vivian and {Vincenzi}, Maria},
        title = "{Uniting the Observed Dynamical Dark Energy Preference with the Discrepancies in $\Omega_m$ and $H_0$ Across Cosmological Probes}",
      journal = {arXiv e-prints},
         year = 2024,
        month = dec,
          eid = {arXiv:2412.04430},
        pages = {arXiv:2412.04430},
          doi = {10.48550/arXiv.2412.04430},
archivePrefix = {arXiv},
       eprint = {2412.04430},
 primaryClass = {astro-ph.CO},
       adsurl = {https://ui.adsabs.harvard.edu/abs/2024arXiv241204430T}
}

@ARTICLE{SuperNNova,
       author = {{M{\"o}ller}, A. and {de Boissi{\`e}re}, T.},
        title = "{SuperNNova: an open-source framework for Bayesian, neural network-based supernova classification}",
      journal = {\mnras},
         year = 2020,
        month = jan,
       volume = {491},
       number = {3},
        pages = {4277-4293},
          doi = {10.1093/mnras/stz3312},
archivePrefix = {arXiv},
       eprint = {1901.06384},
 primaryClass = {astro-ph.IM},
       adsurl = {https://ui.adsabs.harvard.edu/abs/2020MNRAS.491.4277M}
}

@ARTICLE{desi2024vii,
       author = {{DESI Collaboration}},
        title = "{DESI 2024 VII: Cosmological Constraints from the Full-Shape Modeling of Clustering Measurements}",
      journal = {arXiv e-prints},
         year = 2024,
        month = nov,
          eid = {arXiv:2411.12022},
        pages = {arXiv:2411.12022},
          doi = {10.48550/arXiv.2411.12022},
archivePrefix = {arXiv},
       eprint = {2411.12022},
 primaryClass = {astro-ph.CO},
       adsurl = {https://ui.adsabs.harvard.edu/abs/2024arXiv241112022D}
}

@ARTICLE{tensions,
       author = {{Raveri}, Marco and {Doux}, Cyrille},
        title = "{Non-Gaussian estimates of tensions in cosmological parameters}",
      journal = {\prd},
         year = 2021,
        month = aug,
       volume = {104},
       number = {4},
          eid = {043504},
        pages = {043504},
          doi = {10.1103/PhysRevD.104.043504},
archivePrefix = {arXiv},
       eprint = {2105.03324},
 primaryClass = {astro-ph.CO},
       adsurl = {https://ui.adsabs.harvard.edu/abs/2021PhRvD.104d3504R}
}

@ARTICLE{lemosraveri_tensions,
       author = {{Lemos}, P. and {Raveri}, M. and others},
    collaboration =    {DES Collaboration},
        title = "{Assessing tension metrics with dark energy survey and Planck data}",
      journal = {\mnras},
         year = 2021,
        month = aug,
       volume = {505},
       number = {4},
        pages = {6179-6194},
          doi = {10.1093/mnras/stab1670},
archivePrefix = {arXiv},
       eprint = {2012.09554},
 primaryClass = {astro-ph.CO},
       adsurl = {https://ui.adsabs.harvard.edu/abs/2021MNRAS.505.6179L}
}

@article{PhysRevD.99.043506,
     author = {{Raveri}, Marco and {Hu}, Wayne},
        title = "{Concordance and discordance in cosmology}",
      journal = {\prd},
         year = 2019,
        month = feb,
       volume = {99},
       number = {4},
          eid = {043506},
        pages = {043506},
          doi = {10.1103/PhysRevD.99.043506},
archivePrefix = {arXiv},
       eprint = {1806.04649},
 primaryClass = {astro-ph.CO},
       adsurl = {https://ui.adsabs.harvard.edu/abs/2019PhRvD..99d3506R}
}

@ARTICLE{Kelsey23,
       author = {{Kelsey}, L.  and others},         collaboration = {DES Collaboration},
        title = "{Concerning colour: The effect of environment on type Ia supernova colour in the dark energy survey}",
      journal = {\mnras},
         year = 2023,
        month = feb,
       volume = {519},
       number = {2},
        pages = {3046-3063},
          doi = {10.1093/mnras/stac3711},
archivePrefix = {arXiv},
       eprint = {2208.01357},
 primaryClass = {astro-ph.CO},
       adsurl = {https://ui.adsabs.harvard.edu/abs/2023MNRAS.519.3046K}
}

@article{y3-3x2ptkp,
	author = {{DES Collaboration}},
      title = "{Dark Energy Survey Year 3 results: Cosmological constraints from galaxy clustering and weak lensing}",
      journal = {\prd},
         year = 2022,
        month = jan,
       volume = {105},
       number = {2},
          eid = {023520},
        pages = {023520},
          doi = {10.1103/PhysRevD.105.023520},
archivePrefix = {arXiv},
       eprint = {2105.13549},
 primaryClass = {astro-ph.CO},
       adsurl = {https://ui.adsabs.harvard.edu/abs/2022PhRvD.105b3520A}
}

@article{y3-3x2ptkp_ext,
	author = {{DES Collaboration}},
title = "{Dark Energy Survey Year 3 results: Constraints on extensions to {\ensuremath{\Lambda}} CDM with weak lensing and galaxy clustering}",
      journal = {\prd},
         year = 2023,
        month = apr,
       volume = {107},
       number = {8},
          eid = {083504},
        pages = {083504},
          doi = {10.1103/PhysRevD.107.083504},
archivePrefix = {arXiv},
       eprint = {2207.05766},
 primaryClass = {astro-ph.CO},
       adsurl = {https://ui.adsabs.harvard.edu/abs/2023PhRvD.107h3504A}
}

@ARTICLE{des-y1-bao,
       author = {{DES Collaboration}},
        title = "{Dark Energy Survey Year 1 Results: Measurement of the Baryon Acoustic Oscillation scale in the distribution of galaxies to redshift 1}",
      journal = {\mnras},
         year = 2019,
        month = mar,
       volume = {483},
       number = {4},
        pages = {4866-4883},
          doi = {10.1093/mnras/sty3351},
archivePrefix = {arXiv},
       eprint = {1712.06209},
 primaryClass = {astro-ph.CO},
       adsurl = {https://ui.adsabs.harvard.edu/abs/2019MNRAS.483.4866A}
}

@article{y3-cosmicshear1,
        author = {{Amon}, A. and others},         collaboration = {DES Collaboration},
       title = "{Dark Energy Survey Year 3 results: Cosmology from cosmic shear and robustness to data calibration}",
  journal = {\prd},
  year = 2022,
  month = jan,
  volume = {105},
  number = {2},
  eid = {023514},
  pages = {023514},
  doi = {10.1103/PhysRevD.105.023514},
  archivePrefix = {arXiv},
  eprint = {2105.13543},
  primaryClass = {astro-ph.CO},
  adsurl = {https://ui.adsabs.harvard.edu/abs/2022PhRvD.105b3514A}
        }

@article{y3-cosmicshear2,
        author = {{Secco}, L.~F. and {Samuroff}, S. and others},         collaboration = {DES Collaboration},
         title = "{Dark Energy Survey Year 3 results: Cosmology from cosmic shear and robustness to modeling uncertainty}",
 journal = {\prd},
 year = 2022,
 month = jan,
 volume = {105},
 number = {2},
 eid = {023515},
 pages = {023515},
 doi = {10.1103/PhysRevD.105.023515},
 archivePrefix = {arXiv},
 eprint = {2105.13544},
 primaryClass = {astro-ph.CO},
 adsurl = {https://ui.adsabs.harvard.edu/abs/2022PhRvD.105b3515S}
        }

@article{y3-2x2ptaltlensresults,
        author = {{Porredon}, A. and others},         collaboration = {DES Collaboration},
        title = "{Dark Energy Survey Year 3 results: Cosmological constraints from galaxy clustering and galaxy-galaxy lensing using the MagLim lens sample}",
      journal = {\prd},
         year = 2022,
        month = nov,
       volume = {106},
       number = {10},
          eid = {103530},
        pages = {103530},
          doi = {10.1103/PhysRevD.106.103530},
archivePrefix = {arXiv},
       eprint = {2105.13546},
 primaryClass = {astro-ph.CO},
       adsurl = {https://ui.adsabs.harvard.edu/abs/2022PhRvD.106j3530P}	
        }

@ARTICLE{y3-2x2ptredmagic,
       author = {{Pandey}, S. and others},         collaboration = {DES Collaboration},
        title = "{Dark Energy Survey year 3 results: Constraints on cosmological parameters and galaxy-bias models from galaxy clustering and galaxy-galaxy lensing using the redMaGiC sample}",
      journal = {\prd},
         year = 2022,
        month = aug,
       volume = {106},
       number = {4},
          eid = {043520},
        pages = {043520},
          doi = {10.1103/PhysRevD.106.043520},
archivePrefix = {arXiv},
       eprint = {2105.13545},
 primaryClass = {astro-ph.CO},
       adsurl = {https://ui.adsabs.harvard.edu/abs/2022PhRvD.106d3520P}
}

@ARTICLE{y1-clusters-kp,
       author = { {DES Collaboration}},
        title = "{Dark Energy Survey Year 1 Results: Cosmological constraints from cluster abundances and weak lensing}",
      journal = {\prd},
         year = 2020,
        month = jul,
       volume = {102},
       number = {2},
          eid = {023509},
        pages = {023509},
          doi = {10.1103/PhysRevD.102.023509},
archivePrefix = {arXiv},
       eprint = {2002.11124},
 primaryClass = {astro-ph.CO},
       adsurl = {https://ui.adsabs.harvard.edu/abs/2020PhRvD.102b3509A}
}

@ARTICLE{wiseman20,
       author = {{Wiseman}, P. and others},         collaboration = {DES Collaboration},
        title = "{Supernova host galaxies in the dark energy survey: I. Deep coadds, photometry, and stellar masses}",
      journal = {\mnras},
         year = 2020,
        month = jul,
       volume = {495},
       number = {4},
        pages = {4040-4060},
          doi = {10.1093/mnras/staa1302},
archivePrefix = {arXiv},
       eprint = {2001.02640},
 primaryClass = {astro-ph.GA},
       adsurl = {https://ui.adsabs.harvard.edu/abs/2020MNRAS.495.4040W}
}

@ARTICLE{Notari24,
       author = {{Notari}, Alessio and {Redi}, Michele and {Tesi}, Andrea},
        title = "{BAO vs. SN evidence for evolving dark energy}",
      journal = {arXiv e-prints},
         year = 2024,
        month = nov,
          eid = {arXiv:2411.11685},
        pages = {arXiv:2411.11685},
          doi = {10.48550/arXiv.2411.11685},
archivePrefix = {arXiv},
       eprint = {2411.11685},
 primaryClass = {astro-ph.CO},
       adsurl = {https://ui.adsabs.harvard.edu/abs/2024arXiv241111685N}
}

@ARTICLE{DES-SN3YR,
       author = {{DES Collaboration}},
        title = "{First Cosmology Results using Type Ia Supernovae from the Dark Energy Survey: Constraints on Cosmological Parameters}",
      journal = {\apjl},
         year = 2019,
        month = feb,
       volume = {872},
       number = {2},
          eid = {L30},
        pages = {L30},
          doi = {10.3847/2041-8213/ab04fa},
archivePrefix = {arXiv},
       eprint = {1811.02374},
 primaryClass = {astro-ph.CO},
       adsurl = {https://ui.adsabs.harvard.edu/abs/2019ApJ...872L..30A}
}

@article{y3-baosample,
	author = {{Carnero Rosell}, A. and others},         collaboration = {DES Collaboration},
        title = "{Dark Energy Survey Year 3 results: galaxy sample for BAO measurement}",
      journal = {\mnras},
         year = 2022,
        month = jan,
       volume = {509},
       number = {1},
        pages = {778-799},
          doi = {10.1093/mnras/stab2995},
archivePrefix = {arXiv},
       eprint = {2107.05477},
 primaryClass = {astro-ph.CO},
       adsurl = {https://ui.adsabs.harvard.edu/abs/2022MNRAS.509..778C}
}

@article{des-y1-acf,
	adsurl = {https://ui.adsabs.harvard.edu/abs/2018MNRAS.480.3031C},
	archiveprefix = {arXiv},
	author = {{Chan}, K.~C. and others},         collaboration = {DES Collaboration},
	doi = {10.1093/mnras/sty2036},
	eprint = {1801.04390},
	journal = {\mnras},
	month = nov,
	number = {3},
	pages = {3031-3051},
	primaryclass = {astro-ph.CO},
	title = {{BAO from angular clustering: optimization and mitigation of theoretical systematics}},
	volume = {480},
	year = 2018}

@article{Chan21-pcf-method,
	author = {{Chan}, Kwan Chuen and {Ferrero}, Ismael and {Avila}, Santiago and {Ross}, Ashley J. and {Crocce}, Martin and {Gaztanaga}, Enrique},
        title = "{Clustering with general photo-z uncertainties: application to Baryon Acoustic Oscillations}",
      journal = {\mnras},
         year = 2022,
        month = apr,
       volume = {511},
       number = {3},
        pages = {3965-3982},
          doi = {10.1093/mnras/stac340},
archivePrefix = {arXiv},
       eprint = {2110.13332},
 primaryClass = {astro-ph.CO},
       adsurl = {https://ui.adsabs.harvard.edu/abs/2022MNRAS.511.3965C}
}

@ARTICLE{PCF_Y3_BAO,
       author = {{Chan}, K.~C. and {Avila}, S. and {Carnero Rosell}, A. and {Ferrero}, I. and {Elvin-Poole}, J. and {Sanchez}, E. and {Camacho}, H. and {Porredon}, A. and {Crocce}, M. and others},         collaboration = {DES Collaboration},
        title = "{Dark Energy Survey Year 3 results: Measurement of the baryon acoustic oscillations with three-dimensional clustering}",
      journal = {\prd},
         year = 2022,
        month = dec,
       volume = {106},
       number = {12},
          eid = {123502},
        pages = {123502},
          doi = {10.1103/PhysRevD.106.123502},
archivePrefix = {arXiv},
       eprint = {2210.05057},
 primaryClass = {astro-ph.CO},
       adsurl = {https://ui.adsabs.harvard.edu/abs/2022PhRvD.106l3502C}
}

@article{Ross-des-y1-pcf,
	archiveprefix = {arXiv},
	author = {Ross, Ashley J. and others},         collaboration = {DES Collaboration},
	doi = {10.1093/mnras/stx2120},
	eprint = {1705.05442},
	journal = {\mnras},
	number = {4},
	pages = {4456--4468},
	primaryclass = {astro-ph.CO},
	reportnumber = {FERMILAB-PUB-17-162-A-AE, IFT-UAM-CSIC-17-044},
	title = {{Optimized Clustering Estimators for BAO Measurements Accounting for Significant Redshift Uncertainty}},
	volume = {472},
	year = {2017}}

@article{DESY1baomocks,
	archiveprefix = {arXiv},
	author = {{Avila}, S. and others},         collaboration = {DES Collaboration},
	doi = {10.1093/mnras/sty1389},
	eprint = {1712.06232},
	journal = {\mnras},
	number = {1},
	pages = {94--110},
	primaryclass = {astro-ph.CO},
	reportnumber = {FERMILAB-PUB-17-587-A-AD-AE-SCD, IFT-UAM-CSIC-17-124, DES-2017-0292},
	title = {{Dark Energy Survey Year 1 Results: galaxy mock catalogues for BAO}},
	volume = {479},
	year = {2018}}

@article{Camacho-des-y1-aps,
	archiveprefix = {arXiv},
	author = {Camacho, H. and others},         collaboration = {DES Collaboration},
	doi = {10.1093/mnras/stz1514},
	eprint = {1807.10163},
	journal = {\mnras},
	number = {3},
	pages = {3870--3883},
	primaryclass = {astro-ph.CO},
	reportnumber = {FERMILAB-PUB-18-387-AE-CD},
	title = {{Dark Energy Survey Year 1 Results: Measurement of the Galaxy Angular Power Spectrum}},
	volume = {487},
	year = {2019}}

@article{DESY1baosample,
	archiveprefix = {arXiv},
	author = {{Crocce}, M. and others},         collaboration = {DES Collaboration},
	doi = {10.1093/mnras/sty2522},
	eprint = {1712.06211},
	journal = {\mnras},
	number = {2},
	pages = {2807--2822},
	primaryclass = {astro-ph.CO},
	reportnumber = {FERMILAB-PUB-17-585-A-AD-AE-SCD},
	title = {{Dark Energy Survey Year 1 Results: Galaxy Sample for BAO Measurement}},
	volume = {482},
	year = {2019}}

@ARTICLE{y3-baokp,
       author = {{DES Collaboration}},
        title = "{Dark Energy Survey Year 3 results: A 2.7\% measurement of baryon acoustic oscillation distance scale at redshift 0.835}",
      journal = {\prd},
         year = 2022,
        month = feb,
       volume = {105},
       number = {4},
          eid = {043512},
        pages = {043512},
          doi = {10.1103/PhysRevD.105.043512},
archivePrefix = {arXiv},
       eprint = {2107.04646},
 primaryClass = {astro-ph.CO},
       adsurl = {https://ui.adsabs.harvard.edu/abs/2022PhRvD.105d3512A}
}

@misc{vincenzi24,
      title={The Dark Energy Survey Supernova Program: Cosmological Analysis and Systematic Uncertainties}, 
      author={M. Vincenzi and others},         collaboration = {DES Collaboration},
      year={2024},
      eprint={2401.02945},
      archivePrefix={arXiv},
      primaryClass={astro-ph.CO}
}

@ARTICLE{CPL1,
       author = {{Chevallier}, Michel and {Polarski}, David},
        title = "{Accelerating Universes with Scaling Dark Matter}",
      journal = {International Journal of Modern Physics D},
         year = 2001,
        month = jan,
       volume = {10},
       number = {2},
        pages = {213-223},
          doi = {10.1142/S0218271801000822},
archivePrefix = {arXiv},
       eprint = {gr-qc/0009008},
 primaryClass = {gr-qc},
       adsurl = {https://ui.adsabs.harvard.edu/abs/2001IJMPD..10..213C}
}

@ARTICLE{CPL2,
       author = {{Linder}, Eric V.},
        title = "{Exploring the Expansion History of the Universe}",
      journal = {\prl},
         year = 2003,
        month = mar,
       volume = {90},
       number = {9},
          eid = {091301},
        pages = {091301},
          doi = {10.1103/PhysRevLett.90.091301},
archivePrefix = {arXiv},
       eprint = {astro-ph/0208512},
 primaryClass = {astro-ph},
       adsurl = {https://ui.adsabs.harvard.edu/abs/2003PhRvL..90i1301L}
}

@ARTICLE{brout22,
       author = {{Brout}, Dillon and {Taylor}, Georgie and {Scolnic}, Dan and {Wood}, Charlotte M. and {Rose}, Benjamin M. and {Vincenzi}, Maria and {Dwomoh}, Arianna and {Lidman}, Christopher and {Riess}, Adam and {Ali}, Noor and {Qu}, Helen and {Dai}, Mi},
        title = "{The Pantheon+ Analysis: SuperCal-fragilistic Cross Calibration, Retrained SALT2 Light-curve Model, and Calibration Systematic Uncertainty}",
      journal = {\apj},
         year = 2022,
        month = oct,
       volume = {938},
       number = {2},
          eid = {111},
        pages = {111},
          doi = {10.3847/1538-4357/ac8bcc},
archivePrefix = {arXiv},
       eprint = {2112.03864},
 primaryClass = {astro-ph.CO},
       adsurl = {https://ui.adsabs.harvard.edu/abs/2022ApJ...938..111B}
}

@ARTICLE{taylor23,
       author = {{Taylor}, G. and {Jones}, D.~O. and {Popovic}, B. and {Vincenzi}, M. and {Kessler}, R. and {Scolnic}, D. and {Dai}, M. and {Kenworthy}, W.~D. and {Pierel}, J.~D.~R.},
        title = "{SALT2 versus SALT3: updated model surfaces and their impacts on type Ia supernova cosmology}",
      journal = {\mnras},
         year = 2023,
        month = apr,
       volume = {520},
       number = {4},
        pages = {5209-5224},
          doi = {10.1093/mnras/stad320},
archivePrefix = {arXiv},
       eprint = {2301.10644},
 primaryClass = {astro-ph.CO},
       adsurl = {https://ui.adsabs.harvard.edu/abs/2023MNRAS.520.5209T}
}

@ARTICLE{moller22,
       author = {{M{\"o}ller}, A. and {DES Collaboration}
                },
        title = "{The dark energy survey 5-yr photometrically identified type Ia supernovae}",
      journal = {\mnras},
         year = 2022,
        month = aug,
       volume = {514},
       number = {4},
        pages = {5159-5177},
          doi = {10.1093/mnras/stac1691},
archivePrefix = {arXiv},
       eprint = {2201.11142},
 primaryClass = {astro-ph.CO},
       adsurl = {https://ui.adsabs.harvard.edu/abs/2022MNRAS.514.5159M}
}

@ARTICLE{loverde_weiner_neutrinos,
 author = {{Loverde}, Marilena and {Weiner}, Zachary J.},
        title = "{Massive neutrinos and cosmic composition}",
      journal = {\jcap},
         year = 2024,
        month = dec,
       volume = {2024},
       number = {12},
          eid = {048},
        pages = {048},
          doi = {10.1088/1475-7516/2024/12/048},
archivePrefix = {arXiv},
       eprint = {2410.00090},
 primaryClass = {astro-ph.CO},
       adsurl = {https://ui.adsabs.harvard.edu/abs/2024JCAP...12..048L}
}

@ARTICLE{Valcin2020,
       author = {{Valcin}, David and {Bernal}, Jos{\'e} Luis and {Jimenez}, Raul and {Verde}, Licia and {Wandelt}, Benjamin D.},
        title = "{Inferring the age of the universe with globular clusters}",
      journal = {\jcap},
         year = 2020,
        month = dec,
       volume = {2020},
       number = {12},
          eid = {002},
        pages = {002},
          doi = {10.1088/1475-7516/2020/12/002},
archivePrefix = {arXiv},
       eprint = {2007.06594},
 primaryClass = {astro-ph.CO},
       adsurl = {https://ui.adsabs.harvard.edu/abs/2020JCAP...12..002V}
}

@ARTICLE{Haselgrove1956,
       author = {{Haselgrove}, C.~B. and {Hoyle}, F.},
        title = "{A preliminary determination of the age of type II stars}",
      journal = {\mnras},
         year = 1956,
        month = jan,
       volume = {116},
        pages = {527},
          doi = {10.1093/mnras/116.5.527},
       adsurl = {https://ui.adsabs.harvard.edu/abs/1956MNRAS.116..527H}
}

@ARTICLE{Jimenez2019,
       author = {{Jimenez}, Raul and {Cimatti}, Andrea and {Verde}, Licia and {Moresco}, Michele and {Wandelt}, Benjamin},
        title = "{The local and distant Universe: stellar ages and H$_{0}$}",
      journal = {\jcap},
         year = 2019,
        month = mar,
       volume = {2019},
       number = {3},
          eid = {043},
        pages = {043},
          doi = {10.1088/1475-7516/2019/03/043},
archivePrefix = {arXiv},
       eprint = {1902.07081},
 primaryClass = {astro-ph.CO},
       adsurl = {https://ui.adsabs.harvard.edu/abs/2019JCAP...03..043J}
}

@ARTICLE{polychord2,
       author = {{Handley}, W.~J. and {Hobson}, M.~P. and {Lasenby}, A.~N.},
        title = "{POLYCHORD: next-generation nested sampling}",
      journal = {\mnras},
         year = 2015,
        month = nov,
       volume = {453},
       number = {4},
        pages = {4384-4398},
          doi = {10.1093/mnras/stv1911},
archivePrefix = {arXiv},
       eprint = {1506.00171},
 primaryClass = {astro-ph.IM},
       adsurl = {https://ui.adsabs.harvard.edu/abs/2015MNRAS.453.4384H}
}

@article{nautilus,
    author = {Lange, Johannes U},
    title = "{nautilus: boosting Bayesian importance nested sampling with deep learning}",
    journal = {\mnras},
    volume = {525},
    number = {2},
    pages = {3181-3194},
    year = {2023},
    month = {08},
    doi = {10.1093/mnras/stad2441},
    url = {https://doi.org/10.1093/mnras/stad2441},
    eprint = {https://academic.oup.com/mnras/article-pdf/525/2/3181/51331635/stad2441.pdf},
}

@ARTICLE{cosmosis,
       author = {{Zuntz}, J. and {Paterno}, M. and {Jennings}, E. and {Rudd}, D. and {Manzotti}, A. and {Dodelson}, S. and {Bridle}, S. and {Sehrish}, S. and {Kowalkowski}, J.},
        title = "{CosmoSIS: Modular cosmological parameter estimation}",
      journal = {Astronomy and Computing},
         year = 2015,
        month = sep,
       volume = {12},
        pages = {45-59},
          doi = {10.1016/j.ascom.2015.05.005},
archivePrefix = {arXiv},
       eprint = {1409.3409},
 primaryClass = {astro-ph.CO},
       adsurl = {https://ui.adsabs.harvard.edu/abs/2015A&C....12...45Z}
}

@ARTICLE{camb1,
       author = {{Howlett}, Cullan and {Lewis}, Antony and {Hall}, Alex and {Challinor}, Anthony},
        title = "{CMB power spectrum parameter degeneracies in the era of precision cosmology}",
      journal = {\jcap},
         year = 2012,
        month = apr,
       volume = {2012},
       number = {4},
          eid = {027},
        pages = {027},
          doi = {10.1088/1475-7516/2012/04/027},
archivePrefix = {arXiv},
       eprint = {1201.3654},
 primaryClass = {astro-ph.CO},
       adsurl = {https://ui.adsabs.harvard.edu/abs/2012JCAP...04..027H}
}

@ARTICLE{camb2,
       author = {{Lewis}, Antony and {Challinor}, Anthony and {Lasenby}, Anthony},
        title = "{Efficient Computation of Cosmic Microwave Background Anisotropies in Closed Friedmann-Robertson-Walker Models}",
      journal = {\apj},
         year = 2000,
        month = aug,
       volume = {538},
       number = {2},
        pages = {473-476},
          doi = {10.1086/309179},
archivePrefix = {arXiv},
       eprint = {astro-ph/9911177},
 primaryClass = {astro-ph},
       adsurl = {https://ui.adsabs.harvard.edu/abs/2000ApJ...538..473L}
}

@ARTICLE{halofit-takahashi,
       author = {{Takahashi}, Ryuichi and {Sato}, Masanori and {Nishimichi}, Takahiro and {Taruya}, Atsushi and {Oguri}, Masamune},
        title = "{Revising the Halofit Model for the Nonlinear Matter Power Spectrum}",
      journal = {\apj},
         year = 2012,
        month = dec,
       volume = {761},
       number = {2},
          eid = {152},
        pages = {152},
          doi = {10.1088/0004-637X/761/2/152},
archivePrefix = {arXiv},
       eprint = {1208.2701},
 primaryClass = {astro-ph.CO},
       adsurl = {https://ui.adsabs.harvard.edu/abs/2012ApJ...761..152T}
}

@article{halofit-bird,
    author = {{Bird}, Simeon and {Viel}, Matteo and {Haehnelt}, Martin G.},
        title = "{Massive neutrinos and the non-linear matter power spectrum}",
      journal = {\mnras},
         year = 2012,
        month = mar,
       volume = {420},
       number = {3},
        pages = {2551-2561},
          doi = {10.1111/j.1365-2966.2011.20222.x},
archivePrefix = {arXiv},
       eprint = {1109.4416},
 primaryClass = {astro-ph.CO},
       adsurl = {https://ui.adsabs.harvard.edu/abs/2012MNRAS.420.2551B}
}

@ARTICLE{getdist,
       author = {{Lewis}, Antony},
        title = "{GetDist: a Python package for analysing Monte Carlo samples}",
      journal = {arXiv e-prints},
         year = 2019,
        month = oct,
          eid = {arXiv:1910.13970},
        pages = {arXiv:1910.13970},
          doi = {10.48550/arXiv.1910.13970},
archivePrefix = {arXiv},
       eprint = {1910.13970},
 primaryClass = {astro-ph.IM},
       adsurl = {https://ui.adsabs.harvard.edu/abs/2019arXiv191013970L}
}

@ARTICLE{2021Handley_ok,
       author = {{Handley}, Will},
        title = "{Curvature tension: Evidence for a closed universe}",
      journal = {\prd},
         year = 2021,
        month = feb,
       volume = {103},
       number = {4},
          eid = {L041301},
        pages = {L041301},
          doi = {10.1103/PhysRevD.103.L041301},
archivePrefix = {arXiv},
       eprint = {1908.09139},
 primaryClass = {astro-ph.CO},
       adsurl = {https://ui.adsabs.harvard.edu/abs/2021PhRvD.103d1301H}
}

@ARTICLE{2021DiValentino_ok,
       author = {{Di Valentino}, Eleonora and {Melchiorri}, Alessandro and {Silk}, Joseph},
        title = "{Investigating Cosmic Discordance}",
      journal = {\apjl},
         year = 2021,
        month = feb,
       volume = {908},
       number = {1},
          eid = {L9},
        pages = {L9},
          doi = {10.3847/2041-8213/abe1c4},
archivePrefix = {arXiv},
       eprint = {2003.04935},
 primaryClass = {astro-ph.CO},
       adsurl = {https://ui.adsabs.harvard.edu/abs/2021ApJ...908L...9D}
}

@ARTICLE{2020Efstathiou_ok,
       author = {{Efstathiou}, George and {Gratton}, Steven},
        title = "{The evidence for a spatially flat Universe}",
      journal = {\mnras},
         year = 2020,
        month = jul,
       volume = {496},
       number = {1},
        pages = {L91-L95},
          doi = {10.1093/mnrasl/slaa093},
archivePrefix = {arXiv},
       eprint = {2002.06892},
 primaryClass = {astro-ph.CO},
       adsurl = {https://ui.adsabs.harvard.edu/abs/2020MNRAS.496L..91E}
}

@ARTICLE{2022Glanville_ok,
       author = {{Glanville}, Aaron and {Howlett}, Cullan and {Davis}, Tamara},
        title = "{Full-shape galaxy power spectra and the curvature tension}",
      journal = {\mnras},
         year = 2022,
        month = dec,
       volume = {517},
       number = {2},
        pages = {3087-3100},
          doi = {10.1093/mnras/stac2891},
archivePrefix = {arXiv},
       eprint = {2205.05892},
 primaryClass = {astro-ph.CO},
       adsurl = {https://ui.adsabs.harvard.edu/abs/2022MNRAS.517.3087G}
}

@article{Wilks1938,
author = {S. S. Wilks},
title = {{The Large-Sample Distribution of the Likelihood Ratio for Testing Composite Hypotheses}},
volume = {9},
journal = {The Annals of Mathematical Statistics},
number = {1},
publisher = {Institute of Mathematical Statistics},
pages = {60 -- 62},
year = {1938},
doi = {10.1214/aoms/1177732360},
URL = {https://doi.org/10.1214/aoms/1177732360}
}

@PREAMBLE{
 "\providecommand{\noopsort}[1]{}" 
 # "\providecommand{\singleletter}[1]{#1}%" 
}

\end{document}